\documentclass[12pt,preprint]{aastex}
\def\dfplot#1{\plotone{#1}}

\begin{document}

\title{The 2.5 m Telescope of the Sloan Digital Sky Survey} 

\author{
James E. Gunn\altaffilmark{1},
Walter A. Siegmund\altaffilmark{2},
Edward J. Mannery\altaffilmark{2},
Russell E. Owen\altaffilmark{2},
Charles L. Hull\altaffilmark{2},
R. French Leger\altaffilmark{3},
Larry N. Carey\altaffilmark{2},
Gillian R. Knapp\altaffilmark{1},
Donald G. York\altaffilmark{7},
William N. Boroski\altaffilmark{3},
Stephen M. Kent\altaffilmark{3},
Robert H. Lupton\altaffilmark{1},
Constance M. Rockosi\altaffilmark{4},
Michael L. Evans\altaffilmark{2},
Patrick Waddell\altaffilmark{2},
John E. Anderson\altaffilmark{3},
James Annis\altaffilmark{3},
John C. Barentine\altaffilmark{5},
Larry M. Bartoszek\altaffilmark{3},
Steven Bastian\altaffilmark{3},
Stephen B. Bracker\altaffilmark{3},
Howard J. Brewington\altaffilmark{5},
Charles I. Briegel\altaffilmark{3},
Jon Brinkmann\altaffilmark{5},
Yorke J. Brown\altaffilmark{6},
Michael A. Carr\altaffilmark{1},
Paul C. Czarapata\altaffilmark{3},
Craig C. Drennan\altaffilmark{3},
Thomas Dombeck\altaffilmark{7},
Glenn R. Federwitz\altaffilmark{3},
Bruce A. Gillespie\altaffilmark{5},
Carlos Gonzales\altaffilmark{3},
Sten U. Hansen\altaffilmark{3},
Michael Harvanek\altaffilmark{5},
Jeffrey Hayes\altaffilmark{5},
Wendell Jordan\altaffilmark{3},
Ellyne Kinney\altaffilmark{5},
Mark Klaene\altaffilmark{5},
S. J. Kleinman\altaffilmark{5},
Richard G. Kron\altaffilmark{7},
Jurek Kresinski\altaffilmark{5},
Glenn Lee\altaffilmark{3},
Siriluk Limmongkol\altaffilmark{2},
Carl W. Lindenmeyer\altaffilmark{3},
Daniel C. Long\altaffilmark{5},
Craig L. Loomis\altaffilmark{5},
Peregrine M. McGehee\altaffilmark{8},
Paul M. Mantsch\altaffilmark{3},
Eric H. Neilsen Jr.\altaffilmark{3},
Richard M. Neswold\altaffilmark{3},
Peter R. Newman\altaffilmark{5},
Atsuko Nitta\altaffilmark{5},
John Peoples Jr.\altaffilmark{3},
Jeffrey R. Pier\altaffilmark{9},
Peter S. Prieto\altaffilmark{3},
Angela Prosapio\altaffilmark{3},
Claudio Rivetta\altaffilmark{3},
Donald P. Schneider\altaffilmark{10},
Stephanie Snedden\altaffilmark{5},
Shu-i Wang\altaffilmark{7}
}
\altaffiltext{1}{Department of Astrophysical Sciences, Princeton University, 
Princeton, NJ 08544
\label{Princeton}}
\altaffiltext{2}{University of Washington, Department of 
Astronomy, Box 351580, Seattle, WA 98195
\label{Washington}}
\altaffiltext{3}{Fermi National Accelerator Laboratory, P.O. 
Box 500, Batavia, IL 60510
\label{Fermilab}}
\altaffiltext{4}{Lick Observatory, University of California, Santa Cruz,
CA 95064
\label{UCSC}}
\altaffiltext{5}{Apache Point Observatory, 2001 Apache Point Road, PO Box
59, Sunspot, NM 88349-0059
\label{APO}}
\altaffiltext{6}{Yorke J. Brown Scientific and Engineering Consulting, 
96 King Rd. Etna, NH 03750
\label{YJB}}
\altaffiltext{7}{Department of Astronomy and Astrophysics, The University
of Chicago, Chicago, IL 60637
\label{Chicago}}
\altaffiltext{8}{Los Alamos National Laboratory, LANSCE-8, MS H820,
Los Alamos, NM 87545
\label{LANL}}
\altaffiltext{9}{U. S. Naval Observatory, Flagstaff Station, 
10391 West Naval Observatory Road, Flagstaff, Arizona 86001-8521
\label{USNO}}
\altaffiltext{10}{Department of Astronomy and Astrophysics, The Pennsylvania
State University, 525 Davey Laboratory, University Park, PA 16802
\label{Pennstate}}
%

\begin{abstract}

We describe the design, construction, 
and performance of the Sloan Digital Sky Survey
Telescope located at Apache Point Observatory.  The telescope is a modified
two-corrector Ritchey-Chr\'etien design which
has a 2.5-m, $f$/2.25 primary, a 1.08-m secondary, a Gascoigne astigmatism
corrector, and one of a pair of interchangeable highly aspheric correctors
near the focal plane, one for imaging and the other for spectroscopy. 
The final focal
ratio is $f$/5. The telescope is instrumented by
a wide-area, multiband CCD camera and a pair
of fiber-fed double spectrographs.
Novel features of the telescope include:
(1)~A 3$^{\circ}$ diameter (0.65~m) focal plane that has excellent image quality
and small geometrical distortions over a wide wavelength range (3000~\AA\ to
10,600~\AA ) in the imaging mode, and good image quality combined with very
small lateral and longitudinal color errors in the spectroscopic mode.  
The unusual requirement of very low distortion is set by the demands of
time-delay-and-integrate (TDI) imaging.
(2)~Very high precision motion to support open loop
TDI observations; and (3)~A unique wind baffle/enclosure construction to
maximize image quality and minimize construction costs.  The telescope had
first light in May~1998 and began regular survey operations in~2000.
\end{abstract}

\section{Introduction}

The planning for the enterprise that would evolve into the Sloan Digital
Sky Survey (SDSS; York et al.~2000) began in the mid-1980s, when the
marriage of rapid
advances in solid-state detectors and the explosive increase in computational
processing and instrument control
capabilities suggested that it would be possible to carry out
a wide-area digital optical sky survey.  Digital surveys covering tens of square
degrees had indicated the potential power of
operating CCDs in the time-delay-and-integrate (TDI) or ``scanning" mode.
(see, e.g., Schmidt et al.~1986 for early scientific results and a
description of the technique.)
This mode combines extremely high observational efficiencies
(the time spent gathering data often
exceeds~90\% of the available observing time) and excellent calibration
of pixel-to-pixel sensitivity variations (the ``flat field") with the
high quantum efficiency, noise properties, and linearity of CCDs.

To execute a wide-area, multi-band imaging survey of a substantial part
of the celestial sphere, and to create spectroscopic galaxy and quasar samples
that exceed existing ones by an order of magnitude, required a number of
significant innovations: 

\begin{itemize}

\item
(1)~A large optically, 
mechanically, and electronically
complex camera to gather the imaging data and provide suitable astrometric
calibration.  The SDSS camera, which consists of 30 \hbox{2048 $\times$ 2048}
SITe/Tektronix CCDs and 24 \hbox{2048 $\times$ 400} CCDs, is described by
Gunn et al.~(1998). 

\item
(2)~A spectroscopic system that can
simultaneously obtain 640 spectra with broad wavelength coverage (3800-9200\AA)
and can be efficiently configured to each field in the sky.
The SDSS uses two fiber-fed double spectrographs (Uomoto et al.
1999); the first
SDSS spectroscopic observations are presented by Castander et al. (2001).

\item
(3)~A data acquisition system which can reliably store the incoming data
and provide a modicum of real-time analysis for quality control, focus, etc.

\item
(4)~A data processing system that can automatically and rapidly calibrate the
observations, identify objects, and measure their properties from the tens of
gigabytes of data produced in a typical night of imaging observations,
so that spectroscopic targets can be selected within a short time of
the acquisition of the imaging data.
The photometric calibration and object detection and characterization software,
called {\it Photo}, is described by Lupton et al.~(2001, 2002), 
and (in preparation) Lupton (2006); its outputs and their use are
described in the SDSS data release papers (Stoughton et al.~2002,
Abazajian et al.~2003, 2004, 2005, Finkbeiner et al.~2004).
The target selection code which selects objects
from the photometric catalogs for spectroscopic observation is
described in Strauss et. al 2002, Richards et al. 2002, Eisenstein et al. 2001,
and in the data release papers. The
outputs of the spectroscopic pipeline, which reduces the two-dimensional
CCD frames to one-dimensional spectrophotometrically calibrated spectra
and then analyzes the spectra for morphology and radial velocity are
described in the data release papers; there is not yet a published
description of the algorithmic structure of the pipeline.
The astrometric
calibration is described by Pier et al.~(2003).

\item
(5)~A telescope that
possesses a wide angle, extraordinarily low-distortion focal plane and extremely
accurate drives, and allows for rapid changes between imaging and spectroscopic 
modes in order to adapt to changing weather/seeing conditions.

\end{itemize}

The requirements
of TDI imaging present challenging hurdles for any optical/mechanical
design. First, if the sky is to be surveyed in an acceptably short
time, the size of the field must be quite large,
$\rm 3^{\circ}$ for the SDSS camera. That means that if the
instrument is to operate in TDI mode anywhere but the celestial equator
it cannot observe as a transit device but must scan along great circles.
The telescope drive system must be able to maintain precise rates and
direction without guiding information from celestial objects. Second, 
the optics must
not only provide excellent image quality over a large field in the entire
wavelength band covered by CCDs, but also constrain geometric
distortions in order to assure that the star images traverse the CCDs
accurately along their columns and at a rate which is constant over
the whole field.

To accomplish these goals, the SDSS team designed and constructed a dedicated
wide-field 2.5 m telescope.  This telescope, its instruments, the 0.5 m
photometric telescope which provides the photometric calibrations for SDSS
(Hogg et al. 2001, Tucker 2005) and the support facilities are 
located at Apache Point
Observatory (APO), New Mexico (Figure~1).
APO is 1 km south of the Sacramento Peak
site of the National Solar Observatory and is at an elevation of 2800 m.

The SDSS telescope carries out both imaging and multi-object spectroscopic
observations, scheduled and interleaved depending on the quality of the
weather. Thus the telescope must be able not only to support these 
observing modes, but to allow rapid change between imaging and spectroscopy.
The many innovative and unique features of the SDSS telescope design 
(and some of the associated pitfalls) are
the subject of this paper.  
Figure~2 presents an 
image of the telescope. Some of the
unusual aspects of the design are obvious at first glance: the entire telescope
enclosure is a retractable structure, and the telescope's wind and light
baffling is accomplished by a complex structure surrounding the telescope
but mechanically isolated from it.

The next section describes the optical design of the telescope, while \S3
discusses the engineering and mechanical design. The wind and light
baffles are described in \S4, and the telescope enclosure
in \S5. A brief history of the construction, installation,
commissioning and operations of the telescope is given in \S6, where
the future of the telescope is also discussed.
Some of the material in this paper has been presented 
in part in (and is sometimes taken word-for-word from)
the SDSS Project Book and technical papers by Hull,
Limmongkol \& Siegmund (1994), McCall \& Siegmund (1994), Siegmund et al.
(1998), Waddell et al. (1998), Carey et al. (2002), Long
(2002), and Leger et al. (2003a,b).

\section{Optical Design}

\subsection{General Considerations}
Initial desiderata for the project which was to become the SDSS
included the acquisition of order a million redshifts in a 
spatially contiguous volume as nearly spherical as possible, with
photometry in several bands covering the wavelength region accessible from
the ground with silicon CCDs (~3000-10000\AA). Crude optimization 
parameters such as telescope aperture, available field with acceptable
image quality, available optical fibers, and available (or soon-to-be available)
CCDs led to a telescope of approximately 2.5 meters aperture and focal
ratio approximately $f$/5, which provides a good match both to fibers
for spectroscopy (180 $\mu$m $\sim$ 3 arcsec) and to the imaging CCDs
(pixel size 24 $\mu$m $\sim$ 0.4 arcsec). It was desired for several reasons
to conduct the TDI imaging at sidereal rate, and this instrument with
a 2048$^2$ CCD yields an effective integration time of about 54 seconds
and a limiting magnitude of about 23. The optimum extragalactic survey area
for an observatory at moderate northern latitudes is the north
galactic cap. This gives a survey volume whose width and depth are
comparable for any magnitude limit, and avoids galactic obscuration
to the extent possible.
The desired million redshifts
dictates a spectroscopic limit of about r=18
(York et al. 2000; Eisenstein et al. 2001; Strauss et al. 2002). 
Spectra of sources at this brightness 
of sufficient quality should be obtainable from this telescope
in of order one hour with efficient spectrographs.  Since at a given
aperture the time to complete a survey of given solid angle is 
inversely proportional to the field area, it is clear that
the telescope should
have a field which is as large as practicable, and must lend
itself to the use of fiber-fed spectrographs. It was with these
simple considerations that we began to seek a satisfactory system design.

While wide-field optical designs exist
(the Baker-Paul three-mirror design, for
example, and more recent variants, and the Schmidt, of course) they all
suffer from either excessive length for a 2.5-m aperture or an
inaccessible focal plane, which would render the fiber spectroscopy
difficult.  Early design efforts revealed some promise
that a more-or-less conventional
Ritchey-Chr\'etien-like optical system would deliver the requisite
performance at $f$/5. At the time the system
was designed and built 
the detectors of choice 
were the 
SITe/Tektronix TK2048E CCDs, and these chips were used both in the
SDSS camera (Gunn et al. 1998), and in the spectrographs.
With 24~$\mu$m pixels, these CCDs provide excellent focal plane image 
sampling at about $f$/5 with the SDSS image quality requirement of
about 1$''$ full width at half maximum (FWHM).
This combination yields an image scale of~0.40$''$~pixel$^{-1}$, about
2.5~pixels~per~FWHM, and about~1\% total power in a Gaussian star image
beyond the Nyquist frequency.

Efficient fibers are available in diameters of 100 to 600~$\mu$m which
preserve focal ratios faster than about $f$/7, and which suffer
hardly any degradation at focal ratios around $f$/5.  Fibers which
subtend about 3$''$ are needed to cover the bright parts of
galaxies in the brightness range of interest; smaller fibers would
not only degrade the expected signal-to-noise ratio (S/N)
for galaxy spectra, but would introduce
stringent requirements on the astrometric accuracy and mechanical placement
of the fibers.  At the scale at $f$/5, 60.6~$\mu$m~arcsecond$^{-1}$, the
3$''$ angular size corresponds to
180~$\mu$m diameter fibers; these fibers have excellent
optical performance if handled and terminated properly and also yield
good sampling for the 24~$\mu$m pixels of the large Tektronix/SITe imaging
arrays at the demagnified focus of the spectrographs.
The overall focal ratio of the optical system, therefore, was fixed at~$f$/5.0, 
and for both imaging and spectroscopy the instrument point spread functions
(PSFs) are required to be $\rm < 1''$ across the entire $\rm 3^{\circ}$
field.

The final optical design achieved for the SDSS 2.5 m telescope and 
described below is simple and yields
excellent performance, with a focal plane well-matched to the TDI imaging
requirements.  The ray trace image sizes for the imaging final corrector 
are less than~0.6$''$~rms
over the entire 3$^{\circ}$ diameter field at the design wavelength, and
are less than 0.7$''$~rms in any of the five SDSS filter passbands 
($ugriz$; see Fukugita et al.~1996) which have central wavelengths 
that cover the
wavelength range of 3500 to 9300~\AA{}. Geometric distortion is less
than 12 $\mu$m over the whole field, and lateral and longitudinal color
errors have little effect on the image quality.
The lateral color
using the spectroscopic final corrector is very small, of order 
6~$\mu$m  (0.1$''$) over the entire field
over the spectroscopic
wavelength range of 3800 to 9200~\AA{}.  The dominant aberration in the
spectroscopic mode (with its demand for wideband image quality)
is longitudinal color. In the final SDSS telescope design, this
creates images larger than~1$''$~rms
(to be compared to the~3$''$ diameter fibers)
only at extreme field angles at the ends of the spectrum. 

As pointed out in the design report for the Swope and Irenee DuPont
telescopes at Las Campanas by Bowen \& Vaughan~(1973), 
it is possible to design a
Ritchey-Chr\'etien telescope with a flat field by making the curvatures of
the primary and secondary mirrors the same, which yields zero Petzval
curvature in the focal plane.  Since a Gascoigne astigmatism corrector
is required, and since this element introduces a bit of positive field
curvature, the design needs to deviate a little from this prescription,
but only slightly.  This approach results in a final focal ratio of just under
twice the primary ratio, depending a little on the back focal distance,
and requires a very large secondary, about half the primary diameter for the
field sizes obtainable with $f$/4 primaries (about 3$^{\circ}$).  For
large telescopes, one would like to circumvent the limitation of slow
primaries and large secondaries.  These desiderata were met in the
design of the 2.5-meter DuPont telescope with the introduction of only
moderate field curvature. 

\subsection{The Design}
The instrument described here uses the same philosophy as that
of the DuPont telescope, taken to even
faster primary and overall $f$/ratio.  Our requirements are rather
unusual for an astronomical instrument, since we must accomplish TDI imaging
over a large field.  This requires that field distortion be 
carefully controlled, since either a change of scale or a differential 
deviation from conformality of the mapping of the sky onto the focal plane 
across a CCD translates immediately into image degradation.  It is in fact
the case that for large enough fields, there is no satisfactory
circularly symmetric optical design for a flat focal plane, since the
requirement that parallels of latitude map into straight lines, so
stars travel along the straight columns of the CCDs, and the
condition of constant scale along those lines so that the tracking rate
across the device be constant each determine
a unique map and they are different.
For our 3$^{\circ}$ field, such troubles with mapping a sphere onto
the focal plane cause image degradation of the order of 0.15$''$; although
this value is negligible even with 
0.5$''$ images, the errors grow as
the square of the field diameter.  The desired map (onto a flat focal
plane) is one which creates a Mercator-like projection of the sky
onto the focal plane, can be achieved with anamorphic optics;
we did not investigate this option, but the technology of generating
and figuring complex optical elements is such now that this route
would be quite feasible.
The errors for a compromise circularly symmetric design are in
any case not excessive with our field.  
Conventional Ritchey-Chr\'etien designs
have two orders of magnitude too much distortion for this application. 

In addition, conventional fast Ritchey-Chr\'etien designs with
single-element Gascoigne correctors 
have unacceptably
large lateral color, both for this imaging application and
(especially) for 
fiber spectroscopy.  

We were therefore compelled to adopt a somewhat more
complex system, and have evolved a design with a two-element refracting
corrector which has excellent performance.  It makes use of the fact
that the astigmatism correction of a Gascoigne plate increases as the square
of the distance from the focal plane for a given strength, while the
lateral color and distortion only rise linearly.  Thus a pair of lenses,
one of the usual form and of weak power placed some distance from the focal
plane, and another the negative of the usual form with $n$ times the strength of
the first placed 1/$n^{\rm th}$ the distance of the first from the focus, can
correct astigmatism while introducing no lateral color or geometric distortion. 
Since the final corrector is part of the structure of the camera 
(Gunn et al 1998), a second
lens had to be made for the spectrograph in any case, and the two were
optimized somewhat differently--for distortion in the case of the camera,
and lateral color in the case of the spectrograph.

Geometric distortion remains at a level
(12~$\mu$m over the field of the SDSS camera) 
set by the order of the
aspheric used for the second corrector element and can in principle be
removed (or specified) exactly. The lateral color for the spectrographic
configuration is less than
10~$\mu$m peak-to-peak over the entire field throughout
the spectral range of the
spectrograph, and is a negligible contribution to the image diameter
for any filter or field location in the camera. 

The final design has almost zero distortion in the sense that
the radius in the focal plane is proportional, to high accuracy, to the
field angle (not its sine or tangent); zero distortion for most
wide-field imaging is defined for the condition that the radius in the
focal plane is proportional to the {\it tangent\/} of that angle, which
results in faithful representations of figures on {\it planes\/}, but we
wish as faithfully as possible to image figures on a sphere onto a
surface which is almost planar.  For this case a compromise is necessary
between the wishes for constant scale in the sense that meridians have
constant linear separation in the focal plane, and the desire that
parallels of latitude do likewise.  The optimum case depends somewhat on
the aspect ratio of the field and lies somewhere between the designs where
the radius in the
focal plane increases as the sine  and the tangent of the input angle.  For
a square focal plane, which is close to the situation at hand, the case where
the
radius is approximately proportional to the angle itself is best.
The errors can be minimized by clocking different CCDs at
different rates to correspond to the local scale along the columns, but
we have chosen, for reasons of noise reduction and simplicity in the data
system, to clock all CCDs synchronously (Gunn et al.
1998).  Our design for the best compromise
tracking rate results 
in worst-case image smearing along the columns
of 0.06$''$, 3~$\mu$m, or 0.14~pixels over the imaging array.
Points on the sky do not quite follow straight trajectories in the
focal plane, but this error is compensated for by a slight rotation of
the CCDs (0.006$^{\circ}$) near the corners of the field,
and introduces
an error of only~$\approx$~0.24~pixel if uncompensated. 

The design uses a 2.5-m $f$/2.25 primary with a 1.08-m secondary,
which, with its baffles (1.30~m diameter), obscures 27\% of the incoming 
beam.  The central hole in the primary is~1.17~m in diameter; its
baffle is slightly larger, 1.20~m.  Cassegrain
telescopes with fields this large are notoriously difficult to baffle,
but a simplified variant on the ``Venetian blind'' baffling system used in the
DuPont design works well (see \S4).  The optical layout, showing the
baffles and a set of rays from the field edge at 1.5$^\circ$ is
shown in Figure~3.
The output $f$/ratio is~5.0, and the focal plane is~0.76~m
behind the vertex of the primary in order to clear the
primary mirror support cell and allow
space for the instruments.  The telescope is quite short-coupled, with
the secondary only~3.6~m in front of the primary.  The corrector
consists of two aspheric fused quartz elements, as discussed above.  The
first (``Gascoigne'') element is approximately coincident with the vertex
of the primary mirror. There are two interchangeable rear
correctors, a thick one associated with the camera
(and in fact an intimate
part of its mechanical design --- see Gunn et al. 1998), and a
much thinner one for use with the spectrographs.  
The top surface
of the second element in the camera configuration is just 58~mm from the
focus, and for the camera element the back surface of the filters, which
are cemented to the corrector, is 8 mm from the focus. The spectrographic
configuration has 34~mm of working space behind the second corrector.
The 3$^{\circ}$ field
is 0.65~m in diameter; over this field, the focal surface of the
camera configuration is described to reasonable accuracy as a simple
quartic in the radius; it is flat to within about $\pm$0.2~mm
over the inner 0.5~m, and rises rapidly to about 2~mm at the edge.

The CCDs for the imaging camera are mounted to conform to the focal surface,
which requires a tilt of just under a degree at the edge of the field. 
There is one further complication in the design, namely that the CCDs
as produced are slightly convex, with a reasonably well controlled radius
of about 2.2 meters. The best fit plane results in focus errors of about
100$\mu$m rms, which at f/5 corresponds to an image degradation of about
20$\mu$m. We have chosen {\it not\/} to live with this, but instead to 
correct this curvature individually for each CCD with weak field
flatteners cemented to the rear face of the corrector.
This is more to attempt to keep the point-
spread-functions reasonably constant over a chip than a fundamental
discomfort with global focus errors this large, but
for some CCDs in the array there are unavoidable significant
variations (Gunn et al.~1998).

The spectroscopic optical configuration is similar to that of the imaging
camera; the two share the primary,
secondary, and Gascoigne corrector, but the final corrector is
substantially different. The spectroscopic corrector
is much thinner (which substantially reduces
longitudinal color), quite strongly curved,
and a bit farther from the focal plane.
Its design was optimized for lateral color, which is better than 4~$\mu$m~rms
over the entire field, while maintaining polychromatic images better
than 1$''$~rms diameter.
The primary-secondary
spacing is also slightly different for the two instruments; the
required difference lies well
within the secondary focus range. The spectroscopic design
violates the ``telecentric" condition that the focal plane be
perpendicular to the central ray in each image.  This choice means that for
optimum performance the
fibers cannot be placed perpendicular to the focal plane.  Since the SDSS uses
drilled plug-plates for the fibers, the most straightforward
way to deal with the lack of telecentricity is to drill the
plug-plates for the fibers while the plates are deformed slightly
(Owen et al. 1994). This has proven very straightforward and successful.

The camera design has
been optimized for the chosen distribution of filters over the focal
plane. The overall scale is 60.4 microns/arcsecond; one pixel is 0.403
arcseconds = 24 $\mu$m. There
are a total of six glass-air surfaces, the pupil radius is 1250.00~mm with
a~625 mm central obscuration, taken (not quite correctly) at the primary, 
and the first conjugate is at infinity. The telescope's optical designs
for the camera and the spectrograph configurations are given in Table~1
and Table~2 respectively.
In these tables, $c$ are the curvatures,
positive if concave right, and the $k$s are the conic constants ($k = 0$ is a
sphere, $k = -1$ a paraboloid, $k < -1$ a hyperboloid, $-1 <k < 0$ a
prolate ellipsoid, and $k > 0$ an oblate ellipsoid; generally, $k =
-e^2$).  The column labelled ``Space" lists
the spacings in millimeters from the previous surface,
positive if to the right.  The material following the
surface is given in the column labelled ``Glass".  The sign of ``glass"
changes for reflections and is positive
for rightward-moving rays, negative for left.  The quantities
$a_2$, $a_4$, $a_6$, and
$a_8$ are the aspheric coefficients for polynomial aspherics, where the
general form of the surface~is
$$
t = c(h^2 + (k+1)t_c^2)/2 + (a_2 h^2 + a_4 h^4 + a_6 h^6 + a_8 h^8),
$$
where $t_c$ is the solution to the conic surface equation
$$
t_c = c(h^2 + (k+1)t_c^2)/2.
$$
The index of refraction
for fused quartz (fq in the tables) is 1.46415 at 4700~\AA.

The primary is almost hyperbolic, with about a wave and a half of 6$^{\rm th}$
and 8$^{\rm th}$ order flattening at the edge; the secondary is likewise, with
about two waves of 6$^{\rm th}$ order steepening.  

\subsection{The Performance of the Imaging Design}

The discussion of the theoretical optical performance of the camera 
configuration design is
a bit complicated both because of the complexity of the focal plane, with
different filters and field flatteners in different locations,
and because of the
effect of distortion on the final TDI image quality.
A series of simple monochromatic traces of
the camera system without the individual field flatteners is presented
in Table~3.  Here the focus
(distance behind the dummy surface 8,
which is the nominal 8~mm back focal distance behind the last (filter)
element) $f_b$, the image height $h$ at that focus, and the rms image
diameter $\epsilon$ are tabulated for each of eight field angles from the
center to the edge of the imaging field (which is somewhat smaller than
the whole spectroscopic field) for the effective
wavelengths of the five SDSS filters.  The last four field angles
correspond to the outer corners of the outermost CCD in some row, and
angles which are not actually reached at a given color with the camera
are prefixed with a~`$*$'.
The images are
degraded somewhat at the very edge, where the radial field curvature is
maximum, by the finite (flat) area of the chips. This effect 
has been evaluated in detail along with TDI and polychromatic effects in
Gunn et al. (1998), where there is also presented a greyscale `spot diagram'
like that shown in Figure 5 in this paper for the spectroscopic configuration.

The form of the focal surface at 4760~\AA{} is presented in Table~4.  The
quantity $\delta$ is the total longitudinal
focal deviation from a plane
at the indicated angle, ht the height in the focal surface, and lindev the
deviation from a best-fit strict linear proportionality between the 
input angle and the height.
These deviations (maximum of about
4$\mu$m) simply reflect the maximum order of the surface of the corrector; 
it is clear that the
distortion is controlled by the local slope of this surface, and can be
made to vanish (or take on any reasonable form) exactly.  The corrector
is sufficiently close to the focus that there is little repercussion for the
image quality or the lateral color when it is modified slightly. 

The final calculated images produced by the optics and convolved with 0.8
arcsecond FWHM Gaussian seeing (rather optimistic, as it turns out; see below)
are described by Gunn et al. (1998).  Optical distortions cause the
instrumental PSF to vary across the $3^{\circ}$
field, with the quality degrading away from the optical axis - it is for this
reason that the $u$ filters form the central row in the SDSS camera.
At the edge of the field, the FWHM of the
instrumental PSF can vary by up to 15\% from one side of the CCD to
the other. Further, the atmospheric contribution to the PSF is not
constant over the camera array, and significant differences are seen in the
scan direction because of the short exposure time, 54 seconds for a given
point on the sky in TDI mode. As a result, the PSF is a complex and
varying function of both the ``x'' and ``y'' position in the focal plane
and within an individual CCD image, and must be dynamically modeled 
during photometric data reduction using images of bright stars (see
Lupton et al. 2001 and in preparation). Examples of the PSF variation
observed across the camera in excellent seeing are shown in Figure~4.

\subsection{The Performance of the Spectrographic Design}

Table~5 presents data which are relevant to the spectrographic
mode design.  At each of seven field angles from the center to the extreme
edge, the focal properties are given on a surface which represents the
average focal surface over the spectrograph wavelength range of 3900~\AA{}
to~9200 \AA{}.
The entries for the third column for wavelength 5300 \AA{},
which roughly centers the
range of index variations for the spectrograph, give the height of the
focus.  For the other four wavelengths, the third column lists the height
differences
(in the sense of the height at the given wavelength minus the height at
5300 \AA{}), thus
represent lateral color, which is seen to be 10~$\mu$m total, 
\hbox{$\pm 5 \mu$m,} or less over the whole field,
and are even somewhat smaller at the
edges where the images are larger.  The $D$s are longitudinal
deviations from best focus at that wavelength, and the $\epsilon$s are
rms image diameters at the compromise focus.  It is only at the
wavelength extremes and at the extreme edge of the field that the
rms diameters of the compromise images exceed 1 arcsecond; the best focus images
there are substantially less than 1 arcsecond (45 and 40~$\mu$m at
4000~\AA{} and 9000~\AA{}, respectively, at 90$'$ radius), and the
increase is due solely to longitudinal color.  The 72 micron worst-case
rms diameter is still much smaller than the 180 micron fibers, however,
and the effect on throughput is not large; we discuss the issue more
fully below.  The details of the average focal surface are presented
next: the sagitta of the focal surface, the mean height (here just the
average of the 4000 \AA{} height and the 9000 \AA{} one, and presumably
where one drills the fiber hole), the deviation from a linear
relation with the field angle (it is seen here that the different final
corrector form, chosen to yield the best polychromatic images, results
in quite serious distortion, but this is of no importance for the
spectrograph), the direction cosine of the central ray measured from the
direction of the axis, and the {\it difference} between this angle and
the angle which the normal to the focal surface makes with the axis.
This last entry is the angle with which the fiber hole must be drilled
into a plate which conforms to the focal surface.
The maximum value is about 2.0
degrees, compared to the 5.7 degree half-angle input cone at $f$/5.  The
losses, even into the $f$/4 input beam of the spectrograph, are large
enough to be important, and we compensate for it by drilling the
holes into a deformed plate.

As discussed by York et al.~(2000), the highest quality conditions
(photometric, good seeing) are devoted to imaging, while other workable
time is given to fiber spectroscopy; the average seeing for the latter
is typically \hbox{1.5$''$--1.7$''$.}  Differential
refraction at the ends of the spectrum at the maximum zenith angle
allowed for survey observations (55$^{\circ}$)
is just under $\pm 1''$ from the central wavelength
image at the altitude of the site.  With 1.5$''$ Gaussian seeing,
a 3$''$ fiber at the field edge collects 95\% and 92\% of the
light at 4000~\AA{} and 9000~\AA{} from a point source, respectively, 
when centered on the
image, and at worst~65\% and~67\% when decentered by~1$''$.  This
is not substantially worse than the situation in the center of the field
at the central wavelength, where the centered number is~98\% and the
1$''$ offset number is~72\%.  A smaller fraction of the light is collected
for extended objects, of course, but the {\it differential\/} effect
between the center and edge is smaller. 

The point spread function, convolved with 1.5$''$ Gaussian seeing,
for the field angles in Table~5 are shown in Figure~5.
The images are separated by 6$''$ in the mosaic, and the circles 
are 3$''$ in diameter, the input diameter of the fibers.

\subsection{The Optical Elements}

\subsubsection{The Primary Mirror}

The primary mirror borosilicate honeycomb blank was cast by Hextek Corporation,
Tucson, AZ, with final cooling accomplished in Spring, 1993. The casting
technique is similar to that developed at the University of Arizona Mirror
Lab (Angel \& Hill, 1982). 
For a mirror of this size and focal ratio, it was not necessary
to rotate the furnace during casting.  The 
generation of the front plate was carried out by Arizona Technologies
and completed at the Optical Sciences Center (OSC) of the University of 
Arizona, where the final figuring and polishing was also carried out.
The testing during the figuring and polishing phases was accomplished
with the use of null lenses which were fabricated by OSC and verified
using a computer generated hologram (Burge et al. 1994), and 
high-speed phase retrieval testing techniques (Dettmann \& Modisett
1997). Figure 6 shows the mirror blank at OSC, while Figure 7 shows
part of the mirror faceplate, showing the lightweight, very strong
honeycomb structure.  The fabrication tolerance specification for the
mirror figure is in terms of a Kolmogorov seeing-like structure function
and requires that the contribution of the primary mirror
to the image size be FWHM $<$ 0.2 arcseconds. The mirror figure
passed this specification on small scales, but the final 
figure produced degraded performance at large separations
due to about 230 nm of astigmatism. It is necessary
while figuring mirrors of this type with thin faceplates to pressurize
the cells in order to prevent `print-through' from the pressure of the
polishing tool. It is 
believed that the astigmatism arose because one or more of the seals 
associated with the pressure system exerted excessive nonaxisymmetric
forces on the mirror, and it sprang slightly when they were removed.
Simple calculations
demonstrated that this could be corrected with the application of forces
of 20-40 N by the primary mirror mount, resulting in wavefront errors of
less than 100 nm across the entire primary (see Waddell et al. 1998),
and a simple pneumatic system to accomplish this was implemented.
The primary mirror support system and the astigmatism corrector
will be described in \S3.4.2.

The mirror was aluminized at NOAO's Kitt Peak 4 m telescope aluminizing
facility (where it has been regularly re-aluminized annually over
the lifetime of the project). Figure 8 shows the primary mirror
after aluminization and coating.

\subsubsection{The Secondary Mirror}

The SDSS secondary is of diameter 1.1m, almost half that of the
primary, necessary for an unvignetted large focal
plane. The borosilicate  secondary mirror blank was also made by Hextek
corporation using hot gas fusion. The blank was generated to a sphere
by Astronomically Xenogenic Enterprises (AXE) of Tuscon, Arizona, and
the hole drilling, optical generation and edging were completed at
HexTek.  The resulting radius, 7331 $\pm$ 6 mm, was well within the 
specified radius of 7334 $\pm$ 25 mm.

The secondary was then delivered to Steward Observatories Mirror Lab (SOML)
for figuring, polishing, and testing. It was the first such optic fabricated
at a new facility developed by SOML for large secondary optics (Andersen
et al. 1994), and  several innovative techniques were employed. The
mirror was figured using stressed-lap tooling, and the testing was first
done with a swing-arm profilometer (Andersen \& Burge 1995), which resulted
in the control of errors to of the order of one wave. The
secondary mirror at this stage is shown in Figure~9. The surface accuracy
of order 100 nm over the full aperture required more accurate testing
methods, and this mirror was the first figured using the computer-generated
hologram technique (Burge, 1997), which refers the convex surface to an
easily made accurate spherical concave reference. This technique is
enormously simpler, inherently more accurate, and less expensive than  
conventional Hindle sphere testing, and requires only optics the size
of the secondary under test. The technique was very successful for
our mirror.

The polishing cell for the secondary was designed to support the mirror
in two orientations.  During testing, the mirror faces down towards
the test optics.  During polishing, the mirror faces up, and a mass
the same as that of the mirror pulls down on the mirror mount attachment
pads.  In this way, deformations which would be induced by the mirror
supports are polished out.

The fabrication tolerance specification for the secondary mirror figure 
is the same as that for the primary - a wavefront structure function
for 0.2 arcseconds FWHM seeing. Again, the rms wavefront difference was
less than 100 nm across the entire mirror. The secondary mirror was
aluminized at the nearby facilities of the National Solar Observatory, Sunspot
NM, and delivered in 1996.

\subsubsection{The Common Corrector}

The third optical element, the
Gascoigne corrector, is the last one
common to both spectroscopic and imaging modes and is therefore 
called the common corrector. It was manufactured of Corning 7940
fused silica (grade 5F) and figured and polished by Contraves, Inc.
of Pittsburgh, PA.  The lens is 802 mm in diameter and is about 12 mm thick.
Optical testing by Contraves showed that the lens has a turned edge over
a roughly 45$^\circ$ angular sector.
It proved possible to avoid use of most of this region during imaging by
installing the lens in its cell in the telescope (see Figure 10)
with a preferential orientation relative to the camera CCD array (the
corrector rotates with the image rotator carrying the camera), but the
degraded image quality in the upper right of the array in Figure~4, which
is a persistent feature of the system,  may
well be an artifact of this problem. 

Subsequent to polishing, the lens was anti-reflection coated by QSP
Optical Technology, Inc., Santa Ana, CA (now Infinite Optics, Inc., but
hereinafter simply QSP).  
The requirements for these coatings was
that the average reflectance across the 3200 \AA{} to 11000 \AA{}
optical bandpass should be less than 1.5\% per side with peak
reflectance $<$ 2\% per side. This level of performance was 
achieved with a 12-layer coating which is both durable and easily
removed.  Fabrication of the common corrector was completed with
its mounting in an elastomeric bonding ring at the University of
Washington, and the lens was delivered to APO in November 1997.

\subsubsection{The Final Correctors}

The {\it imaging corrector} is a very complex thick lens (45 mm central
thickness with almost a centimeter asphere on its front surface; the
rear surface is planar)  which 
forms the structural element of the SDSS camera upon which the detectors
are mechanically mounted. It is thus an integral part of, and is mounted
and dismounted with, the camera (see Gunn et al. 1998, which also discusses
the complex coating on this element). We would therefore have had to make
two correctors in any case, and this gave us the opportunity to 
optimize the spectroscopic lens for its application.

The {\it spectroscopic
corrector} is also made of Corning 7940 fused silica, grade 5F.
It was figured and polished by Tinsley Laboratories Inc. (Richmond, CA)
using proprietary computer control techniques and is 727 mm in diameter.
It is quite strongly curved, in contrast to the imaging corrector, and
much thinner, 10 mm thick in the center.

The figuring requirements were for peak-to-valley slope errors to be less than
150 microradians for spatial frequencies less than 160 mm over 95\% of the
clear aperture, with linear increase in slope errors allowed for larger
spatial scales up to 600 microradians.
The small spatial scale errors are required to be less than 250 microradians
peak-to-valley across the entire face of the lens. As delivered by  Tinsley,
the aspheric side of the lens has slope errors of less than 50 microradians 
peak-to-valley over spatial frequencies below 160 mm and less than 10
microradians for larger spatial scales. Slope errors for the spherical face
are about ten times better than this.

The smoothness over small areas was measured using a 250 mm diameter check
plate positioned on the corrector. Microroughness at very small scales
(1 and 5 mm) was measured to be less than 11 \AA{} rms using a Chapman
MP-2000 microscope with a Nomarski objective.

The lens was then AR-coated.  The average reflectance per
side across the 3900 \AA{} to 9100 \AA{} spectroscopic bandpass was 
required to be less than 0.8\% with peak reflectance less than 1.4 \%.
This level of performance was achieved with a durable 12-layer coating
designed and applied by QSP.
The finished lens is shown in Figure~11.

The history of the installation of the 2.5 m optics, and their performance,
is summarized in \S6.

\section{Mechanical Design}

\subsection{General Considerations}

The mechanical design of the {\it telescope} was driven by the desire to obtain
the highest quality images possible and the requirement of highly accurate
and stable motion.  
The telescope is an altitude-azimuth design similar to the Apache
Point Observatory (APO) and Wisconsin, Indiana, Yale, NOAO (WIYN) 3.5 m 
telescopes (Mannery et al.~1986a,b, Gunnels~1990a,
Johns \& Pilachowski~1990). As discussed in numerous places in this chapter,
the 2.5 m design is strongly influenced by that of the 3.5 m.  
This design takes full advantage of
lightweight mirror technology resulting in a telescope with low
inertia, low friction, and mechanical simplicity.

We elected both for cost reasons and for thermal performance to use
a roll-away {\it enclosure} (\S5). During observations of course, this
design greatly increases the telescope's
exposure to both wind buffeting and stray light contamination compared with
the conventional dome configuration. This problem is dealt with by
protecting the telescope from the wind and
stray light by means of an independently mounted and driven baffle which
is coaxial with and encloses the telescope (\S4). 

\subsection{Structure}

The telescope optics support structure (OSS) consists of the primary
support structure (PSS) and the secondary truss (Figures~12 and 13). 
The PSS is a steel weldment that supports the primary mirror and couples
the OSS to the fork.  The one-piece construction of the PSS has a higher
stiffness-to-weight ratio and is lower in cost than the more traditional
detachable mirror cell.  The secondary space truss controls five of
the degrees of freedom of the secondary mirror directly.  With
adequate tension in the secondary vanes (see Figure~13),
the rotation mode of the
secondary about its optical axis can be kept above 10~Hz.  The square
secondary frame is efficient at resisting this tension.

The eight metering elements of the secondary truss are graphite 
fiber reinforced epoxy tubes.  This material has about 2.3
times the stiffness to mass ratio of steel.  This confers the
following benefits:

\begin{itemize}
\item A substantial amount of mass is removed from the truss without
degrading its static deflection or lowest natural frequency.  

\item The reduced moment of inertia reduces the susceptibility of the
telescope to wind-induced tracking errors.

\item The diameters of the truss elements are reduced without lowering
their natural frequency. This, in turn, decreases the wind loads on the
truss.

\item The reduced mass of the truss moves the center of gravity of the OSS
forward.  This allows the altitude bearings to be located lower on the OSS
and increased the clearance for instruments mounted behind the PSS. 
\end{itemize}

Tubes and other linear structural shapes of graphite fiber reinforced 
epoxy have a much lower coefficient of thermal expansion than steel in 
the long direction, resulting in another benefit of improved metering of 
the primary/secondary separation with temperature changes.

The telescope was constructed by L\&F Industries, Huntingdon Park, CA.
The largest pieces of the structure, and their masses, are: primary mirror
support with pillow block (5215 kg); fork assembly (4308 kg); rotating
floor framing (3084 kg); and secondary truss with cage assembly (499 kg).
The moment of inertia about the azimuth axis is 33855 $\rm kg~m^2$ at
zenith and 34594 $\rm kg~m^2$ at the horizon, while the moment of 
inertia of the optics support structure (OSS) about the altitude axis
is 10405 $\rm kg~m^2$.  These quantities are important for the telescope
control system.

\subsection{Bearings and Drives}

The moving mass of the SDSS 2.5-m telescope is 15,500 kg, which is
light enough that exotic bearing technology is not required. We
have chosen to use precision rolling-element bearings and friction drives
throughout.

A pair of 2.54-m diameter, hardened and ground drive segments are
mounted on the sides of the PSS next to the fork. The measured high 
frequency (greater than eight cycles/revolution) runout of the drive 
segments is less than 100~nm~rms.  Motor driven capstans, friction-coupled 
to each disk, provide balanced altitude drive torques and minimize drive
torque-induced distortions
of the PSS (Gunnels~1990b).

The telescope azimuth structure consists of the fork and the azimuth
cone.  At the apex of the azimuth cone is a spherical roller bearing
that supports the weight of the telescope.  At the upper end of the cone
is a 2.54~m diameter disk, with a hardened and ground outer surface.
Its high frequency runout is 220~nm~rms. This disk is guided by four
roller assemblies, two of which are motor driven.  These rollers, with 
the bearing at the cone apex, define the telescope
azimuth axis.  

Rolling element bearings are used for each axis.  These bearings require
little maintenance, are low in friction, and generate negligible heat
during operation.  The measured high frequency radial run-out of the 
spherical roller bearing used as the lower azimuth bearing for the SDSS 
2.5-m telescope is less than 310 nm rms. This corresponds to a contribution 
of 23~milliarcsecond to the rms tracking error for the telescope. The high 
frequency radial run-out of the altitude bearings is 51 nm rms. They are 
estimated to contribute less than 6~milliarcsecond rms tracking error. In
addition, the frequencies of these errors during normal tracking are
small enough that they are completely dominated by atmospheric effects,
so the astrometric effects are small and the effect on image quality
completely negligible.

The telescope drive assemblies are pushed against their respective drive
disks by radial links. The contact force must be large enough to transfer the
necessary drive torque to the telescope via friction without slipping, 
which would certainly damage the drive surface, but must be less
than the force that would permanently deform the disk or roller. In the
case of the azimuth axis, the radial link must limit the contact force during
seismic accelerations, and the links must provide an extremely stiff link
between the azimuth drive housings and the telescope pier to allow high control
system bandwidths and resist wind-induced tracking error.

The azimuth drives are preloaded against the azimuth disk drive through
a series of springs: the material of the frame itself, a set of soft springs
and a set of hard springs. The soft springs are Belleville washers
which provide a preload for
the initial assembly, producing up to 2600 lb of force on the azimuth
drive housings. The hard springs are also Belleville washers 
and produce up to 12000 lb of force. The system is so configured that
it does not float; the preloads are taken up by the elasticity of the
frame itself, but large forces such as might result from seismic 
activity or accidents can be absorbed by the springs if the forces are
large enough to unseat them.

Incremental encoders are friction coupled to the large disks. Readily
available Heidenhain encoders (type ROD~800) 
with a reduction ratio of roughly 25:1 and reliable interpolation 
produce 14~milliarcsecond resolution 
on
the sky and allow slew rates higher than~4$^\circ sec^{-1}$.  Absolute
axis encoding is provided by a series of optical tape encoder segments, also
from Heidenhain, for which only the fiducial signals are used. These
fiducials provide a reasonably stable angular reference frame to keep the
absolute zeros and scale of the high-accuracy incremental encoders over the
interval from one pointing model on the sky to the next, typically a month
or two. The instrument rotator uses a friction drive identical to the
telescope axes, but its position is measured with a continuous Heidenhain
optical tape. Each axis (and the rotator) is
controlled by a digital PID servo control (Schier~1990),
which is more fully discussed below in \S3.8.

\subsection{Optics: Support, Motion Control and Thermal Control}

\subsubsection{High Level Mirror Control}

The support and position controls for both the primary and secondary
mirrors allow the mirrors up to 12 mm of precisely positioned axial motion
as well as limited tilt and transverse motion. The secondary position controls
are necessary to maintain focus and collimation as the telescope moves
across the sky and as the temperature changes. The collimation changes as
a function of zenith angle are determined annually and applied open-loop
with completely satisfactory performance. The large axial motion of the
primary is necessary to change the image scale at the focal surface in 
spectroscopic mode to 
compensate
for thermal expansion or contraction of the aluminum fiber plug plates
and to compensate partially for observing with a plate
at a different airmass from the one for which it was designed.

In order to control the image scale of the telescope, it is necessary to
monitor the distance from the focal plane to the vertex of the primary
mirror very accurately (25 microns). The load path connecting
these locations is very stiff and we have excellent control
of the temperature uniformity of this material as part of the primary mirror
temperature control system.

Position control commands to both the primary and secondary motion controllers
are supplied by the telescope control computer (\S3.8), which automatically
adjusts the secondary mirror position for collimation and focus based on the
current pointing altitude. The position of the 
primary mirror does not change significantly with altitude.
and only the secondary mirror requires this
automatic adjustment procedure. The corrections are applied every few seconds.

The imager operating software incorporates a focus loop. This takes
advantage of the fact that the imaging camera is always used in
drift-scan mode, so new star image size data from the dedicated focus
CCDs are constantly available. These produce images symmetrically placed 
in focus inside and outside of the nominal focus for the camera array;
a detailed description of how they are used to generate the control signals
can be found in 
Gunn et al. (1998), but no details were given
in that paper of the as-delivered performance of the servo system,
since there was at that time no data.

Figure~14 shows the performance of the focus servo for a
typical run. We estimate the error in the focus position from both the
leading- and trailing- focus CCDs; the mean of these two estimates has
been adjusted to lie very close to the focal position of the main
array. The telescope's secondary is adjusted, via a classical 
P(roportional)I(ntegral)D(erivative), hereinafter simply PID,
servo loop, to follow the error signal.  The loop has very low gain
and is very slow 
as it must be stable even in fields of
low stellar density when the error signal is updated only infrequently;
since the focus changes only very slowly in response to thermal
variations in the optics and structure and to differential 
deflections during tracking, this is quite satisfactory.
It should be noted that the variable time delay in the error signal 
actually makes this a rather \emph{non-}classical servo loop, 
but it appears to work well in practice.
Careful inspection will show that the telescope focus slightly lags the 
estimated true focus, a result of the small value of the integral gain.
The contribution to the image quality as measured on the main imaging array 
from focus errors is negligible (at most $\sim 0.3$ arcsec added in quadrature,
and usually much less).

In spectroscopic mode, 
we use aluminum plug-plates to position the optical
fibers for the multifiber spectrograph, and the plates are drilled
for the predicted temperature of use.  However, since the ambient
temperature may not have the predicted value, the
image scale is matched to the fibers just before each 
spectroscopic observation, using the observed focal plane positions of
the multiple guide stars on each spectroscopic plate (see \S3.8).  
This is done
by translating the primary axially and refocusing the secondary.  A
10$^{\circ}$C temperature mismatch can be corrected by translations of 2.4
mm and 2.0 mm for the primary and secondary respectively. The focus
itself is not automated in spectroscopic mode, but is monitored and
adjusted by the observers. It is not nearly so critical for spectroscopy,
of course, because loss of critical
focus only introduces small inefficiencies in data collection, not
serious degradation of data quality. In addition, exploration of focus
during an exposure is possible and in fact is routinely done using the
guide star images. The guide images are archived, so one has in principle
a record of the seeing through each exposure.

As the telescope changes elevation from zenith to
horizon, the secondary will sag about 600 $\mu$m with respect to the
primary optical axis.  This decollimation is corrected by actively 
translating the secondary so that its vertex remains on the optical axis 
of the primary and correcting the tilt of the secondary as necessary, both
according to an annual calibration of these deflections. 

\subsubsection{Primary Mirror Support and Control}

The primary mirror (Figure~6) is supported on air pistons using elastomeric,
low-friction rolling diaphragm air cylinders manufactured by Marsh Bellofram
for both axial and transverse supports. 
There are 48 such cylinders distributed on the
back surface of the mirror for the axial supports.
Three stiff load cells serve as axial hard-points.  Simple servo systems
act to control the pressure provided to those air pistons in the
120$^{\circ}$ sector associated with each load cell (each of which
is supported by sixteen cylinders with all cylinders pressurized by
a single pneumatic servo-controlled valve), so that the
unsupported mirror weight applied to the load cell is less than 10 N. 
Each axial hard-point is positioned axially with a motor driven lead screw.
This allows control of primary piston and tilt. In the transverse direction,
the mirror is supported by 16 air cylinders all plumbed in parallel. The
cylinders are mounted on posts, each of which extends through a hole in the 
mirror backplate into a mirror cell (cf. Figure~7). This allows the transverse
loading on the mirror to be applied between the front and back face at 
approximately the center of gravity. To accomodate the 12 mm axial motion of 
the primary, these cylinders are fitted with rollers which bear on stainless 
steel spreader plates. These plates in turn transfer the load into the 
webbing of the cell near the front and back surfaces of the mirror.

Each of the three axial cylinder arrays, as well as the transverse cylinder
array, is controlled by a fast-response pneumatic servo control system. Each
servo-control loop is closed around a load cell which provides the control
signal to modulate pressure in the respective cylinder array. The load cell is
part of a stiff, stepper motor driven, linear actuator assembly, and
is actually used as a very high-resolution  {\it position} sensor.
Each of the assemblies is hard-mounted to the primary support structure
and contacts the mirror surface, three axially on the back surface and one
transversely on the outer radius. The system is designed for a nominal
7N force from the mirror. The servo is a two-level system
consisting of a very fast (~500Hz) inner pressure loop which keeps
the system pressure at its commanded level via a fast Data Instruments
SA-series pressure sensor and
a fast continuous-flow three port (pressure, vacuum, and output) 
Dy-Val PC-2 proportional valve (both with $\sim$ 1 kHz bandwidth), 
and an outer position loop, which is closed around the load cell and has
a bandwidth of approximately 1 Hz. The inner loop needs to be fast
because the natural frequency of the mirror on the belloframs is about
10 Hz, and this must be stabilized.
 
The primary's lateral position and rotation about the z-axis are
controlled by two lateral links attached to the back surface of the
mirror near the top and bottom. The links are 
oriented horizontally to prevent their carrying any of the mirror's weight
independent of telescope orientation, and the other end of each link
is attached to a stepper-motor driven linear actuator.

The linear actuators incorporate a 40 thread/inch actuator rod threaded
through the axis of a 200 step/revolution Eastern Air Devices stepper motor,
giving a linear displacement of 2.5 $\mu$m per step.
The motors are driven by a six-channel
programmable motion controller manufactured by Galil Motion Control Inc.,
and microstepping motor drivers (Intelligent Motion Systems Inc.),
Microstepping is used for smoothness of motion, but
the move always ends on an integral step so that the motors can be powered
off after each move to reduce heat generation (especially necessary 
because of their close vicinity to the primary mirror and the optical path).
Each actuator has an associated Mitotoyu linear encoder
which independently monitors the position of
the primary relative to the primary support structure to a resolution
of about 1 $\mu$m.

The E6 borosilicate glass used in the primary mirror is a low thermal
expansion material with an expansion coefficient of 
$2.8 \times 10^{-6}/^{\circ}/$C . 
To prevent thermal distortion of the mirror and mirror
seeing from significantly degrading image quality, the temperature of
the mirror must be uniform to 0.2$^{\circ}$/C.  This figure is larger than that
established for other telescope using similar mirrors because the
required image quality of this wide-field optical system is only moderate.
In addition, the front surface of the mirror must be
maintained within a few tenths of a degree of the ambient air temperature.
Several active temperature control systems have been developed to perform
these tasks (Johns and Pilachowski 1990; Siegmund et al. 1990; Lloyd-Hart
1990).  We use a variant of the rather simple system that is implemented
on the ARC 3.5-m mirror (Hull, Siegmund \& Long 1994). 
Ambient air from above the primary mirror is used for
ventilation. This ventilating air is drawn down through the central hole 
of the primary mirror and from around its periphery into a plenum
behind the mirror before
being drawn up into the hexagonal interior voids (Figure~7) of the
mirror through holes in the back plate.  There, it is drawn upward to the
back of the face plate and radially inward through a small gap between
the perimeter of a hexagonal air baffle and the back of the face plate,
across the back of the face plate to the center of the air baffle
and down an exhaust tube to the interior of the mirror platform. It
was intended originally that the suction required to establish this
flow be established by a large axial blower which exhausts the room beneath
the telescope, but this proved to be inadequate, and 
has been replaced by a pair of low-vibration centrifugal blowers mounted in the 
lower part of the telescope fork.

We have installed in the primary and secondary mirrors and on the telescope
and support structure a temperature measurement system based on individually
calibrated National Semiconductor LM234 temperature sensors. 
These devices
act as current sources with nominal currents of 1 $\mu$a per K, and can be
calibrated to and are stable to about 0.1C. There are 115 such sensors in
all, 40 on the primary, 20 on the secondary, and the remainder on
the telescope structure, wind baffle, and at various critical places in the 
enclosure. 
The signals from these are multiplexed,
digitized to 12 bits, and stored by the Telescope Performance Monitor
(TPM) computers, which also monitor and archive telemetry from all the other 
telescope systems (see \S3.8).

Temperature changes and, most dramatically, temperature differences between
the face and back plate of a mirror, affect its curvature. This is most strongly
felt in focus, but it also causes a scale
change at the focal surface which contributes to astrometric error if not
corrected in the analysis of the image data. No attempt is made to
vary the ventilation, and the temperature sensing system has been most
valuable not as an element of the control system but to inform us about
the many original
inadequacies in the thermal environment and ventilation of the telescope
and optics, and thus to focus our attention on what we needed to improve.

As described in \S2.5.1, the primary 
is inherently astigmatic.  Rather than re-generate and polish the
mirror, its figure is recovered and the astigmatism corrected by applying
small forces the outer edge of the mirror. This is accomplished by small 
double
acting air cylinders, 12 in all, mounted every $\rm 30^{\circ}$ on the 
back surface near the outer radius. These air cylinder are independently
able to apply an axial load of up to $\pm$ 40 N on the mirror, and the
adjustment is done manually. To minimize hysteresis during the adjustment
and piston movement of the primary, ultra-low-hysteresis  Airpot Corporation
``Airpel'' cylinders are used. 
These devices use a loosely-fitting graphite piston
inside a glass cylinder with no elastomeric seals, are essentially frictionless,
and perform very well.

\subsubsection{Secondary Mirror Support and Control}

The 2.5-meter secondary support borrows its general philosophy and much
of its detailed design from
that of the ARC 3.5-meter telescope. No cell in the classical sense
is used, the notion being that a lightweight mirror should be supported
in as lightweight a manner as possible to maintain the advantage of 
low mass on the thermal and mechanical properties of the telescope. The
mirror is protected by a lightweight baffle system (\S4.4.3 below) but
is attached directly to a cubical spaceframe structure via its position
actuators and central support.

Three linear actuator assemblies mounted in parallel and positioned at
$\rm 120^{\circ}$ separation provide the axial
support plus axial and tip/tilt position control for the secondary
mirror. The assemblies are attached to a space frame supported by tension
rods attached to the square frame affixed to the telescope's carbon
fiber truss (Figure~13). Each actuator assembly consists of a stepper motor
mounted to an 80:1 harmonic drive speed reducer which drives a 40 thread/inch
high precision drive screw. The
axial step size is 53 nm, which gives a resolution of 7.8 milliarcseconds
on the focal plane.  The lead screw/bearing assembly is specified to have
an accuracy of 250 nm. In practice, changing loads and direction reversals
lead to occasional errors of the order of 500-700 nm. To correct these,
a piezo-electric subassembly is mounted in series 
with the mechanical actuator,  incorporated into a flexure element which 
allows for the transverse motion of the mirror relative to the frame
on which the actuators are mounted. At telescope operating angles above 
altitude $\rm 30^{\circ}$, all three actuators are subjected to tensile 
loading. Since piezo-electric actuators function best under compressive
loading, the piezo electric actuators are mounted in a cage assembly
that reverses the direction of the load vector. The piezo actuators, low-
voltage devices from Physik Instrumente GmbH, have a total range of
$\pm1.5 \mu$m. The position of the secondary is controlled by a two-level
servo loop, the outer, coarse level controlling the screws and the inner,
fine level the piezos. Monitoring is via three linear Heidenhain encoders
with a resolution of about 5 nm.
The secondary mirror is supported from these actuators by three aluminum 
whiffletrees,
which spread the load from each actuator to three flexural attachment locations
on the back face of the mirror.  
A spring-actuated load limit mechanism was (belatedly, see \S6.1) 
incorporated
into the connection between the actuator assemblies and the whiffles to 
protect the mirror from the possibility of an actuator overload.
The secondary mirror actuator assembly is shown in Figure~15.

Transverse support and motion control of the secondary are handled by a central
shaft mounted through a spherical bearing mounted on the space frame. 
One end of the shaft extends through the 
back face of the mirror into another spherical bearing mounted in its center 
cell, while the other end extends 
through the frame to the opposite side, where it is attached through 
flexures to a pair
of stepper-motor driven linear actuators mounted at 45 degrees to the vertical
in the x-y plane. To
accomodate the secondary's required 12 mm of axial motion, a precision 
linear ball bearing is mounted into the inside diameter of the mirror's
spherical bearing.  This system was intended as a quick interim solution to
replace the original mechanically inadequate flexure-based system while
a new robust flexure system was designed, but lack of time and resources
prevented the new system from being implemented. While the performance of the
present system is adequate most of the time, there are serious problems
with hysteresis and stick-slip in the spherical bearings when the direction
of motion is reversed, which cause isolated astrometric anomalies. The
position of the mirror is monitored by 4 (3 piston, 1 transverse) 
Mitotoyu linear gauges similar to the ones used for the primary. 
A single horizontally oriented lateral link
between the mirror and the frame prevents rotation of the mirror
about the z-axis. 

The temperature control requirement for the secondary mirror to avoid degrading
image quality also corresponds to a uniformity of 0.2$^{\circ}$C, 
and this has
proven difficult to achieve. A free convection scheme using tubes into the 
secondary was planned but never implemented. Since it has
no cell, the mirror is essentially in free air, 
and originally with its back free to radiate
to the sky. Since there are electronics in the secondary support cube, the
radiative environment was different in the center of the mirror from the edges,
and overcooling of the back at the edges was initially a serious problem. 
This has since been brought under control by aluminizing the back of the
mirror and installing a radiation shield which interferes very little with
the airflow over the back of the mirror; the performance is now
satisfactory, but the thermal time constants are still larger than
we would like. The original plan was to use a lightweighted low-expansion
(Zerodur or ULE) mirror, which was abandoned due to its higher price.
It is clear, however, that the borosilicate compromise was not a good one
and quite possibly in the end not even economically favorable.

\subsection{The Instrument Rotator and Collimation}

\subsubsection{The Rotator}

All altitude-azimuth telescopes require an instrument rotator. The
SDSS instrument rotator covers the back of the
mirror cell and is mounted to the primary support structure.
The spectrographs, camera and fiber plug-plate cartridges
mount to this rotator.  Since the spectrographs corotate
with the plug-plates, this eliminates most of the flexing of the
fibers that might occur during an integration and results in much
better sky subtraction. Also, as the fibers are less than 2 meters long, 
additional benefits include reduction of light loss and materials cost.
The rotator is slightly over 2.7 meters in diameter and weighs approximately
550 kg. The spectrographs, each weighing 270 kg, are permanently
mounted to the rotator. During spectroscopic operations, the rotator
carries an additional 145 kg (the spectroscopic fiber cartridge
and plug plate) and during imaging the additional 450 kg of the 
imaging camera. The rotator design incorporates an outer and inner ring 
structure (the inner ring holds the imaging camera) to minimize the
communication of radial forces to the instruments. It depends on
the primary support structure for structural integrity.

The angular accuracy required for the rotator is reduced from that needed
for the axes by the ratio of the telescope focal length to the field
radius, a factor of 34.  This degree of accuracy ($0.1''$)
is quite straightforward
to achieve, but is by no means a negligible task. The bearing is a Rotek 
four-point contact ball bearing. Its high frequency (greater than eight 
cycles/revolution) lateral runout was measured at about 170 nm rms per 
axis at an altitude of $0^{\circ}$. The encoder and friction drive for the 
rotator are similar to those used for the axes. The drive disk for the 
rotator is about 2.80 meters in diameter and its high frequency runout 
is 1.0 $\mu$m rms. Position accuracy and feedback is provided by a
Heidenhain optical tape and read-head system.

\subsubsection{Collimation}

Collimation for fast, wide-field telescopes is critical to their 
performance, and this is certainly true for the SDSS telescope. The
problem is complicated by the fact that the telescope has moving
optical elements and detectors. A little thought will demonstrate that
it is the axis of the instrument rotator which {\it must} define the
optical axis, and this simple notion makes the whole collimation
process relatively simple. When the telescope was assembled, and each
time it is reassembled after the primary mirror has been removed for
aluminizing, the mirror is centered according to the information
about the optical center provided by OSC. The tilt of the mirror is
determined by demanding that the reading of a depth micrometer carried by a 
radial arm affixed rigidly to the image rotator and touching the face of
the mirror near its edge does not change as the rotator is swung
through 360$^\circ$. It is henceforth assumed that the primary's
vertex is orthogonal to the optical axis, and adjustments to the
primary's x and y position are always accompanied by appropriate
changes in tilt to make this remain true. When the truss and secondary
are installed, a mark on the optical center of the 
secondary is centered using the length
adjustment on the secondary support vanes for coarse adjustment and
the centering motors on the secondary central support for fine. An
alignment telescope mounted on the rotator at the focus and itself adjusted by
rotating the rotator is utilized to make this measurement. The
deflection in secondary centration and tilt as a function of telescope 
elevation is measured using this setup and recorded; it is compensated
for by the control system in such a fashion as to keep the secondary
centered and normal to the optical axis during observations.

With these precautions the only adjustment made in normal operations
is to refine the centering of the primary to remove residual coma,
always doing so keeping the surface at the vertex normal to the
optical axis. This has worked well in practice, though really
only very well after getting the primary mirror thermal problems
under control.

For a telescope operating in TDI mode, adjustments fully as important
as collimation are those required to ensure that the stars track
accurately along the CCD columns and that the tracking rate is consistent
with the vertical charge transfer rate. After some experimentation, 
a simple scheme was evolved to accomplish this. At some arbitrary point
in a scan, the vertical clocks in the CCD are turned off for a precise
short time interval and then turned back on. This results in double
images for each star on any of the arrays at the time this was done.
Centroids for those images are obtained. If the camera is properly
rotationally aligned, the two will lie exactly in the same column, and
if the tracking rate is correct, the displacement will be exactly the 
number of vertical clock cycles skipped. Seeing effects are large enough
that the results of many star images must be averaged to obtain
meaningful results. When this is done, the crossing positions and time
differences in the leading and trailing astrometric CCD arrays can be 
calibrated, and it is through these data that the rotation and scale
are monitored in normal operation (\S3.8)

\subsection{Tracking}

The wind baffle (\S4) reduces wind loading on the telescope OSS by a factor of
ten. With care taken in the design of drive and encoding systems, bearings,
and the structure, the result is a telescope with very low
wind-induced tracking error that performs at the level necessary to achieve
our goals for image quality and astrometric accuracy.  

Crude Komogorov turbulence calculations indicate that we should be able
to achieve of order 30 milliarcsec (mas) astrometric accuracy if
limited by the atmosphere alone. The primary mechanical limit to 
the accuracy of pointing and tracking is the runout of the encoder
rollers, which is of the same order, 30-35 mas. The wind loading
is of very minimal importance nearly all the time; the wind-induced
servo errors averaged over the effective exposure time are less than 
1 encoder unit, about 14 mas on both elevation and azimuth. Comparison
with the UCAC catalog (Zacharias et al. 2004), 
to which we tie our astrometry over most of the
survey area, indicates end-to-end absolute astrometric accuracy of about
70 mas most of the time (Pier et al. 2003). We know that the system
stability, and in particular that of the camera focal plane, are
negligible contributors to this error, and it seems likely that
most of it is due to the so-called ``anomalous refraction'' effects
arising presumably from long-wavelength pressure disturbances in the
atmosphere.
There is some evidence that the accuracy can be improved substantially
by more sophisticated processing which makes explicit use of the observed
correlation structure in the atmosphere, the known error spectrum
of the encoders, and existing archived telemetry information which can exclude
egregious hysteresis events in the secondary support,
but there are no plans at the moment to pursue this.

\subsection{Instrument Change}

It is necessary in the operation of the survey to be able to change
the fiber cartridges on the spectrograph quickly and be able to switch
back and forth from spectroscopy to imaging, 
which involves the removal/installation of the
spectroscopic corrector lens and installation/removal of the camera, both
quickly and with minimal danger to the instruments.

The camera lives when it is off the telescope in the ``doghouse'', an
enclosure permanently mounted to the rotating floor (see \S5). It
sits upon a cart with flanged wheels on rails, and can easily be
rolled under the telescope. A hydraulic lift then picks the camera
and its cart up and presses it against a set of trefoil kinematic mounts
on the instrument rotator. It is held in place by a set of three
over-center pneumatic latches which are disabled as soon as the
camera is mounted. An additional three identical safety latches
are latched at the same time; these provide almost no support but
are present for safety. An additional pair of ``paranoia'' latches
is provided which engage automatically as the cart lowers away.
The empty cart is stored in the doghouse during imaging operations.

When the camera is removed for spectroscopic operations, the 
spectroscopic corrector, normally stored in the ``cathouse'', an
enclosure which is built into the roll-off building, again on its
special cart, is wheeled under the telescope, lifted into place
with the lift, and latched with three of the six pneumatic latches.
There are also automatic paranoia latches for the corrector.

The spectroscopic fiber cartridges are installed using a special cart
which can carry two, one position for the `spent' one just completed
and one for the new one. This cart can be placed under the telescope
accurately in either position. Again the lift is used to install
the cartridge and the other set of latches used to latch it into
place. The same set of kinematic mounts which serve the camera
are used for the cartridges. Yet another set of pneumatic latches 
built into the
spectrographs latch the fiber slit assemblies, built into but
only attached via springs and guides to the cartridge, against a set
of kinematic mounts in the spectrographs.

The instrument latches were built with quite severe space constraints
and require higher pressures than are provided by the observatory
compressed air system; this is provided by a pneumatic booster pump
in the enclosure.

Instrument change is an operation which is performed very often,
performed at night with very subdued lighting, must be done quickly, 
and is intrinsically dangerous
to a set of quite expensive and difficult (or impossible) to replace
equipment. There is a sophisticated interlock system (see \S3.8)
one of whose main functions is to attempt to prevent egregious
errors (such as attempting to drop something or ramming the camera
through the spectroscopic corrector) in this process, but we depend
as much or more on the conscienciousness, care, and training of the
observers as on any such system.

\subsection{Telescope Control Software and Hardware}

Software has been at least as big a challenge for SDSS as is the hardware.  All
SDSS software - the observing software briefly discussed here, the data
acquisition software, and the data reduction and archiving software -
is managed by the open source package CVS (Concurrent Versioning 
System), an invaluable utility which is one of the truly vital
components of the system. Problems and bugs are handled by by the 
public-domain GNATS
problem reporting data base system, where the description, analysis, 
correction and resolution
of every problem, requests for enhanced capability and software failures
are tracked and archived. (The same problem reporting system is used
for all aspects of the projects, not just the software.)

At the lowest level the telescope motion is controlled by an MEI V6U/DSP
16-bit digital PID servo system, which accepts a stream of real-time
quintuplets of position, velocity, acceleration, jerk, and time at 20 Hz, 
position feedback information from the incremental axis encoders, and absolute
fiducials from the tape segments. The output
from the MEI is amplified by a set of modified and repackaged 
Glentec GA4552-1 linear power amplifiers,
which supply current to the DC axis servo motors; these act as torque
transducers to the telescope and rotator. There are five channels, two
each for the altitude and azimuth axes and one for the rotator. The
redundant channels on the telescope axes are used simply to detect
encoder errors; the system is wired so that the pairs of motors on the
axes deliver identical torques (or at any rate are delivered identical
currents) in response to servo errors derived from the primary encoders.
The five motors are all identical, and are Danaher/Kollmorgan QT-7801 
48 ft-lb DC servo motors driving the 100 mm diameter drive capstans directly.

The SDSS telescope is also equipped with an interlock system of 
considerable complexity, built around an Allen-Bradley SLC504 
industrial programmable logic control (PLC) system. 
This system monitors and sets limits on almost all telescope functions,
including allowable movement of the telescope in the enclosure, soft
limits on telescope motion in the open, velocity limits, drive current
limits, various
conditions on the hydraulic instrument lift, the instrument latches,
and instrument and telescope configuration
during instrument change, contact between the telescope and wind baffle,
slip detection on the friction axis drives, 
emergency stop switches, etc. Most of the interfaces with mechanisms
on the telescope are done with PLC modules. 
The PLC system is autonomous in its basic functions, as it must
be for safety, but communicates with the rest of the control system.

At the next level up are two processors which handle motion control and
general telemetry/monitoring, the MCP
(Motion Control Processor) and the TPM (Telescope Performance Monitor),
both implemented via Motorola MVE162 VME processors. Their tasks
are very much as suggested by their names: The MCP generates the 
real-time control data required by the MEI servo system, integrates the
fiducial information into the data stream from the incremental axis
encoders to produce absolute axis positions, institutes its own notion
of velocity and acceleration limits and a few other complex interlock
issues, serves as a
interface between the interlocks and higher-level control software,
and controls (via the PLC) many auxiliary devices such as the flat-field 
screen, the flat-field lamps, etc.

The TPM is a real-time and archival performance monitoring system 
(McGehee et al. 2002) using
two MVE162 processor cards, which receives
data from the interlock system, the axis servos,
and independently from many other subsystems.
The first processor, which shares a VME crate with the MCP,
monitors the MCP and PLC data via a shared-memory interface with the MCP, 
status of the LN2
pressure system, the primary mirror support servo, the drive amplifiers,
wind speed and direction, and the drive slip detection hardware. 
The second monitors all serial communications, including the thermometer
network, the scales which monitor the weight of the 180-liter LN2 supply
dewars, and the mirror support actuator Galil controller.
The TPM runs the 
Experimental Physics and Industrial Control System (EPICS) 
control/monitoring software system and can be queried for the
state of any of the subsystems which it monitors in real time, and in
addition writes data to disk which are permanently archived. The
archives can be accessed in much the same way as the real-time
displays are generated.

The telescope control computer (TCC), a DEC Alpha system running VMS,
is the astronomical part of the
control system; it is essentially identical with its counterpart for
the 3.5-meter telescope, even though all of the lower-level hardware
is completely different. It
handles all of the interface between telescope motion and the observers,
keeps a telescope pointing model, takes care of all the tedious (and
complex) coordinate
and tracking issues, etc. Since the SDSS telescope images in TDI mode,
one of the things it must do, and does, exquisitely well is to track along
an arbitrary great circle on the celestial sphere at constant apparent rate. It
computes and feeds (position, velocity, time) triplets to the
MCP, which interpolates in this set and feeds the servo hardware. It
accepts offsets in any one of a number of coordinate systems to move
the telescope differentially for pointing corrections. This facility
is used for target acquisition and automatic guiding in spectroscopic mode.

At the highest level is a Unix machine called sdsshost, running
control interface operations software mostly coded in Tcl/TK.
The main control interface is called IOP (for imaging operations)
and is supplemented by SOP (spectroscopic operations), a set of tasks
specific to observing in spectroscopic mode. These programs
communicate with the data acquisition computers specific to the
instruments and to the TCC, MCP, and TPM. Tasks include data archiving,
guiding (for spectroscopy), record keeping, general instrument control,
calibrations, running and keeping the system log via a logging package
called  murmur developed at Fermilab, running a routine called
Watcher which gathers data from all the subsystems and warns of
dangerous and general out-of-spec conditions on all the monitored
subsystems, and an interface to the interlock system which allows the
observers to find quickly where in the complex interlock tree a
condition exists which has triggered an interlock. This host machine
is currently a Silicon Graphics Challenge, chosen many years ago because
of its ability to interface more-or-less directly with the VME 
instrument computers. The entire data acquisition system (including
sdsshost), which has
become something of a maintenance liability,
will be replaced in the summer of 2005 with more modern
machines.

Several software routines work with the data stream to provide instantaneous
feedback on the data quality. Since the SDSS imaging along great circles runs
``open loop'' - that is, the tracking rate of the telescope and the
TDI readout rate of the imaging camera must be the same but are not
connected by feedback loops - the rates of the tracking and image rotator
must be precisely monitored and recorded. The SDSS accomplishes this
during imaging mode by examining the imaging data as they are taken.
The SDSS imaging camera contains two 
arrays of CCDs - as well as the imaging CCDs, a leading and trailing set
of shorter exposure CCDs acquires $r$ band data over the dynamic range
straddling the imaging camera saturation threshold of about 14th magnitude
and that of the astrometric catalogues, thereby providing astrometric
calibration of the SDSS imaging.  During imaging observations, 
an astrometric routine compares position information from the
astrometric arrays, both leading and trailing, with the on-line
astrometric catalogues FASST and UCAC (Zacharias et al. 2004; see also Pier et
al. 2003 and Gunn et al. 1998), thereby measuring the tracking rate and 
the lateral position of the imaging camera with respect to the 
predefined scan path, and the image rotator offset. A second routine
compares the positions of {\it all} stars seen in the leading
and trailing astrometric arrays (not just the stars in the astrometric
catalogues), providing better measures of the tracking rate (via the 
time it takes a given star to cross the imaging array and be detected
in the trailing astrometric array) and of the field rotation (measured
by whether a star crossed the same CCD column in the leading and trailing
astrometric chips).

The variation of the imaging PSF is also monitored by measuring the Gaussian
widths of stars detected by the data acquisition system.
This diagnostic gives information
on whether the telescope is in focus, but more importantly monitors the
seeing. SDSS imaging is done in pristine weather with good seeing, 
while spectroscopy uses any usable conditions, so
it is important to closely monitor the image quality during imaging
observations. It is also important, of course, to know when it is 
appropriate to switch to imaging; these decisions are made partly on the
basis of measured image diameters of the spectroscopic guide stars,
but heavily supplemented by images from an all-sky 10 micron 
camera which is very sensitive to cloud cover (Hogg et al. 2001) 
and a differential image
motion monitor (DIMM) which continuously monitors the seeing.

During spectroscopic observations (typically three 15 minute pointed 
observations for each plate) 
the necessary information on tracking and guiding is
provided by real-time analysis of the data from the spectroscopic guider,
a small independent CCD camera which images the output of 11 coherent
fiber bundles placed on the images of relatively bright guide stars.
Information on image smearing and on cloud cover is provided indirectly by
monitoring the signal-to-noise ratio of the spectroscopic exposures. The
quality of the night itself is provided by the observations taken automatically
with the auxiliary 0.5 m Photometric
Telescope (Hogg et al. 2001; Tucker at el. 2005), and the aforementioned
10-micron all-sky camera.

\section{The Wind and Light Baffles}

\subsection{General Considerations}

The SDSS telescope uses a roll-off enclosure which is very compact and has a
low cross-section for wind loading. This greatly reduces both the mass and
the cost of the enclosure base. The low mass of the enclosure base in
turn greatly reduces thermal differentials between the structure and
the ambient air, leading to much-improved near-field seeing performance.
The properties of the enclosure will be described in \S5 immediately
below. However (see Figures 1 and 2), the telescope is completely exposed to
the wind and to light sources during observing, and is protected by
a set of external and internal light baffles. The requirements for
a high-accuracy, wide-field photometric survey telescope are quite
stringent and somewhat different from those for normal astronomical
telescopes, and was the focus of a quite considerable effort
in the design of the project. We thought it therfore worthwhile to go into some 
detail in the description of the baffle design and implementation.

The baffle sytem was specified to provide
effective baffling of the crescent moon at positions of $\rm 30^{\circ}$
or more from the optical axis and of the lights from the nearby town
of Alamogordo, and especially those of Holloman Air Force Base, both of which
lie below and to the west of the SDSS site. This
translates into a specification for the point source normalized
irradiance transmission (PSNIT) - the ratio of the stray light irradiance
at the focal surface to the incident irradiance from a point source - 
of less that 2$\rm \times 10^{-6}$ for sources more than $\rm 30^{\circ}$
off-axis.  For sources closer to the optical axis, higher values of the 
PSNIT are permitted, but it is desirable that the focal surface illumination
be uniform. 

There are two main techniques for tackling the problems of stray and scattered
light. The first is to paint all relevant structural surfaces black; suitable
aerospace paints, such as the Lord Aeroglaze Z306 which we use, 
have a hemispherical reflectance of less than 5\% for
normal incidence, but this rises to over 15\% for grazing incidence
(Pompea \& Breult 1995).
The second is to block scattering paths; and here, edge scatter is a 
concern.  In the following discussion, {\it critical objects} are those
visible from the focal surface either directly or by transmission or reflection
by the optical elements. Elimination to the extent possible 
of critical objects and reducing and lowering the reflectivities
of those crucial to the structure which remain is the crucial 
step of light baffling.  In the following {\it Illuminated objects} 
are those illuminated by
a light source directly or via the optical elements.

\subsection{The Baffle System}

The baffles for the SDSS system were designed using the following criteria,
which apply for the worse-case combination of component and installation
tolerances and flexure for elevations between $\rm 30^{\circ}$
and $\rm 90^{\circ}$:

\begin{itemize}

\item
All direct rays to the focal surface must be blocked, since even a small
amount of direct illumination will cause photometric errors.

\item
Variable vignetting of the $\rm 2.72^{\circ}$ diameter photometric field
of view due, for example, to gravity or wind loads, must be much
less than 1\%.

\item
Variable vignetting of the $\rm 3.02^{\circ}$ diameter astrometric
field of view (see Gunn et al. 1998) must be less than 2\%.

\end{itemize}

The SDSS telescope is protected by a {\it wind
baffle} that closely surrounds the telescope but has a separate
low-precision drive system and transfers wind loads to the stationary
portion of the telescope building.  The wind baffle has a square
cross-section that fits closely around the square secondary frame of the
telescope, and is constructed to efficiently serve as the external 
{\it light baffle}. In addition,
the telescope structure has the usual internal light baffles protecting 
the secondary and primary mirrors.  However, the optical system 
includes two transmitting elements which must also be baffled. To avoid
excessive central obstruction of the entrance pupil, a conical baffle
is suspended approximately midway between the primary and secondary
mirrors (Siegmund et al. 1998).

The baffle system for the SDSS telescope is shown in Figure 3.
The sky-facing end of the wind baffle contains an
annular opening formed by
a central disk and a panel with a circular opening (both supported by the
wind baffle frame).  This opening provides clearance for light from the
3$^{\circ}$ field of view to reach
the telescope entrance pupil. The wind baffle 
blocks light rays that would otherwise have to be
intercepted by the other baffles and prevents direct illumination of the
primary mirror by sources more than
27$^{\circ}$ from the optical axis.

The inner baffles consist of the secondary baffle (in front of the
secondary mirror), the primary baffle (extending through the primary 
center hole), and the conical baffle (suspended between the primary and
secondary mirrors). Most conventional two-mirror telscopes have no analog of
the conical baffle, but a similar arrangement has been used at the
Meyer 1.5 meter telescope at Palomar and a much more complex but
analogous design on the 2.5 meter Las Campanas telescope. Extra
baffles of this general sort are necessary in fast, wide-field
Cassegrain designs with large secondaries, 
to avoid an unacceptably large central 
obstruction of the entrance pupil.

The primary and secondary baffles each consist of a stack of annuli.
Short struts connect each annulus to the next one in the stack and 
control spacing and centering. The outer surface of each strut 
contains a vane to block near-grazing scattered light paths. The
annular baffle design facilitates the air circulation near the optics 
needed to bring them to ambient temperature, is light weight,
is easy to fabricate, and has low wind resistance.

The annulus spacing for both mirror baffles is closer further away
from the mirror, and each baffle is terminated by a truncated cone (Figure 
3). The cone surfaces are vaned to eliminate near-grazing light paths.
The conical baffle is also vaned to minimize its cross section and
avoid excessive direct light blockage.

Of the two optical configurations, imaging has the
more stringent requirements, so the telescope baffle design is
optimized for the imaging configuration.

The baffle design was modified after construction of the optics.
In the original telescope design, the inner edge of the entrance pupil
is defined by the tip of the secondary baffle. However, the inside
diameter of the primary mirror as built was larger than designed because
of damage incurred during surface generation (see \S6.1), and instead the
inner edge of the entrance pupil is defined by the 1.20 m
outer diameter of the primary baffle annulus nearest the primary 
mirror, at least for small field angles. At field angles larger than
about 38 arcminutes the effective pupil becomes quite complex, as the
shadow of the secondary baffle emerges from the central primary baffle.
The 2.52 m diameter of the outer aperture stop was chosen to
mask the surface of the primary near its outer edge because of a slightly
turned edge 
which occurred during polishing. The stop is 
located just above the primary mirror and is integrated with the 
supports for the primary mirror seismic 
mirror cushions.

The secondary baffle must be large enough to mask the edge of the secondary. 
It is also desirable that the baffle and secondary supports not be
critical objects. The 1.28 m diameter of the secondary mirror baffle tip
was minimized consistent with these criteria - if the diameter of the 
baffle tip were increased, the conical baffle would have to become shorter and
moved towards the primary mirror. While this would decrease the central
vignetting, the central obstruction dominates such that the total blockage
increases, even at the field edge.

\subsection{Performance of the 2.5 m Baffle System}

Analysis of the entire baffle design using Breault Research's
Advanced System Analysis Package (ASAP)
indicates quite satisfactory performance.
This design achieves over the whole range of
incidence angles greater than about 30$^{\circ}$ off axis ratios of incident
flux to flux in the focal plane (PSNIT) better than $2\times 10^{-6}$. 
This translates, for example, to a scattering contribution to the night
sky in the focal plane of about 26.0 mag/arsec$^2$ by a quarter moon 40
degrees off axis.  This is 2\% of the {\it dark\/} night sky, and less than
half a percent of the moonlit night sky.  

Stray light analysis of the as-constructed 2.5 m telescope was performed
using Breault's  APART analysis program 
(Breault 1995). For source angles of $\rm 25^{\circ}$
and less, the most important critical object is the directly-illuminated
aperture stop located just above the primary mirror. For angles greater
than $\rm 25^{\circ}$, the most important critical objects are the 
interiors of the primary and conical baffles, which are illuminated
by light scattered from the interior of the wind baffle. At angles
of $\rm 30^{\circ}$ or more, critical objects are no longer illuminated.
The resulting PSNIT is shown in Figure 16.

The fast focal ratio and wide field of view of the SDSS 2.5 m telescope 
precludes the typical mirror baffling system, in which the primary
baffle is contained within the shadow of the secondary baffle -
the resulting central obstruction would be more than 50\%. The SDSS
baffle system produces a much lower central obstruction, at the cost 
of minor blockage by the conical baffle and some differential vignetting
with field angle (which does not occur in the typical mirror baffling
system on other telescopes). The resulting total obstruction is 28.6\% on-axis
and 31.8\% at the field edge, resulting in differential vignetting 
of 3.2\%. Most of the obstruction is due to the secondary baffle and 
most of the differential vignetting to the conical baffle (Figure 17).

\subsection{Construction of the Baffles: Details}

\subsubsection{The Wind and Light Baffle}

The requirement for the wind baffle is that it should reduce the wind load
on the telescope by a factor of 10. 

The sides of the wind baffle (Figure 3), fabricated under contract with 
CVE Machining, are covered with wind-permeable panels.  The
panels have 25\% equivalent open area, and
consist of interlocking ``C''
cross-section elements.  Light paths through the panel require scattering
from a minimum of two surfaces. Thus, with suitable coatings, the panel can
be made quite light-opaque.

Water tunnel studies indicate that with this 25\% porosity, the
flow speed around the telescope secondary is reduced to about 1/3 of the
free stream flow speed while the flushing time for fluid inside the baffle
is still a very rapid 15 to 30 seconds. Flow
passing through the channels in the panel diffuses rapidly, i.e. on
scales of 0.1 m. No spatially persistent high-velocity
jets that might cause wind-induced tracking error are observed.

Centered on the telescope azimuth axis and flush with the telescope
enclosure floor and the top of the telescope fork base is a motor-driven 
circular floor panel that follows the motions of the azimuth axis of 
the telescope.  It
supports the wind baffle altitude drive, drives it in azimuth and provides
rotating floor space around the telescope for the storage of the camera
when the spectroscopic system is on the telescope. This structure is
suspended from a ring built into the building and is vibrationally
isolated from the telescope pier.

Traditional telescope enclosures act as cavity radiators. The net radiation
imbalance with a clear sky is roughly 100 watts/m$^2$ of horizontal
projected area. In a well-designed telescope enclosure with low thermal
inertia and minimal heat sources, this power (kilowatts for a typical slit
size) comes from conduction from the air within the telescope chamber,
i.e. via the production of colder than ambient air. This cold air can
cause image degradation should it enter the telescope light path.

The wind baffle reduces radiative coupling of the telescope OSS to the sky
by minimizing the area of the opening at the end of the telescope. The
outer surfaces of the wind baffle are polished aluminum. Wind baffle surfaces 
which must be black in the visible to absorb scattered light are painted
with Lord Aeroglaze Z306; this coating has very high thermal emissivity
but the areas involved which are well coupled to the sky are small.

The control system for each of the two axes of the wind baffle is 
implemented using a Sumitomo SS-6100 AC PID servo system controlling
A830-GTMT10 high-power AC servomotors, three on the azimuth axis
gear-driving the support wheels for the rotating floor, and one
on the elevation driving the wind baffle via a chain drive.

The wind baffle is referenced
to the telescope by a pair of Balluff BTL-5-G11 magnetostrictive sensors,
which serve to keep the windbaffle centered on the telescope to of 
order a millimeter. The windbaffle can keep up with the 
telescope adequately at the maximum accelerations we use, which
are 0.1 deg/sec$^2$ in elevation and 0.2 deg/sec$^2$ in azimuth.

Mounted on the sky-facing end of the wind baffle are eight segments
made from honeycomb aluminum panels and driven by DC gearmotors
which can be opened completely out of the light path or closed to
almost completely cover the opening in the baffle. These are coated
on their lower (when closed) surfaces with Labsphere's Duraflect,
which provides
an accurately Lambertian reflective surface over the entire wavelength range
of interest to the SDSS. These are illuminated by projectors mounted
in the corners of the wind baffle just above the PSS for spectroscopic flatfield
calibration and wavelength calibration. To this end there are four
sets of three projectors, the central one of each set a quartz-iodine
flatfield lamp, flanked by a mercury-cadmium lamp on one side and a 
neon lamp on the other. The projectors have a condensing system and
an imaging system which projects a pupil segment on the screen to
minimize stray paths through the optics. There is in addition a low-power 
module consisting of a small filtered quartz-iodine lamp and a
UV LED on each lamp cluster which are used to produce reference flat fields
for the camera. These are used only to
monitor changes in the response of the imaging CCDs.

\subsubsection{The Primary Baffle}

The internal baffles - the conical baffle mandrel and the primary and secondary
baffles - were fabricated by Machinists Inc., Seattle, WA.

The tip of the primary baffle (Figure 18) is a machined aluminum truncated cone
with vanes. The balance of the baffle consists of progressively larger
aluminum annuli with sharp inner and outer edges. The spacing of the annuli 
increases towards the primary mirror.  The outer edges of the annuli are 
directly illuminated from the sky and by the secondary mirror. The edge
bevel faces the secondary. 
The finished primary baffle assembly weighs
63 kg, and the ellipticity  of the upper edge of the
primary baffle tip is less than 1 mm. The assembled primary baffle is shown in
Figure 19.

The primary mirror aperture stop is located just above the primary mirror
and defines the outer edge of the entrance pupil. The edge bevel faces
the primary mirror. Integrated with the primary stop are the 
elastomeric retention cushions which protect the
primary mirror during seismic accelerations, particularly when pointed at
the horizon (the original intended stow position for the telescope, see
Figure~12).

\subsubsection{The Secondary Baffle}

The design of the secondary baffle is similar to that of the primary
baffle, but is inverted and shorter. The baffle tip is a machined
aluminum truncated cone with vanes (Figure 20). Its tip is more 
conical than the rest of the cone and blocks the view of the baffle exterior
from the focal surface. The balance of the baffle consists of smaller diameter
aluminum annuli with sharp outer and cylindrical inner edges. The spacing
of the annuli increases towards the secondary mirror. The outer edges of the
annuli are illuminated directly from the sky. The edge bevel faces the sky and 
is not a critical object.  The inner edges of the lower surfaces are 
illuminated by the converging beam from the primary mirror. The cylindrical
inner surface is not illuminated. The finished secondary baffle assembly weighs
25 kg and the maximum diameter of the lower edge of
the secondary baffle tip is 1.28m. 

\subsubsection{The Conical Baffle}

The conical baffle design is illustrated in Figure 21.
It is the most challenging of the baffles to fabricate: it is supported
by small (1.5mm) steel rods from the telescope truss so must be lightweight; 
to resist
tension from the rods, it must have good bending stiffness; to minimize
on-axis light blockage, its central thickness must be minimized; to minimize
vignetting at the edge of the field of view (Figure 3), the thickness 
near its lower and upper edges must be minimized; and vanes must be added
to the baffle surfaces to interrupt grazing scattered light paths.
These constraints lead to a design with 
a relatively deep central vane that provides the baffle with bending 
stiffness and anchor points for the supports (Figure 21). Extending
above and below the central vane are 1.6 mm thick conical segments with
half-angles of $\rm 10.45^{\circ}$ and $\rm 8.83^{\circ}$ respectively.

The conical baffle was fabricated by Quality Composites Inc. (QCI) of
Sandy, Utah. It was the largest part of this sort ever attempted by QCI,
and three aspects required development: (1) fabrication of the inside vanes
so that the tips were smooth and uniform and so that good adhesion
to the cone was achieved: (2) fabrication of the central stiffening vane
and installation of the stainless steel inserts to which the support rods 
attach: and (3) design of the outside vanes.  These are 750 micron bare acrylic
optical fibers tacked to the surface at 80 mm intervals with superglue 
(cyanoacrylate adhesive). Subsequently, a fillet of epoxy was applied between
the surface and either side of the fiber.

The baffle was formed on a machined aluminum mandrel out of graphite
fiber reinforced epoxy. Grooves were machined into the mandrel and formed
the vanes on the interior surface. The mandrel was made in three pieces
so that it could be collapsed inside the finished baffle and removed.
Its weight was minimized, which reduced the thermal inertia and speeded 
the curing of the epoxy. 

The finished conical baffle weighs only 9.8 kg, and all specifications
were met or exceeded. In particular, the out-of-roundness
was 3 mm or less from both the upper and lower edges of the baffle,
the surfaces are smooth and uniform, and the fibers are strongly bonded by
the adhesive.

The completed telescope baffles are shown in Figure 22

\section{The Enclosure}

The SDSS telescope site is about 90 meters (300 feet) south of the
existing 3.5-m telescope enclosure at the Apache Point Observatory,
about 20 meters (70 feet) west-southwest of the ridge top,
i.e. in the prevailing upwind direction, and allows the telescope to
be located at the same level as the ridge top and still be above the
trees, 9 to 12 meters high (see Figure~1). The telescope is built
on a platform constructed out over the slope of the mountain, allowing
access to the telescope at ridge ground level and the location of the telescope
azimuth support, bearing and drive systems
below ridge ground level, thereby greatly reducing the 
height of the structure. This platform faces west, into the prevailing 
laminar airflow. A roll-off enclosure covers the telescope when it is
not observing and is rolled off downwind of the telescope, 
eliminating wake turbulence from the enclosure itself. The structure of
the SDSS telescope and its platform in this way also allows for adequate
storage for the accessories: the camera, spectroscopic plates etc.
A support building on the ridge top, near the telescope, is used for plate
plugging and the storage of plug-plate cartridges.  With the support
building level with the telescope, plug-plate cartridges can be wheeled
easily between the support building and the telescope. Care must be
taken, however, to ensure that warm air generated by the drive mechanisms 
beneath the telescope is adequately vented and does not degrade the 
seeing.

The telescope enclosure is a roll-away rectangular frame structure mounted
on wheels, seen in Figure~1; a side and end view are shown in Figure 23.
During
observations, it is rolled 60 feet
downwind from the telescope to the top of the ridge.  Large doors
on either end of the enclosure are opened during this operation to
prevent interference with the telescope and to reduce the wind load in
the direction of motion. 

The SDSS telescope enclosure is designed with several specific needs in
mind. During spectroscopic observations, it is necessary to change
plug-plates cartridges about once per hour, and the plug-plate cartridges
have a mass of about 100 kg.  For this, and other reasons,
the telescope is mounted so that we have access to it via
a level track, yet still have sufficient ground clearance not to degrade
the seeing.

A roll-off enclosure has several advantages. The thermal ones
have been recognized for some time, but others include:

\begin{itemize}
\item The telescope enclosure is considerably smaller and less
expensive at than a conventional enclosure since it does not need to
accommodate the entire volume swept out by the telescope. 

\item The telescope enclosure wake is minimal, especially when open. It is
important to avoid buildings with large wakes at sites (like APO) with
other telescopes, since air overcooled by conduction to the ground is
transported throughout the wake volume by turbulence. If the light path to a
telescope passes through such a wake, image quality degradation will
result. 

\item
The surface area obstruction to the telescope's field of view is 
minimized.

\item When realuminization of the primary mirror is required, a hoist
mounted to the ridge beam serves to lift the primary mirror out of the
telescope.  The enclosure is used to transport the mirror over the bed
of a truck waiting at the ridge top.  This eliminates the need to
provide a large pathway for the primary mirror within the structure of
a conventional enclosure.

\item Traditional dome-induced seeing is eliminated.  Sources of
heat, radiative cooling, and thermal inertia are reduced and the
flushing of the volume surrounding the telescope is improved.
The enclosure design was studied carefully by water-tunnel experiments
on scale models of the SDSS platform, telescope, enclosure and prevailing
wind pattern (Forbes et al. 1991; Comfort et al. 1994).

\item The visual impact of the building is reduced as compared with
a traditional telescope enclosure because of its reduced size.

\item The tallest trees at the site are located just below the ridge
top and near the telescope site.  If the telescope were placed at the
ridge top, it would have been downwind of these trees and the turbulence they
produce.

\item With the enclosure track built up the slope, only the columns
near the telescope need be long.  Much of the track is near ground
level, allowing better access for maintenance.

\end{itemize}

The SDSS enclosure does have some disadvantages, including:

\begin{itemize}

\item
The telescope is unprotected if the building cannot be closed. However,
the bridge crane components are a mature technology and are very reliable,
and in the event of power failure we are backed up both by a sitewide
generator facility and a local generator.

\item
The telescope is not protected from the wind or stray light, and this must be
addressed separately. This is done by the very effective and elegant
wind baffle solution discussed in the previous section.

\item
Locating the telescope off the top of the mountain is really only appropriate
for sites that have prevailing winds from one general direction and that
have broad summits where the windward edge is likely to have the best image
quality.

\item
It was not possible with the initial very low-profile design to move 
the telescope through its full range of motion
with the telescope enclosure closed. In particular it was not possible
to point the telescope to the zenith, where one must go to
change or more than superficially access the instruments. This is
discussed further below, but the short summary is that we were forced
to redesign and rebuild the enclosure to address this issue
(cf. Figure~12). 

\item
It is much more difficult to seal the moving building against inclement weather
and pests than is the case with a conventional dome. This is a problem
we have not completely satisfactorily solved at this writing.

\end{itemize}

The floor surrounding the telescope is well coupled to the night sky and is
in the stagnation region behind the leading edge of the floor.
Consequently, some attention was paid to minimizing its emissivity to
prevent it from cooling too much. As originally constructed,
the floor was covered with 6 mm thick, 
sanded aluminum plate. The subfloor is a composite structure of
plywood, rigid foam, and corrugated steel sheet. The thermal
performance is satisfactory except in occasional conditions of very
low wind speeds.

The enclosure itself is constructed of 
steel framing sheathed with six inch thick panels made of 
prestressed steel with rigid foam cores. It is provided with both air
conditioning and refrigeration systems to control the daytime indoor 
temperatures, and a blower system (see below) to control temperatures
during nighttime operation.

Storage for tools, equipment, and cryogens is always a problem near a
telescope. It is undesirable that such items be in the telescope chamber
since they store heat and interrupt air flow. However, it is inconvenient
if they are too far from the telescope. The roll-off enclosure provides a
good solution. A storage platform is located on the north wall of the
enclosure. Items stored on this platform are located conveniently to the
telescope when the enclosure is closed, but away from the telescope for
operation. A special cart for the spectroscopic corrector is housed
in the structure for this platform as well.

There is a room underneath the telescope the primary purpose of which
is to enclose the electronics racks which hang from the rotating floor
and to house the mechanism for the rotating floor. The racks house
the TPM, the MCP, the interlock system PLC, UPS units for all the telescope
systems, a 24V DC power distribution system, the hydraulic pump for
the lift used for instrument change, a rack of support electronics
for the camera, a large thermoelectric fluid chiller and associated
pumps and power supplies for heat exchange with the flat-field lamps and 
electronics mounted on the camera and spectrographs,
and all of the communication electronics
for the telescope.   The room also provides
some workspace. It is vented by a large axial blower
which moves about 7000 cfm (sea-level equivalent) at low differential
pressures, which is how it is used currently. The mirror ventilation 
is accomplished by two centrifugal blowers moving a total of 2500 cfm 
mounted in the telescope fork
and exhausting into this space.
The room is provided with louvers on the
west end, opposite the blower, to facilitate rapid laminar exchange of air
and elimination of the heat generated by the telescope electronics.
The heat is dumped at ground level 
about a hundred feet east (and generally downwind)
of the telescope. The seeing with occasional easterly winds is usually 
very poor
anyway, so there is little disadvantage in this relatively simple arrangement.
The current configuration dissipates about 4 kW into the air 
in this room. If the coupling
to the flow were perfect this would result in about a 1C temperature
rise in the room. The observed average temperature rise in the structure 
is about twice this,
but the insulation and radiation to the night sky keeps the exterior of the
room and the telescope floor at or slightly below ambient.

A number of features address personnel safety. Controls are designed so
that the operator must be at the end toward which the enclosure is moving.
A
warning bell sounds and a time delay passes before any movement occurs. The
push button must be held down continuously for motion to occur; if the
button is released, the enclosure stops. The perimeter of the platform
surrounding the telescope is protected by a railing. The outdoor portion of
the deck is covered with galvanized steel grating to provide personnel
safety but minimize snow buildup and heat retention.

It was originally believed that the lack of ability to move the telescope 
to the zenith inside the enclosure
would not be a
serious problem
because APO typically has good weather and work could be done
with the enclosure rolled off. However in practice the time so
available for working on the telescope and instruments  proved woefully 
insufficient, and personnel movement inside the enclosure was so 
restricted that little could be done when the telescope was 
inside the enclosure. This was sufficiently serious that the
enclosure is certainly responsible for much
of the difficulty, delay, and unanticipated expense encountered in 
commissioning. It was
especially serious because during bad weather, the telescope could
not be accessed night {\it or} day and in good weather the telescope 
was severely heated
by the sun when it was in the open in the daytime --- and the mirror
ventilation system, as discussed above, was insufficient to the task,
so the performance at night was very poor. In any event, the enclosure
was rebuilt in the fall and winter  of 1999-2000 by adding a central 
raised section to allow moving the telescope to the zenith.
This construction was successful but fraught
with difficulty; the composite steel panels, ordered to match the
ones in the rest of the structure, could not in the end be obtained
on any satisfactory schedule, and when they finally were obtained from
a vendor outside the USA they were sized on the metric system and
were {\it almost} the right size. These problems pushed the
construction into the winter and severely tested the temporary
enclosure constructed to protect the telescope during this 
period. The successful completion was accomplished to the delight and relief
of everyone involved and enormous advantage to the progress
of the project. 

\section{Installation, First Light, and the Future}

The description of the SDSS 2.5m telescope in the previous sections focusses
on some of the many powerful and imaginative aspects of the design that have
made possible the high quality of the SDSS data. Many of the techniques
used in the design, construction and testing of the SDSS hardware were 
deployed here for the first time. This section provides a brief,
incomplete and rather anecdotal history
of the construction of the 2.5 m up to first light, as well as 
a discussion of future plans for the telescope. The installation of
the SDSS telescope proceeded in the following order: site preparation
and enclosure; telescope
bearings, pillow blocks and co-rotating floor panel; 
azimuth bearing; telescope; optics; interior baffles; wind baffle;
interlocks and control system; first light with the completed imaging
camera; and finally installation and testing of the spectroscopic 
system. Improvement of the thermal environment became a high-priority 
task as soon as the system was far enough along to realize that we
had serious problems, but was not really finished until the survey
was well underway.

\subsection{Optics}

The design, construction and testing of the SDSS 2.5 m optics - the 2.5m
primary mirror, the 1.1m secondary mirror, the common first-element
Gascoigne lens, and the imaging and spectroscopic lenses, are described
in \S2 and by Gunn et al. (1998). As delivered, the optics were of
excellent quality, with each element meeting or exceeding specification.
The
optical surfaces and axes were accurately measured with respect to mechanical
references so that mechanical collimation of the telescope was straightforward.
In particular, the null lens which tested the primary mirror
was verified using a computer-generated hologram (Burge et al. 1994),
while interferometric testing of the secondary 
used an innovative and powerful
holographic test against an almost-matching sphere (Burge 1997). The 2.5 m
secondary was the first such
mirror to be so tested. It is worth recalling that
the time period when the SDSS large optics were under construction
closely followed the disastrous discovery right after launch that
the primary mirror of the {\it Hubble Space Telescope} had been 
(beautifully) polished to the wrong figure using an untested null lens,
causing severe spherical aberration (Allen 1990). In this climate, even 
more care was taken and more redundant testing carried out than normal.

At this writing (June 2005), when the SDSS has been underway very successfully 
for five years, it is easy to forget the sometimes heart-stopping
misadventures which befell both mirrors. The primary mirror
was among the first parts of the telescope to begin construction,
informed by much experience that the construction of the primary mirror
is the rate-determining item in the construction of any telescope, and
was the first large cast blank to be made by Hextek corporation. Casting began
in July 1992, but this first attempt failed during annealing, and when
the oven was opened the mirror was found to have cracks. The causes
were identified and corrected, and it was realized that casting need
not be redone {\it ab initio} -  reheating 
and re-annealing the mirror was likely to work. The mirror was reheated in 
January 1993 and re-annealed. This was successful: the cracks fused and
the resulting blank was found to be of high quality and to have low
internal stresses, a property which proved vital during the loss of 
the next of the mirror's nine lives.

The blank was delivered to Arizona Technologies, Inc., for optical
generation under subcontract from the Optical Sciences Center, who
had overall responsibility for figuring the mirror. 
During this process, the blank was seriously
fractured by the generating tool when there was an accidental partial
power failure - the turntable holding the blank continued to rotate
and the feed screw for the grinding wheel continued to advance, but the
power to the grinding wheel itself failed.  Fortunately, the fracture was
close to the inner, very large, hole in the blank. The fracture was
about 85 mm outside the edge of the center hole, quite cylindrical, 
and went through the front surface and into the ribs below (cf. Figure~7).
In some ribs, the cracks continued down to the back plate but did not
pass through it.  The proposed fix, which proved successful, was to
remove the inner annulus of the mirror containing the damaged elements
- the inner skirt and inner portions of the face- and back-plates
and ribs. This accident necessitated
the redesign of the conical light baffle (\S4.4.4).
Optical generation was then completed at the Optical Sciences Center,
University of Arizona, where the final figuring, polishing and testing
phases were also carried out. The mirror was polished to about 100 nm
but as noted in \S2 and \S3,
the mirror was slightly stressed during polishing and the relaxed mirror
had 1/2 wave of astigmatism.  Rather than starting over, the
mirror mount was configured to slightly bend the mirror and bring it 
into tolerance. The primary was aluminized at KPNO, and 
delivered to APO in July 1996. It is annually cleaned and 
realuminized (cf. Figure~8). The mirror has suffered no further
crises and has performed in an exemplary manner.

The secondary mirror was also seriously damaged during operation
by its focus actuator (Figure~15) very shortly after the installation
of a new secondary support system, the one described in \S3.4.3.
The original one, a more-or-less direct copy of that on the 3.5 meter,
proved mechanically inadequate for our considerably larger and heavier
secondary. A dimension error was made in the design 
which, when translated into hardware, allowed
the central support shaft to contact the glass when the mirror actuators
were fully retracted. It did, and
on October 19 1999, about four days
after the end of the dark run, the mirror was discovered to have several 
circular cracks near its center around its axis of symmetry. 
The mirror was stabilized by
drilling several holes to intercept the cracks and shipped for
testing to the Steward Observatory Mirror Laboratory. It
was determined that the cracks had indeed been caused by 
excessive force exerted
normal to the mirror surface along the axis of the central support
pin.
The support structure was modified (to the proper dimensions) with 
a new control system containing
a force limit switch which directly interrupts power to the actuators
in the event of excessive force on the mirror.

SOML determined that, miraculously, the figure of the mirror was
essentially unchanged and that the  mirror could be repaired by removing 
its center section. 
Fortunately, the center of the SDSS secondary
mirror is not part of the optical path because of the large central
obscuration, but {\it is} used for collimation, so a thin E6 meniscus
of the correct curvature was made and bonded to the center of the mirror
for this purpose, and the optical center mark very carefully transferred.
The mirror was
repaired and realuminized at SOML and returned to APO at the end of
January 2000. The telescope was back in operation on Feb 15 2000 and there has
been no further trouble. The telescope down time was not a total loss ---
other work, such as the rebuilding of the telescope enclosure and the
installation of the flatfield screens and a set of Hartmann test screens --- 
was accomplished during
this time. The SOML tests had not been able to verify that the figure
of the secondary on large scales was unchanged, and in the summer of 2000
the mirror was shipped to Lick and tested on their precision profilometer 
(Miller, Hilyard, \& Mast 1998) and found to be essentially unchanged.

The first corrector lens
is relatively straightforward, and was generated, figured and polished using a 
null test under contract with Contraves Inc. It was delivered in March
1997. The lens met our specifications except for a region some 45 degrees
in angular extent by about 100 mm or so width 
at the edge, as has been mentioned before. When we
shipped the secondary to Lick for profilometry we sent this element as 
well, but the lens was too thin to support properly and the tests were
somewhat indeterminate.

The second corrector for imaging mode is very aspheric, with an aspheric
sagitta of more than 8 millimeters, but the accuracy required is not
very high by optical standards and the surface was monitored with
sufficient accuracy with a simple profilometer during generation,
polish, and figuring. The lens was completed in the spring of 1996
by Loomis custom optics. The fabrication
included drilling the CCD optical bench mounting holes (see Gunn
et al. 1998) in the flat side of this corrector,  and was concluded by the
application of a very complex
striped antireflection coating which matches the camera photometric 
bands.

The second corrector for spectroscopic mode was generated and figured 
under contract 
with Tinsley Laboratories, and was delivered in June of 1996. All of the
refractive correctors were coated by QSP.

The individual optical elements were each supposed to contribute less than
0.2 arcseconds FWHM to the image size, and all either passed this 
specification or came close, either on delivery or after some correction
(the primary astigmatism). The early overall performance, however, was
determined by thermal issues, discussed in more detail elsewhere, 
which required a great deal of time and effort to solve. The performance
is now quite satisfactory, but a detailed post-mortem to illuminate where
the remaining errors originate, and what relation, if any, they have
to the input error budget has not been done. 

One issue which has not been discussed at all is ghosting in the imaging
mode. All refracting surfaces are anti-reflection coated, selectively for
the relevant imaging bands where practical. As a result, the only significant
ghosts arise from reflections from the CCD surface, off some optical surface,
and back to the detector, and of these the only important ones are reflections
from the filter surface, about 8 mm from the CCD. The worst of these are
in the u band, partly because the antireflection coating on the UV CCDs is
relatively poor and (mostly) because the coating on the filters does double
duty as a red-leak suppressant and an antireflection coat, and the second
function is not very well performed. Since in TDI mode the form of the
ghosts is determined only by the position of the star in one dimension, it
was originally planned to implement subtraction of these primary ghosts
in the processing pipeline, but that has never been implemented. The fraction
of sky in the survey area 
disturbed by these ghosts is much less than one percent and the
fraction of sky recoverable without exquisite subtraction of the wings of
the star responsible for the ghost much smaller yet. This level of ghosting
would doubtless be problematical at lower latitude, but fortunately has
not been a problem for us.

\subsection{Telescope}

The enclosure was completed in 1994, ready to receive the telescope. The
telescope structure, including the corotating floor panel, was constructed
by L\&F Industries. As various pieces of the assembly were completed,
they were inspected and the entire telescope structure assembled for
final testing.

The first part of the telescope to be installed, the azimuth drive
assembly, arrived at APO on August 10 1995, was installed on August 15, and
tested on August 20 (see Figures~24 and 25).  The wind baffle azimuth 
drive assembly
was installed on September 19 1995, and the 2.5 m telescope itself arrived on 
October 4 1995.  The telescope was installed (with dummy optics for 
balance tests) by October 10 1995 with the assistance of engineers from
L\&F Industries.  Figure 26 shows a close up of the back of the telescope
during construction, showing the primary mirror cell, the installed
image rotator, and the fork.

The rotator, at installation driven by a small geared-down DC servomotor
and a rotary encoder like the ones on the axes, failed to function 
satisfactorily and was too limber to support the spectrographs adequately.
When it was discovered as well that the drive surface had at some
point been seriously damaged, it was decided to redesign the system,
and the current system with tape encoder and a drive motor system
essentially identical to that on the other axes was installed in 
the spring of 1999.

The baffles and optics were constructed and installed 
throughout 1996-7, and the imaging camera arrived at APO on October
22 1997 after being driven across the country from Princeton by
Roberts White Glove Express.
The camera was first tested with the SDSS Data Acquisition system on
December 12 1997 in the APO clean room, achieving instrumental ``first light''. 
The original plan had been to test the telescope
tracking, drive and interlock systems using a single CCD test 
(``engineering'') camera, but the arrival of the full survey mosaic 
camera before the
telescope was quite finished allowed the initial testing and data
acquisition to be performed with the mosaic camera itself. This was done
by drift-scanning with the telescope stationary at the celestial equator 
(a natural great circle) because
the telescope drives were not yet ready,
and first light was obtained in the early morning of May 9 1998. 
That night happened to be close to full moon, and the baffles had not
yet been installed, but the run demonstrated that
the optics, camera, and data acquisition were working essentially 
perfectly. High-quality imaging was obtained during the next dark run
(May 27 1998) and allowed the first science results to be obtained from
SDSS (Fan et al. 1998). SDSS first 
light was officially announced on June 8 1998 at the
June AAS meeting. The first SDSS journal paper, by Fan et al. (1999),
reported the discovery of several high-redshift quasars, including the 
first new ``highest redshift'' quasar in more than 8 years.
The next 18 months saw much work on telescope tracking
and control, rebuilding the enclosure, the installation and commissioning
of the spectroscopic systems, and improvement of the thermal 
environment. This last has been one of the most difficult tasks we
have faced, as is apparent from some of the earlier discussion;
it required
the better part of two years to reach a state in which the progress of the
survey was determined essentially entirely by the weather and not the 
thermal state of the telescope. We were in the end successful, however,
and for the past three years have been able to take advantage of
satisfactory weather conditions when and if they occur with no
concern that the telescope might not be in satisfactory thermal
equilibrium.
A recent image of the SDSS telescope in operation with the new enclosure
is shown in Figure~27. 
Operations began in earnest in Spring 2000 and to mark this
the SDSS dedication was held on October 5, 2000. The rest, as they say, is 
history.

\subsection{The Future}

At this writing, June 2005, the last dark run of SDSS-I has
just been completed. 
Originally, the SDSS was expected to obtain a filled survey covering
$10^4$ square degrees of the
North Galactic Cap (NGC), with uniform spectral coverage also, about $10^6$
galaxies and $10^5$ quasars (York et al. 2000), in five years of operation.
The statistics of data acquisition during the first years of operation
showed, however, that this was unlikely to be accomplished due to slightly
greater inefficiencies in data gathering - which were corrected during the
first two years - and to weather fluctuations.  Accordingly, the strategy
in the final years of the SDSS was changed. The original expectation
was that the imaging survey of the northern cap 
(this is the rate-determining step
in the entire SDSS) would be finished well ahead of time and that all
subsequent 
available clear time could be devoted to finishing the spectroscopy. The
actual rate of data acquisition, however, suggested two alternative
strategies: (1) finishing in five years with incomplete coverage of the
NGC consisting of two large, separated areas; or (2) acquiring the 
funds to extend the SDSS beyond five years and finish the northern cap.
The latter option was adopted, driven by the evident legacy value of SDSS,
and at this writing, the five year survey (SDSS-I) is finished and the 
extension (SDSS-II) is about to begin.

SDSS-II consists of four programs, three science surveys and an
effort to enhance the photometric accuracy:

\begin{itemize}

\item
{\it Legacy:}

This program will finish a contiguous northern survey area (somewhat
smaller than the original $10^4$ square degrees) with both imaging and 
spectroscopy.
The imaging is nearly complete, with only a small area at low right 
ascensions remaining to be done. The entire ``ubercalibrated'' (see below)
imaging map of the sky and all of the spectroscopy will be made public with 
SDSS Data Release 8 in late 2008.

\item
{\it SEGUE}:

The SDSS was designed primarily as an extragalactic survey, but
has made important contributions in many other areas.
One of the major SDSS contributions to non-extragalactic
science has been the discovery of very large structures in the Galactic
halo (e.g. Yanny et al. 2003). Given the ability of the SDSS telescope and
imaging camera to sweep the sky in great circles, a program which
runs mostly in the Fall, the SEGUE project (Newberg et al. 2003) 
has been designed
to obtain imaging (mostly) along stripes of constant Galactic longitude
and to measure spectra of stars selected from the imaging in a fine
enough grid that the metallicity, populations and kinematics of both the smooth
and structured halo and thick disks can be measured. These stripes will
pass through the Galactic plane, providing data for studies of young
stellar populations and the interstellar medium (Finkbeiner et al.
2004b).

\item
{\it Supernova Search}:  

This program builds upon the repeat imaging
of the SDSS equatorial stripe in the Southern Survey (York et al. 2000).
The Supernova program is designed
to scan 200 square degrees of Fall sky along the equator every other
night (when clear), regardless of whether or not the night is photometric.
This program exploits the fact that once an area has been imaged on a clear
night and photometric calibrations obtained for all objects in that area,
the photometric-quality observations
can be used to ``bootstrap'' the photometric calibrations on
non-photometric nights, and transient objects can be found by image-subtraction
techniques. This was demonstrated for supernovae using SDSS data by
Miknaitis and Krisciunas (2001), and descendents of their code will be used
for the new survey.
The primary science goal is to detect and measure some
200 Type Ia supernovae in the rather neglected redshift range z = 0.1 to 0.35 
(Frieman et al. 2004). The redshifts will be obtained via arrangements with
several teams using large telescopes. The imaging
data will also be valuable for
a wide range of other variability studies (proper motions, photometric
variability, etc.).

\item{\it Photometric Ubercalibration}

The SDSS imaging camera saturates at about 14th magnitude,
and all well-calibrated photometric
standards are much brighter than this. To date,
SDSS photometric calibration had been carried out using two smaller 
telescopes to set up a network of bright and faint standards (Hogg
et al. 2001; Smith et al. 2002; Tucker et al. 2005) through which the 
SDSS camera scans. This effort has resulted in remarkably accurate
photometry (Ivezi\'c et al. 2004) but structure along the SDSS scans
can be seen at the 1-2\% level (e.g. Blanton et al. 2005). Accordingly,
the SDSS is now engaged in a major effort to move to a system in which
the many redundancies and overlaps in the sky coverage of the SDSS
are exploited to achieve a self-consistent photometric calibration
which is smooth and accurate all scales,
including redetermination of a set of self-consistent flat fields
for each detector.
Part of the effort involves special photometric scans at four times 
the sidereal rate (and with the imaging data binned by the data 
acquisition system accordingly)
along great circles which cross both the Northern and Southern SDSS
surveys in a spoke-and-rim pattern (this pattern is dubbed ``Apache
Wheel''). Work to date (e.g. Blanton et al. 2005, which does not yet
include the Apache Wheel data)
indicates that stripe-to-stripe variations
become almost undetectable, the photometric accuracy approaches 1\%,
and outliers are greatly reduced in number. The scientific payoff
for this effort should be enormous. The development of photometric 
redshift techniques means that huge accurately calibrated samples can
be made which probe large-scale structure on the largest scales
(see, for example, Schlegel et al. 2003, Budavari et al. 2003, 
Hogg et al. 2005, 
Eisenstein et al. 2005); the removal of calibration
structure from the data removes this contaminant from statistical
analyses such as that of the power spectrum; an extremely accurate 
Galactic extinction map can be made (Finkbeiner et al. 2004a): 
rare objects (such as
low-metallicity stars) whose broad-band colors differ but subtly from those
of more common stellar populations can be selected much more reliably and
efficiently; and, together with the large area of sky covered by
SDSS-I and SDSS-II, this effort provides photometric calibration
to support many other observing efforts (as one example, see Smith
et al. 2003), and including planned large-scale successors to the imaging
part of SDSS such as LSST and Pan-STARRS.
\end{itemize}

SDSS-II is a three-year program. And beyond summer 2008? The SDSS
hardware at APO is currently working superbly (only the 
data acquisition system needs
to be replaced, and this will occur in late summer 2005), the 
seeing is as good as allowed by the site, and the data taking is at
maximum possible efficiency.  Several possible projects, which will
make use of the unique capabilities of this 
telescope, are being investigated and explored,
including Galactic plane studies, various studies of bright stars, a
faint quasar survey and an extensive planet search program. At this writing
nothing is decided beyond 2008, but it is likely that
the SDSS 2.5 m telescope will continue to be operated to make major
contributions to astronomical surveys.

\bigskip
We are very grateful to Michael Strauss and \v{Z}eljko Ivezi\'c for
useful and helpful comments on the manuscript.
Funding for the creation and distribution of the SDSS Archive has been 
provided by the Alfred P. Sloan Foundation, the Participating 
Institutions, the National Aeronautics and Space Administration, the
National Science Foundation, the U.S. Department of Energy, the
Japanese Monbukagakusho, and the Max Planck Society. The SDSS Web site
is http://www.sdss.org/. The SDSS is managed by the Astrophysical
Research Consortium (ARC) for the Participating Institutions. The
Participating Institutions are The University of Chicago, Fermilab, 
the Institute for Advanced Study, the Japan Participation Group,
The Johns Hopkins University, the Korean Scientist Group, Los Alamos
National Laboratory, the Max-Planck-Institute for Astronomy (MPIA), 
the Max-Planck-Institute for Astrophysics (MPA), New Mexico State 
University, University of Ohio, University of Pittsburgh, 
University of Portsmouth,
Princeton University, the United States Naval Observatory, and 
the University of Washington.

\newpage

\bibliographystyle{unsrt}
\bibliography{gsrp}

\clearpage

\halign{\hskip 12 pt
\hfil # & \hfil # & \hfil $#$ & \hfil # \hfil & \hfil # & \hfil # & \hfil # &
 \hfil # & \hfil $#$ & \hfil # \cr
\multispan{10}{\hfil TABLE 1 \hfil} \cr
\noalign{\medskip}
\multispan{10}{\hfil The Optical Design for the SDSS Telescope: Camera
    Mode \hfil} \cr
\noalign{\bigskip\hrule\smallskip\hrule\medskip}
 & & \hfil $Space$ \hfil & & & & & & & \hfil Clear \hfil \cr
\hfil Sur \hfil & \hfil c \hfil & \hfil $mm$ \hfil & \hfil Glass \hfil &
    \hfil $a_2$ \hfil & \hfil  $a_4$ \hfil & \hfil $a_6$ \hfil &
    \hfil $a_8$ \hfil & \hfil  k \hfil & \hfil Diam \hfil \cr
\noalign{\medskip\hrule\bigskip}
1& $-$8.889e-5 &0.0      &-air& & & 3.81e-22 & $-$1.52e-29 &$-$1.285 & 2500\cr
2& $-$1.390e-4 &-3646.14 &air & & &1.79e-19& & -11.97 & 1080\cr
3&             & 3621.59 &fq  & 2.321e-5&  $-$1.173e-10&
$-$7.87e-17& 1.59e-22&\ &722\cr
4& &12.0&air&\ & \ &\ &\ &\ & 721\cr
5& &714.00 & fq & $-$2.732e-4 & 2.056e-9 & $-$5.81e-15 & 1.75e-20 & &657\cr
6& &45.00& bk7 & & & & & &652\cr
7& &5.00&air& & & & & &651\cr
8& &8.00&air& & & & & &651\cr
\noalign{\medskip\hrule}}

\bigskip\noindent
Column Notes: $c$ = curvature, $k$ = conic constants.  See text for
details.

\bigskip\bigskip\bigskip

\halign{\hskip 12 pt
\hfil # & \hfil # & \hfil $#$ & \hfil # \hfil & \hfil # & \hfil # & \hfil # &
 \hfil # & \hfil $#$ & \hfil # \cr
\multispan{10}{\hfil TABLE 2 \hfil} \cr
\noalign{\medskip}
\multispan{10}{\hfil The Optical Design for the SDSS Telescope: Spectroscopic
    Mode \hfil} \cr
\noalign{\bigskip\hrule\smallskip\hrule\medskip}
 & & \hfil $Space$ \hfil & & & & & & & \hfil Clear \hfil \cr
\hfil Sur \hfil & \hfil c \hfil & \hfil $mm$ \hfil & \hfil Glass \hfil &
    \hfil $a_2$ \hfil & \hfil  $a_4$ \hfil & \hfil $a_6$ \hfil &
    \hfil $a_8$ \hfil & \hfil  k \hfil & \hfil Diam \hfil \cr
\noalign{\medskip\hrule\bigskip}
1& $-$8.889e-5&0.0&-air& & &3.81e-22&$-$1.52e-29&-1.285&2500\cr
2& $-$1.390e-4&$-$3644.46&air& & &1.79e-19& &-11.97&1080\cr
3& &3619.91&fq&2.321e-5&
        $-$1.173e-10&$-$7.87e-17&1.59e-22& &722\cr
4& &12.0&air& & & & & &721\cr
5& $-$4.307e-4&672.64&fq& & & & & &657\cr
6& &10.00&air&$-$7.747e-5& $-$4.123e-10&$-$6.53e-15&5.23e-20& &656\cr
7& &86.61&air& & & & & &653\cr
\noalign{\medskip\hrule}}

\bigskip\noindent
Column Notes: $c$ = curvature, $k$ = conic constants.  See text for
details.
\clearpage
\halign{\hskip 12 pt
\hfil #  & \hfil #  & \hfil $#$ & \hfil $#$  & \hfil # & \hfil # \hfil &
\hfil #  & \hfil #  & \hfil $#$ & \hfil $#$  & \hfil # \cr
\multispan{11}{\hfil TABLE 3 \hfil} \cr
\noalign{\medskip}
\multispan{11}{\hfil Telescope Optical Performance
   for the Imaging Mode: Parameters for Best Focus \hfil} \cr
\noalign{\bigskip\hrule\smallskip\hrule\medskip}
\hfil Angle \hfil & \hfil $\lambda$ \hfil & \hfil f_b \hfil &
   \hfil h \hfil & \hfil $\epsilon$ \hfil & &
\hfil Angle \hfil & \hfil $\lambda$ \hfil & \hfil f_b \hfil &
   \hfil h \hfil & \hfil $\epsilon$ \hfil \cr
\hfil arcmin \hfil & \hfil \AA \hfil & \hfil $mm$ \hfil & \hfil $mm$ \hfil &
   \hfil mm \hfil & \ \ \ \ \ \ \ \ &
\hfil arcmin \hfil & \hfil \AA \hfil & \hfil $mm$ \hfil & \hfil $mm$ \hfil &
   \hfil mm \hfil \cr
\noalign{\medskip\hrule\bigskip}
  0.00  &  3540 &  -0.322 &   -0.000 & 0.017 &&
  0.00  &  7690 &  -0.396 &   -0.000 & 0.018\cr
 30.00  &  3540 &   0.031 & -108.435 & 0.014 &&
 30.00  &  7690 &  -0.187 & -108.465 & 0.018\cr
 45.00  &  3540 &   0.267 & -162.666 & 0.013 &&
 45.00  &  7690 &  -0.124 & -162.700 & 0.020\cr
 60.00  &  3540 &   0.260 & -216.895 & 0.019 &&
 60.00  &  7690 &  -0.361 & -216.927 & 0.024\cr
 70.00  &  3540 &  -0.014 & -253.053 & 0.027 &&
 70.00  &  7690 &  -0.815 & -253.081 & 0.029\cr
*73.00  &  3540 &  -0.156 & -263.902 & 0.030 &&
 73.00  &  7690 &  -1.015 & -263.929 & 0.031\cr
*82.00  &  3540 &  -0.803 & -296.461 & 0.037 &&
*82.00  &  7690 &  -1.841 & -296.485 & 0.038\cr
*90.00  &  3540 &  -1.731 & -325.417 & 0.043 &&
*90.00  &  7690 &  -2.934 & -325.442 & 0.048\cr
\phantom{1}\cr
  0.00  &  4760 &  -0.361 &   -0.000 & 0.018 & &
  0.00  &  9250 &  -0.405 &   -0.000 & 0.018\cr
 30.00  &  4760 &  -0.088 & -108.452 & 0.015 &&
 30.00  &  9250 &  -0.212 & -108.468 & 0.019\cr
 45.00  &  4760 &   0.052 & -162.685 & 0.013 &&
 45.00  &  9250 &  -0.168 & -162.704 & 0.022\cr
 60.00  &  4760 &  -0.084 & -216.913 & 0.010 &&
 60.00  &  9250 &  -0.431 & -216.931 & 0.028\cr
 70.00  &  4760 &  -0.458 & -253.069 & 0.011 &&
 70.00  &  9250 &  -0.905 & -253.084 & 0.034\cr
 73.00  &  4760 &  -0.632 & -263.918 & 0.011 &&
 73.00  &  9250 &  -1.111 & -263.932 & 0.036\cr
 82.00  &  4760 &  -1.380 & -296.474 & 0.015 &&
*82.00  &  9250 &  -1.957 & -296.488 & 0.045\cr
*90.00  &  4760 &  -2.399 & -325.431 & 0.021 &&
*90.00  &  9250 &  -3.069 & -325.444 & 0.057\cr
\phantom{1}\cr
  0.00  &  6280 &  -0.384 &   -0.000 & 0.018\cr
 30.00  &  6280 &  -0.154 & -108.461 & 0.017\cr
 45.00  &  6280 &  -0.065 & -162.695 & 0.017\cr
 60.00  &  6280 &  -0.268 & -216.922 & 0.019\cr
 70.00  &  6280 &  -0.696 & -253.077 & 0.022\cr
 73.00  &  6280 &  -0.887 & -263.925 & 0.023\cr
 82.00  &  6280 &  -1.687 & -296.481 & 0.029\cr
*90.00  &  6280 &  -2.756 & -325.438 & 0.037\cr
\noalign{\medskip\hrule}}

\bigskip
$^*$ These field angles/color values are not reached for the
photometric array.
\clearpage
\halign{\hskip 12 pt
\hfil # & \hfil $#$ & \hfil $#$ & \hfil $#$ \cr
\multispan4{\hfil TABLE 4 \hfil} \cr
\noalign{\medskip}
\multispan4{\hfil Imaging Focal Surface ($\lambda$ = 4760 \AA ) \hfil} \cr
\noalign{\bigskip\hrule\smallskip\hrule\medskip}
\hfil Angle \hfil & \hfil \delta \hfil & \hfil $ht$ \hfil &
   \hfil $lindev$ \hfil \cr
\hfil arcmin \hfil & \hfil $mm$ \hfil & \hfil $mm$ \hfil & \hfil $mm$ \hfil \cr
\noalign{\medskip\hrule\bigskip}
  0.0 &  -0.361 &    -0.000 &  -0.000\cr 
 30.0 &  -0.088 &  -108.452 &   0.005\cr 
 45.0 &   0.052 &  -162.685 &   0.003\cr  
 60.0 &  -0.084 &  -216.913 &   0.009\cr  
 70.0 &  -0.458 &  -253.069 &   0.012\cr  
 73.0 &  -0.632 &  -263.918 &   0.011\cr  
 82.0 &  -1.380 &  -296.474 &  -0.001\cr  
 90.0 &  -2.399 &  -325.431 &  -0.027\cr  
\noalign{\medskip\hrule}}

\noindent
Note: Scale=3.61519 mm arcmin$^{-1}$

\clearpage
\halign{\hskip 12 pt
\hfil # & \hfil #  & \hfil $#$ & \hfil $#$ & \hfil # & \hfil # & # &
\hfil # & \hfil #  & \hfil $#$ & \hfil $#$ & \hfil # & \hfil # \cr
\multispan{13}{\hfil TABLE 5 \hfil} \cr
\noalign{\medskip}
\multispan{13}{\hfil Telescope Optical Performance for the Spectrographic
Mode: Average Focus \hfil} \cr
\noalign{\bigskip\hrule\smallskip\hrule\medskip}
\hfil Angle & \hfil $\lambda$ \hfil & \hfil f_b \hfil & \hfil h/dh \hfil &
  \hfil $D$ \hfil & \hfil $\epsilon$ \hfil & \ \ \ \ \ &
\hfil Angle & \hfil $\lambda$ \hfil & \hfil f_b \hfil & \hfil h/dh \hfil &
  \hfil $D$ \hfil & \hfil $\epsilon$ \hfil \cr
\hfil arcmin \hfil & \hfil \AA \hfil & \hfil $mm$ \hfil & \hfil $mm$ \hfil &
   \hfil mm \hfil & \hfil mm \hfil & &
\hfil arcmin \hfil & \hfil \AA \hfil & \hfil $mm$ \hfil & \hfil $mm$ \hfil &
   \hfil mm \hfil & \hfil mm \hfil \cr
\noalign{\medskip\hrule\bigskip}
  0.00&  4000& -0.007&   0.000&-0.135&0.036 &&
  0.00&  6500& -0.007&  -0.000& 0.062&0.031\cr
 30.00&  4000& -0.143&   0.004&-0.081&0.030 &&
 30.00&  6500& -0.143&  -0.002& 0.037&0.027\cr
 45.00&  4000& -0.424&   0.005&-0.015&0.025 &&
 45.00&  6500& -0.424&  -0.002& 0.007&0.024\cr
 60.00&  4000& -0.978&   0.005& 0.076&0.028 &&
 60.00&  6500& -0.978&  -0.002&-0.035&0.029\cr
 70.00&  4000& -1.536&   0.004& 0.148&0.036 &&
 70.00&  6500& -1.536&  -0.002&-0.068&0.034\cr
 80.00&  4000& -2.265&   0.002& 0.231&0.049 &&
 80.00&  6500& -2.265&  -0.001&-0.106&0.036\cr
 90.00&  4000& -3.203&  -0.004& 0.325&0.065 &&
 90.00&  6500& -3.203&   0.002&-0.149&0.040\cr
\phantom{1}\cr
  0.00&  4600& -0.007&  -0.000&-0.058&0.031 &&
  0.00&  9000& -0.007&   0.000& 0.131&0.036\cr
 30.00&  4600& -0.143&   0.002&-0.035&0.027 &&
 30.00&  9000& -0.143&  -0.004& 0.078&0.029\cr
 45.00&  4600& -0.424&   0.002&-0.006&0.024 &&
 45.00&  9000& -0.424&  -0.004& 0.014&0.026\cr
 60.00&  4600& -0.978&   0.002& 0.033&0.025 &&
 60.00&  9000& -0.978&  -0.004&-0.074&0.036\cr
 70.00&  4600& -1.536&   0.002& 0.065&0.027 &&
 70.00&  9000& -1.536&  -0.004&-0.145&0.046\cr
 80.00&  4600& -2.265&   0.001& 0.101&0.030 &&
 80.00&  9000& -2.265&  -0.002&-0.226&0.056\cr
 90.00&  4600& -3.203&  -0.001& 0.141&0.035 &&
 90.00&  9000& -3.203&   0.003&-0.317&0.068\cr
\phantom{1}\cr
  0.00&  5300& -0.007&   0.000& 0.000&0.029\cr
 30.00&  5300& -0.143&-108.818& 0.000&0.026\cr
 45.00&  5300& -0.424&-163.322& 0.000&0.024\cr
 60.00&  5300& -0.978&-217.855& 0.000&0.025\cr
 70.00&  5300& -1.536&-254.241& 0.000&0.027\cr
 80.00&  5300& -2.265&-290.713& 0.000&0.026\cr
 90.00&  5300& -3.203&-327.372& 0.000&0.025\cr
\noalign{\medskip\hrule}}
\clearpage
\halign{\hskip 12 pt
\hfil # & \hfil $#$ & \hfil $#$ & \hfil $#$ & \hfil $#$ & \hfil $#$ \cr
\multispan6{\hfil TABLE 6 \hfil} \cr
\noalign{\medskip}
\multispan6{\hfil Average Focal Surface \hfil} \cr
\noalign{\medskip} 
\noalign{\bigskip\hrule\smallskip\hrule\medskip}
\hfil Angle \hfil & \hfil $Focus$ \hfil & \hfil ht \hfil & \hfil hlindev \hfil &
   \hfil yp \hfil & \hfil dyp \hfil \cr
\hfil arcmin \hfil & \hfil $mm$ \hfil & \hfil $mm$ \hfil & \hfil $mm$ \hfil &
 \hfil $rad$ \hfil & \hfil $rad$ \hfil \cr                             
\noalign{\medskip\hrule\bigskip} \cr
 0.0 &   -0.006&     0.000&    0.000&   0.0000&  0.0000\cr
30.0 &   -0.143&  -108.818&    0.184&  -0.0280& -0.0247\cr
45.0 &   -0.423&  -163.322&    0.185&  -0.0396& -0.0324\cr
60.0 &   -0.978&  -217.855&    0.158&  -0.0477& -0.0347\cr
70.0 &   -1.536&  -254.241&    0.113&  -0.0508& -0.0329\cr
80.0 &   -2.265&  -290.713&   -0.017&  -0.0523& -0.0293\cr
90.0 &   -3.203&  -327.372&   -0.331&  -0.0538& -0.0258\cr
\noalign{\medskip\hrule}}

\noindent
Note: scale=3.62730 mm arcmin$^{-1}$; mean exit pupil at $-$5174 mm

\clearpage
\begin{figure}[tb]
\figurenum{1}
\dfplot{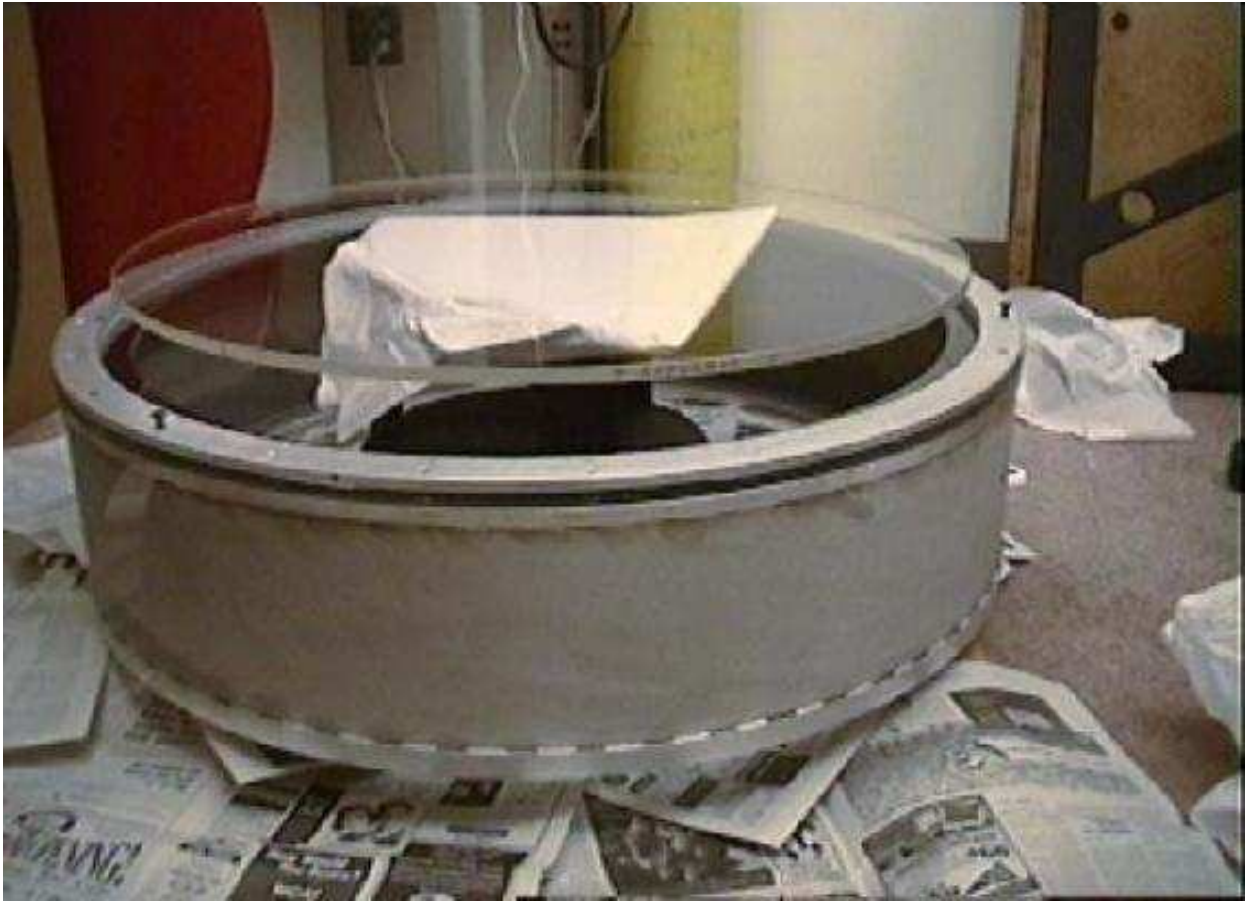}
\figcaption{Apache Point Observatory, facing north. The SDSS telescope on
its west-facing platform is in the foreground.  Note the people standing
on the platform behind the telescope to get an idea of the scale. The
roll-away enclosure is retracted and can be seen on the right. Note the
azimuth bearing and the support structure below the platform. The small round 
dome below the enclosure houses the 0.5 m Photometric Telescope, and the ARC
3.5 m telescope is in the background. White Sands National Monument can be seen
to the north-west.
with the SDSS telescope in the foreground. 
\label{fig1}
}
\end{figure}
\clearpage
%
\begin{figure}[tb]
\figurenum{2}
\dfplot{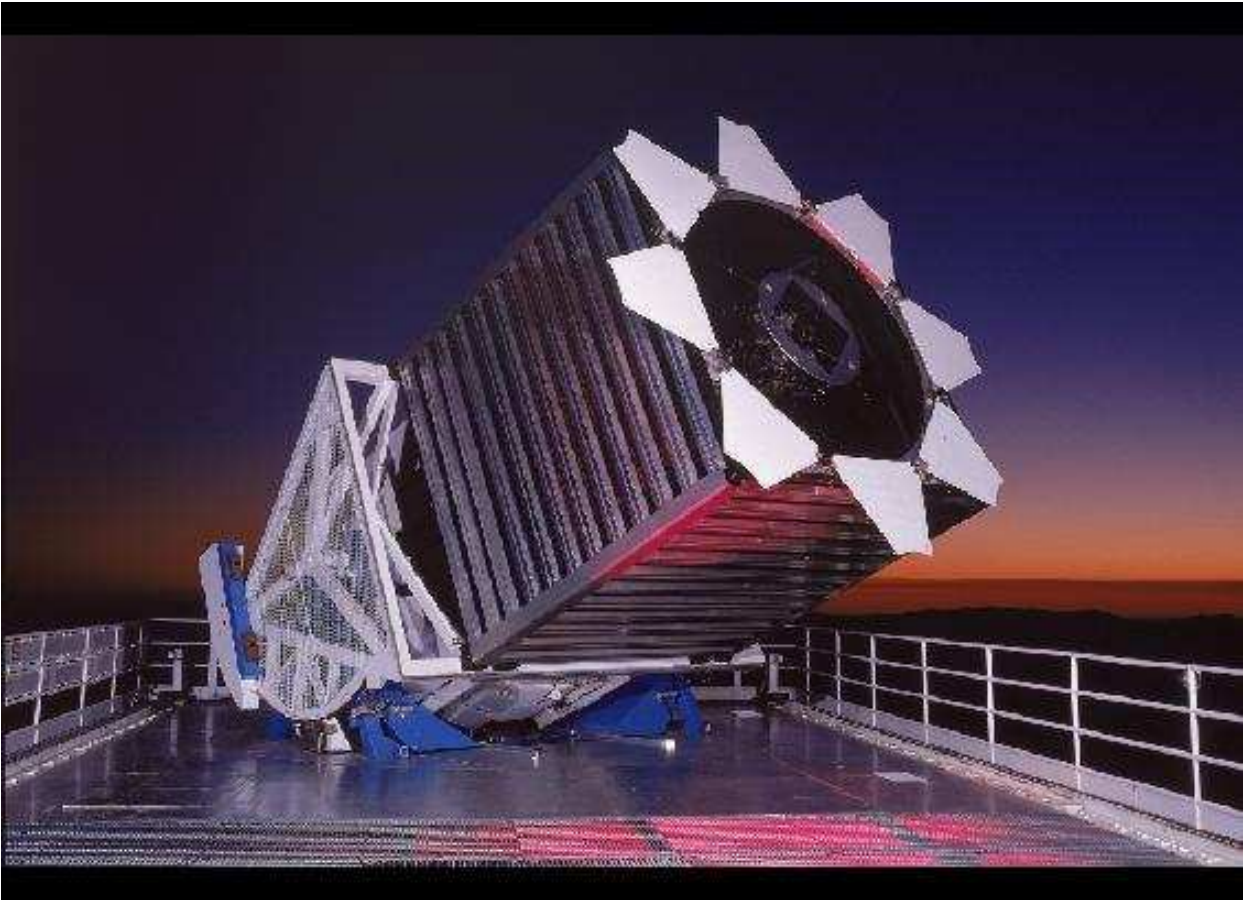}
\figcaption{The SDSS 2.5 m at sunset, ready for a night's observing. The
picture is taken facing west, with the rolled-off telescope enclosure
behind the viewer. The telescope is enclosed in its rectangular wind
baffle. 
\label{fig2}
}
\end{figure}
\clearpage
%
\begin{figure}[tb]
\figurenum{3}
\epsscale{0.5}
\dfplot{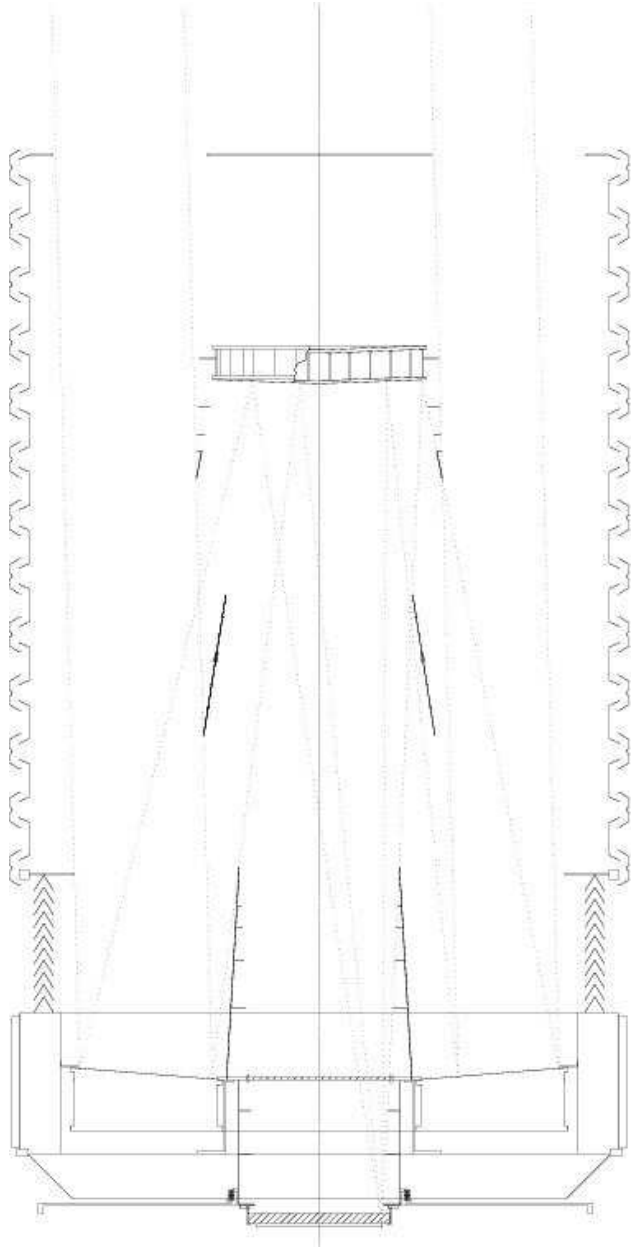}
\figcaption{SDSS 2.5 m telescope optical configuration, showing the
light baffles and a set of rays from the extreme field edge.  
The first transmissive corrector
(the ``Gascoigne'', or ``common'', corrector) is nearly coincident
with the vertex of the primary mirror. The second transmissive corrector,
which forms the main structural element of the SDSS mosaic camera
(Gunn et al. 1998) is just before the focal service. In spectroscopic
mode, the imaging camera, including this corrector, is dismounted and
replaced with the spectroscopic corrector and the spectroscopic plug
plate and fiber harnesses.
The baffle system
consists of the wind baffle (outside the beam to the primary mirror)
and the secondary, conical and primary baffles (inside). The conical
baffle is suspended about halfway between the primary and secondary.
Light enters the wind baffle through the annular opening at the
top. The outer
interlocking ``C''-shaped baffles which form the upper tube are carried
by an independently mounted and driven wind baffle mechanism and take
the major wind loads on the telescope. 
\label{fig3}
}
\end{figure}
\clearpage
%
\begin{figure}[tb]
\figurenum{4}
\epsscale{0.8}
\dfplot{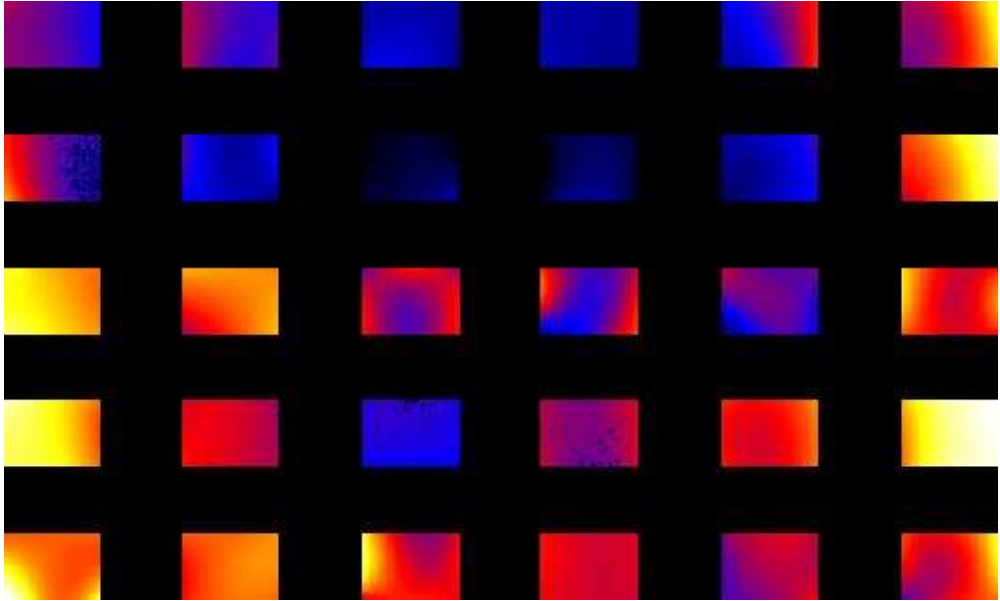}
\figcaption{Instantaneous FWHM of the imaging PSF as a function of position
in the camera.  Each square shows the data for one of the photometric CCDs.
The stretch is linear, from blue to white. 
This image was taken in steady
sub-arcsecond seeing, and the PSF variation across the $3^{\circ}$ field
due to the optics can clearly be seen. The arrangement of the images
is the same as that in the camera; the camera columns are 1-6, left to
right; rows are r, i, u, z, g top to bottom; the motion of a star's image
is from top to bottom in this diagram. Since the imaging is done in TDI
mode, if the seeing were absolutely steady there should be no vertical 
variation in this figure within a given chip, and indeed there is not
as much vertical structure as there is horizontal, but the temporal
seeing variations still make themselves obvious even under such good
conditions as these. The stretch in these images is $\pm20\%$ about the
median seeing, which is 0.9 arcsecond. Black is smallest, then blue
through yellow to white; white is largest.
Figure courtesy \v{Z}eljko Ivezi\'c.
\label{fig4}
}
\end{figure}
\clearpage
%
\begin{figure}[tb]
\figurenum{5}
\epsscale{0.85}
\dfplot{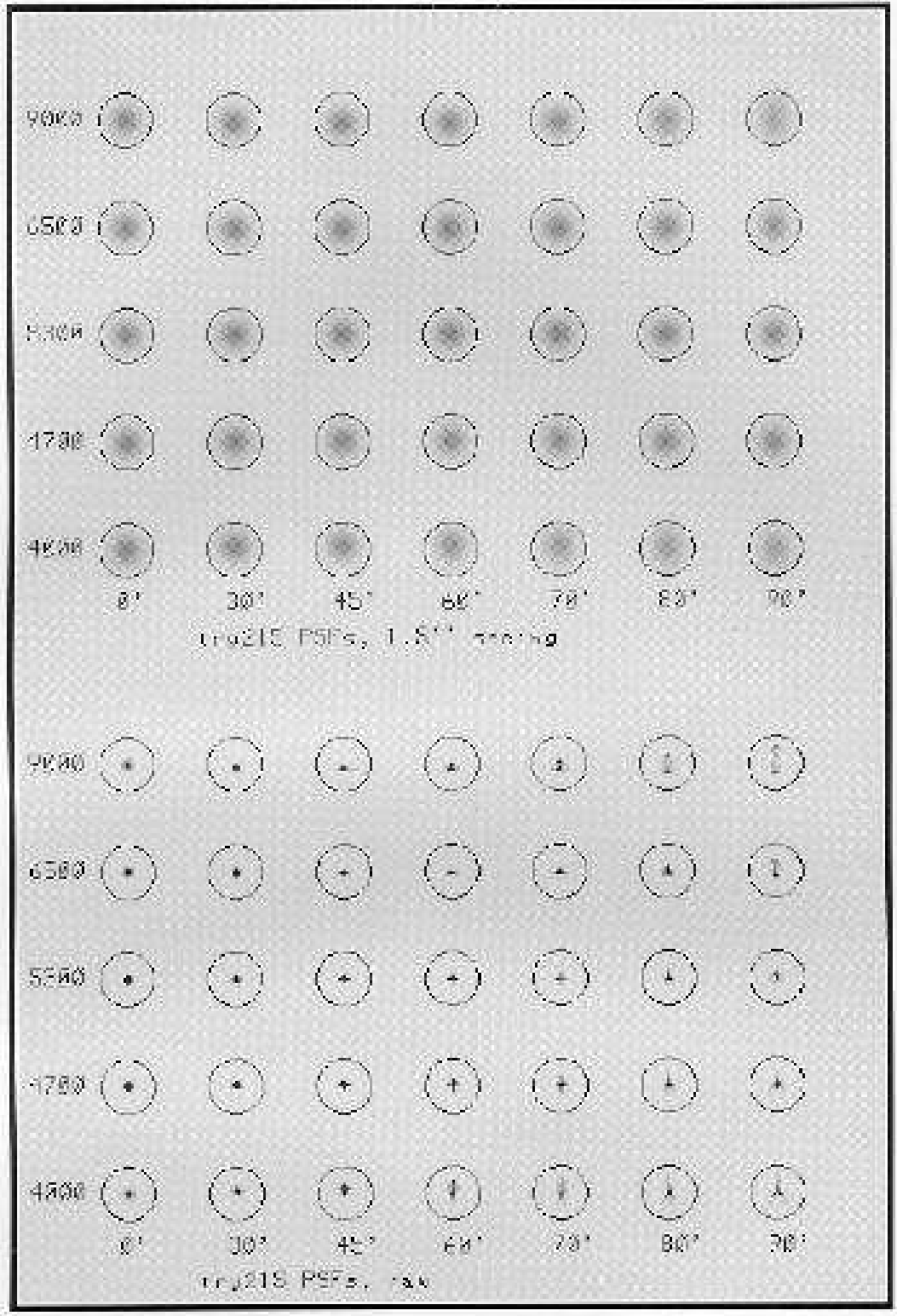}
\figcaption{Spectroscopic PSFs for the SDSS telescope. These are as seen 
in the spectroscopic configuration
on the mean focal surface. The field angles are (from left to right)
0, 30, 45, 60, 70, 80, and 90 arcminutes, and the wavelengths from top
to bottom 4000, 4700, 5300, 6500, and 9000 \AA{}. A corresponding diagram
showing the PSFs in the imaging configuration can be found in Gunn et. al
1998. Please note in both cases that these are the results of raytrace
calculations from a perfect as-designed optical system.
\label{fig5}
}
\end{figure}
\clearpage
%
\begin{figure}[tb]
\figurenum{6}
\dfplot{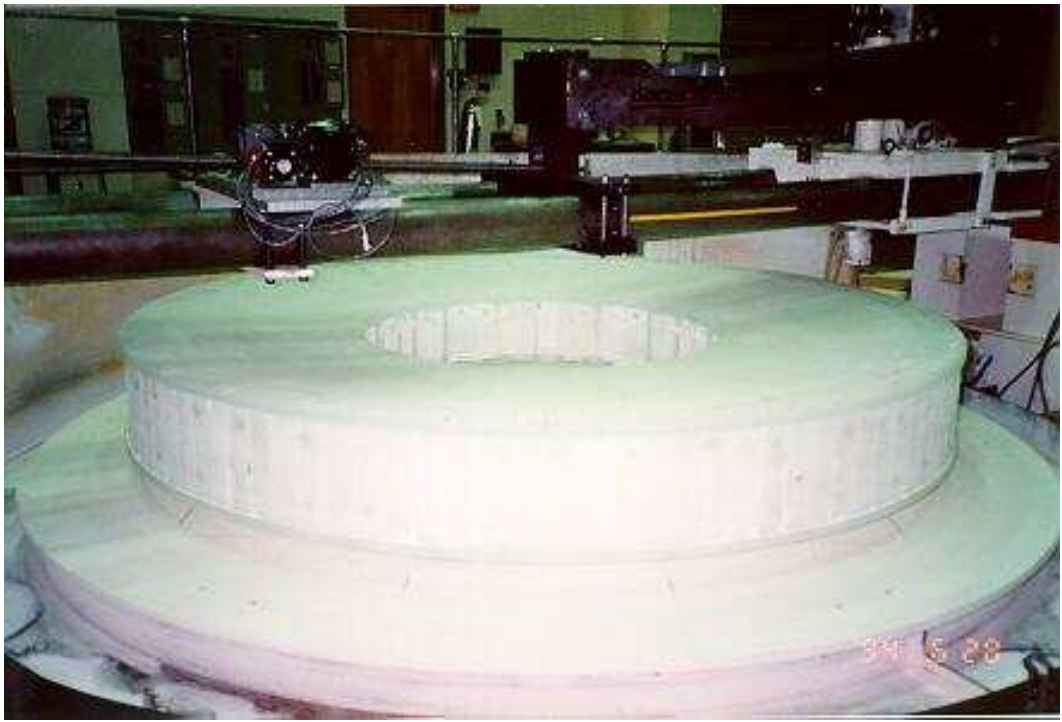}
\figcaption{Primary mirror at the Optical Sciences Center.  The mirror was
cast of Ohara E6 borosilicate glass by the Hextek Corporation.
\label{fig6}
}
\end{figure}
\clearpage
%
\begin{figure}[tb]
\figurenum{7}
\dfplot{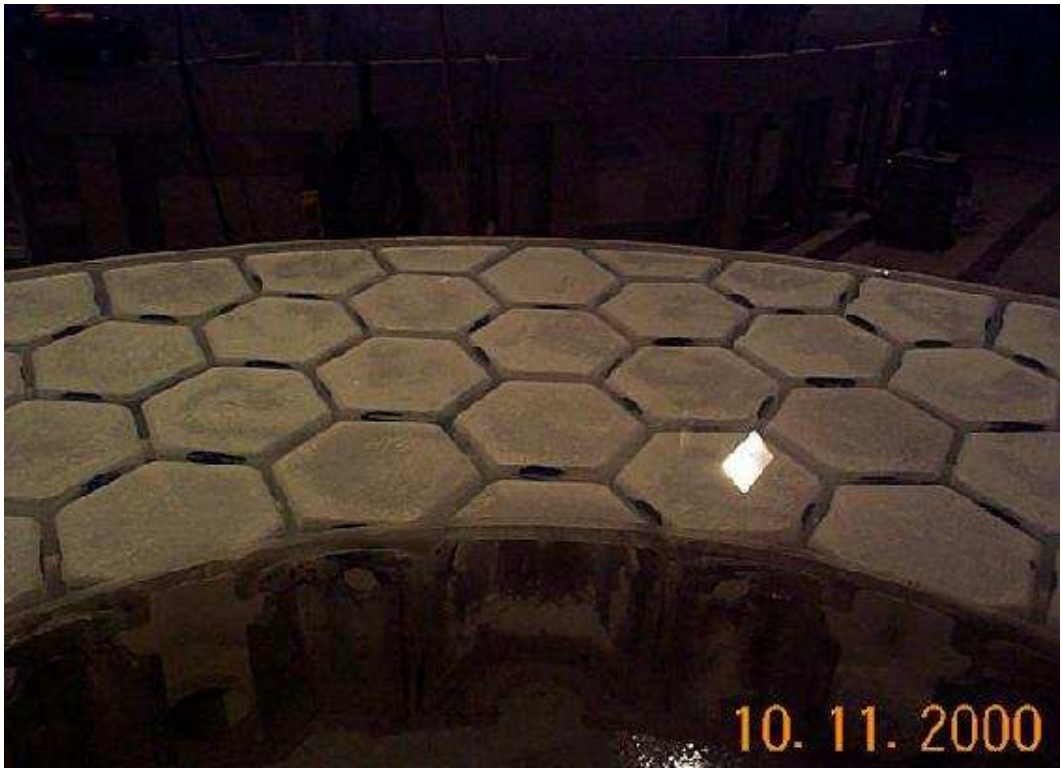}
\figcaption{Faceplate of the SDSS 2.5 m primary mirror, showing the hexagonal
honeycomb structure. 
\label{fig7}
}
\end{figure}
\clearpage
%
\begin{figure}[tb]
\figurenum{8}
\dfplot{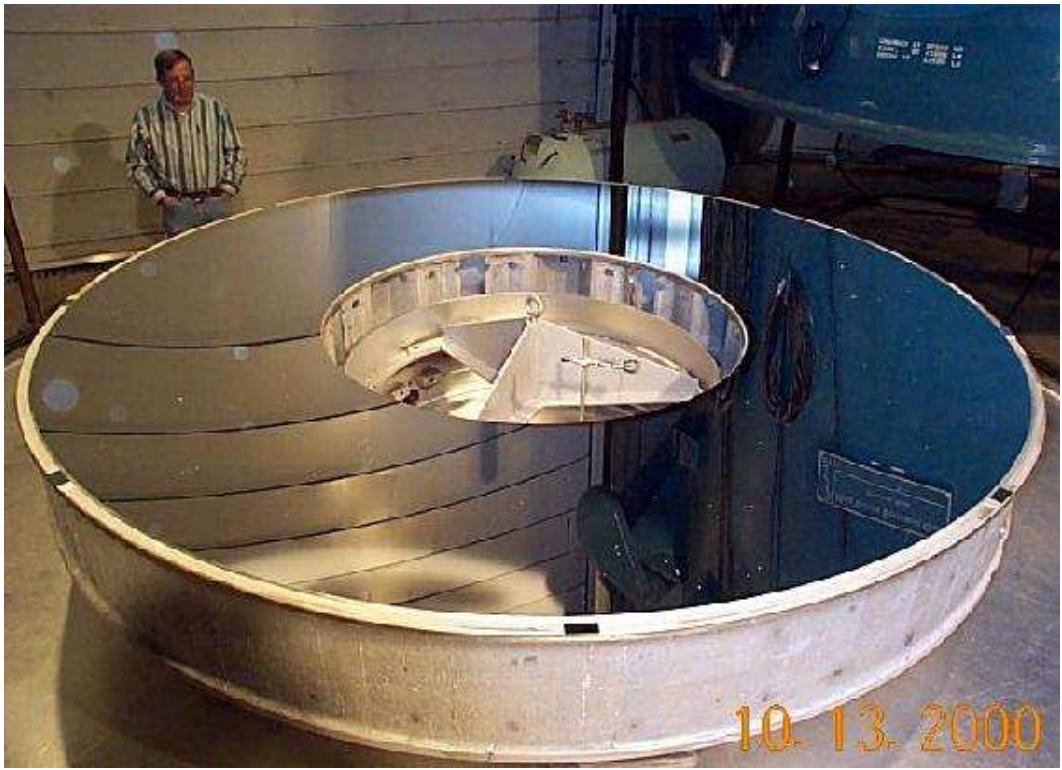}
\figcaption{SDSS 2.5 m primary mirror immediately after re-aluminizing
at the laboratories of Kitt Peak National Observatory, October 13
2000. The first science-grade data were taken after this. The 
very large central hole is necessary to accomodate the large field,
but is even somewhat larger than originally planned (see \S6.1).
\label{fig8}
}
\end{figure}
\clearpage
%
\begin{figure}[tb]
\figurenum{9}
\dfplot{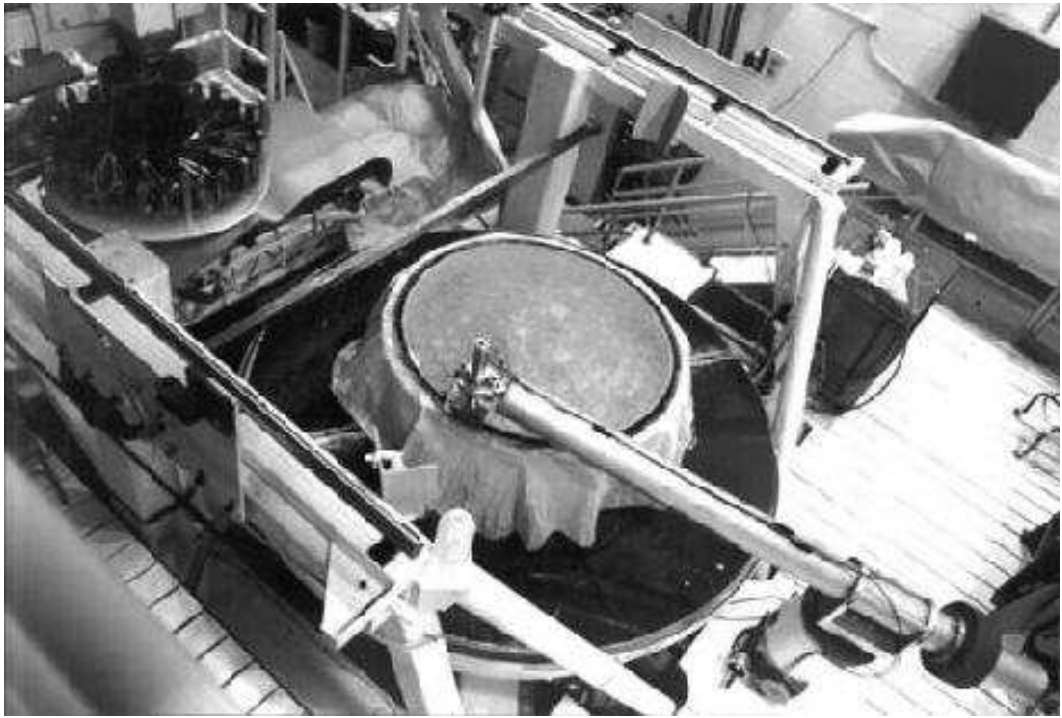}
\figcaption{The secondary mirror of the SDSS 2.5 m telescope on the
polishing turntable at the Steward Observatory Mirror Laboratory. 
The swing-arm profilometer is seen at the lower right, and the stressed-lap
polisher at the upper left.
\label{fig9}
}
\end{figure}
\clearpage
%
\begin{figure}[tb]
\figurenum{10}
\epsscale{0.8}
\dfplot{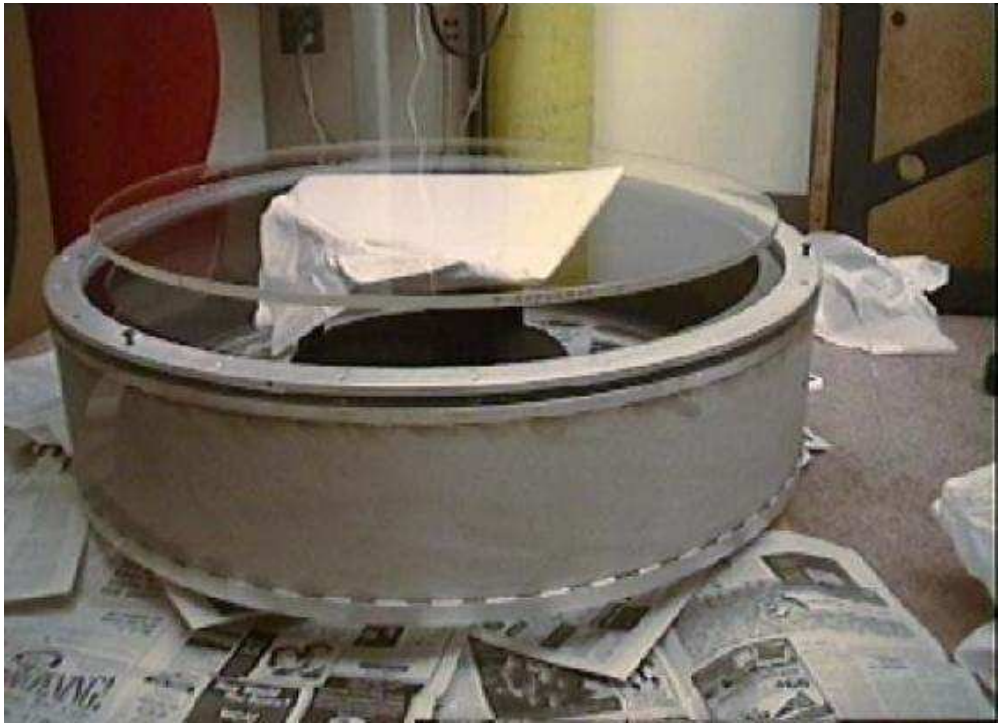}
\figcaption{The Gascoigne corrector of the SDSS telescope (also known
as the ``common corrector'' because it is used in both imaging and 
spectroscopic modes) about to be mounted in its cell. 
\label{fig10}
}
\end{figure}
\clearpage
%
\begin{figure}[tb]
\figurenum{11}
\epsscale{0.8}
\dfplot{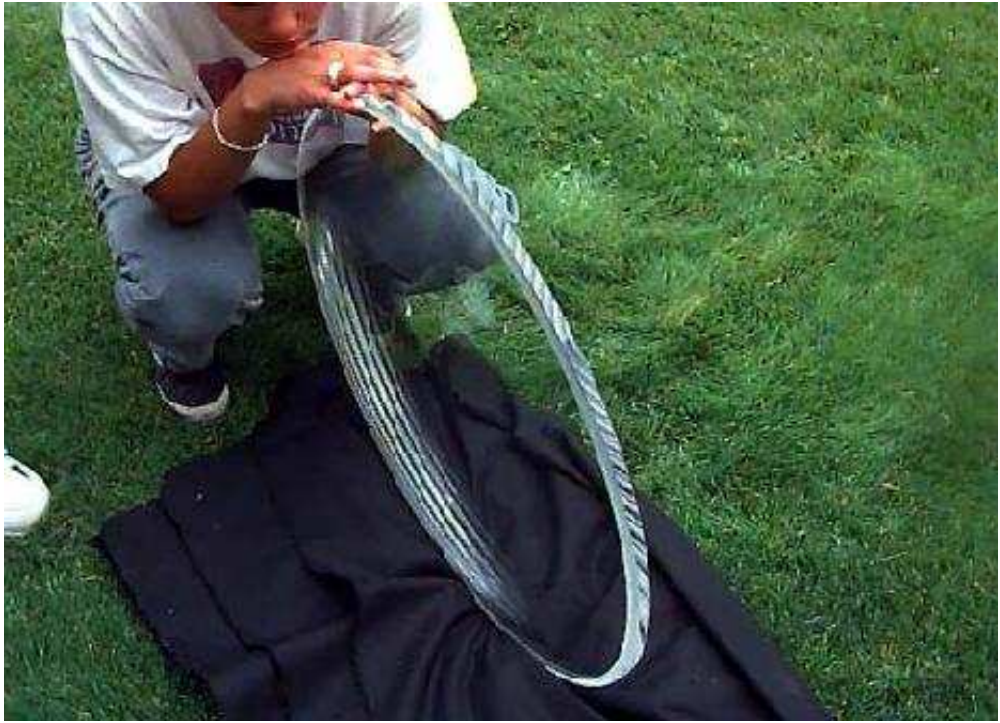}
\figcaption{Second component of the corrector for the SDSS telescope for
use in spectroscopic mode.  The strong banding seen is due to multiple
internal reflections. 
\label{fig11}
}
\end{figure}
\clearpage
%
\begin{figure}[tb]
\figurenum{12}
\epsscale{1.1}
\dfplot{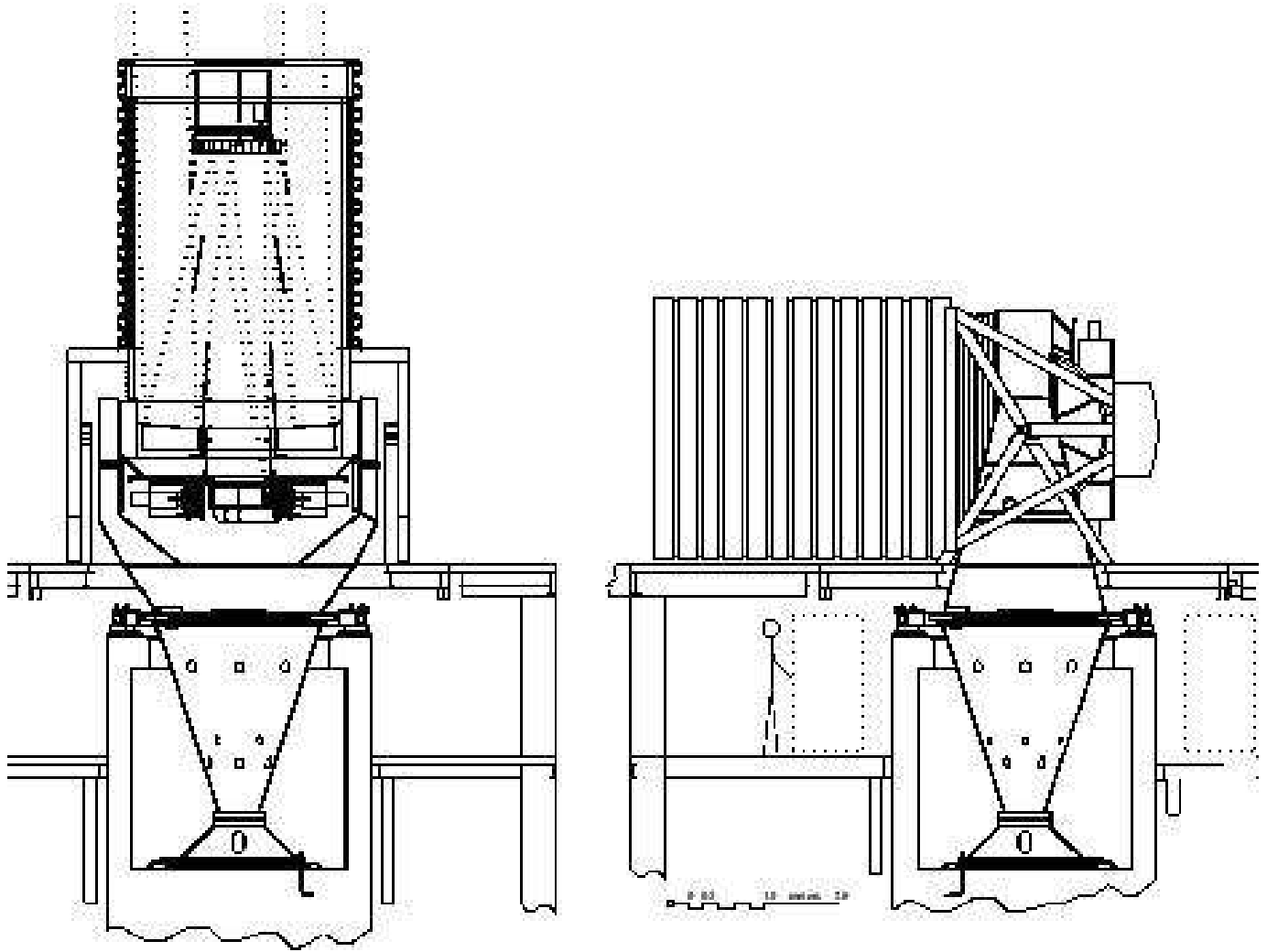}
\figcaption{Two views of the 2.5-meter telescope. The mechanical
design is essentially a scaled version of the WIYN 3.5-m instrument. The
wind baffle and the light baffles are shown in relation to rays from the
edge of the 3$^{\circ}$ field of view.
\label{fig12}
}
\end{figure}
\clearpage
%
\begin{figure}[tb]
\figurenum{13}
\dfplot{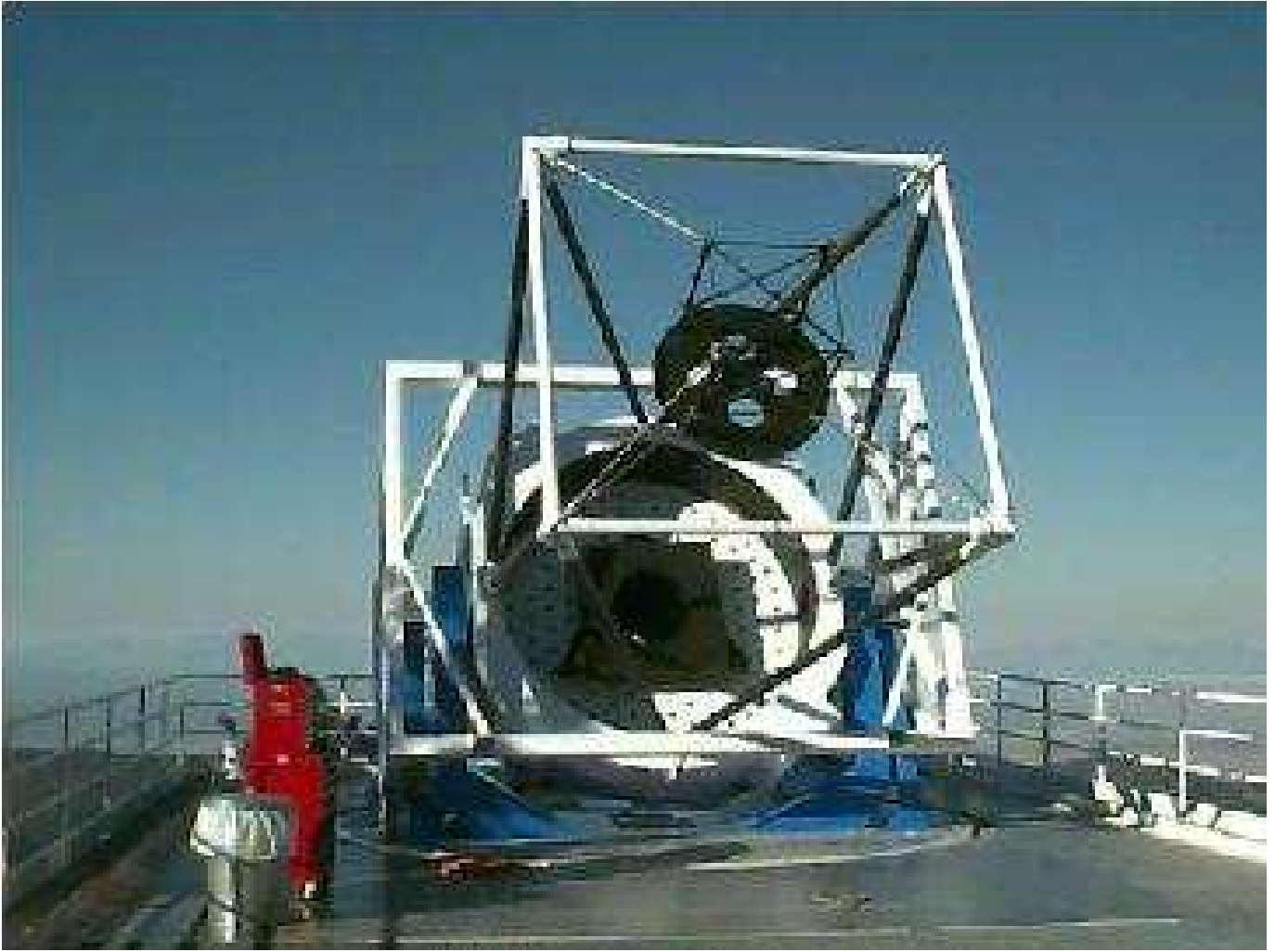}
\figcaption{The 2.5-m telescope as installed in October, 1995. Centered on the 
telescope azimuth axis is a circular floor panel that rotates with the 
telescope. Access to the telescope is along a horizontal ramp from the 
support building
through the telescope enclosure (at the right edge of the photo) which 
is in its open position.  The secondary truss (black) is graphite 
fiber epoxy, while the secondary frame (white) is steel.
\label{fig13}
}
\end{figure}
\clearpage
\begin{figure}[tb]
\figurenum{14}
\dfplot{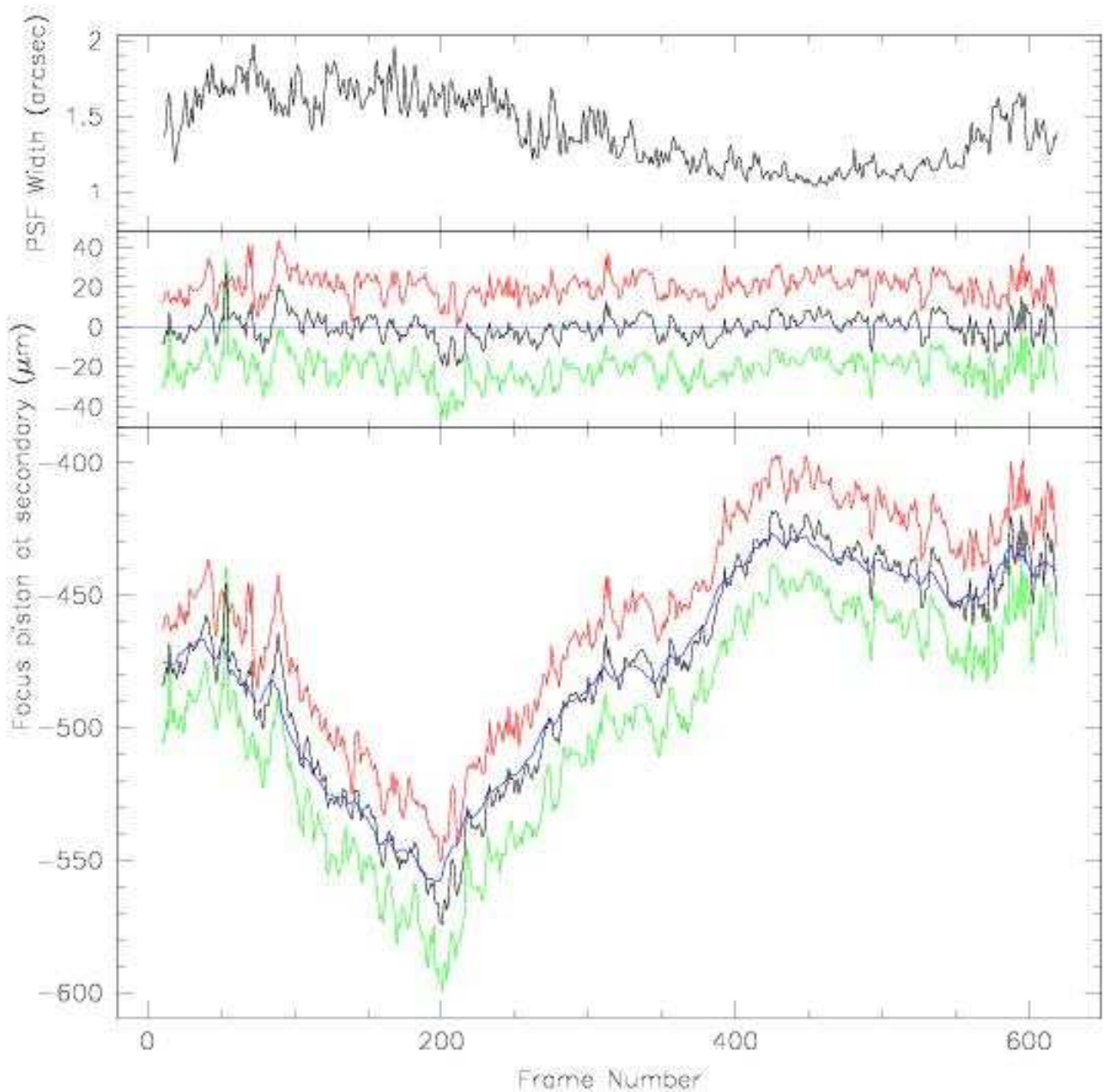}
\figcaption{The performance of the focus loop for run 3836 (MJD 52729). The 
abscissa covers about 6 hours of a single scan.
The bottom panel shows the estimated focus position from the leading 
(upper, red) and trailing (lower, green) focus CCDs.  Their average 
is shown in black, and the position of the telescope focus is shown in blue.
The middle panel is similar to the lower panel, but now only the 
estimated focus error is shown.
A focus error at the secondary of 10$\mu$m translates
to about 40$\mu$m at the focal plane.  With an $f/5$ beam and a scale 
of 60.6 $\mu$m/arcsec,
this corresponds to adding a component with equivalent Gaussian FHWM of 
0.31'' to the seeing disk.
The upper panel shows the FHWM of the seeing, as measured from the r3 
CCD.
}
\label{fig14}
\end{figure}
\clearpage
\begin{figure}[tb]
\figurenum{15}
\dfplot{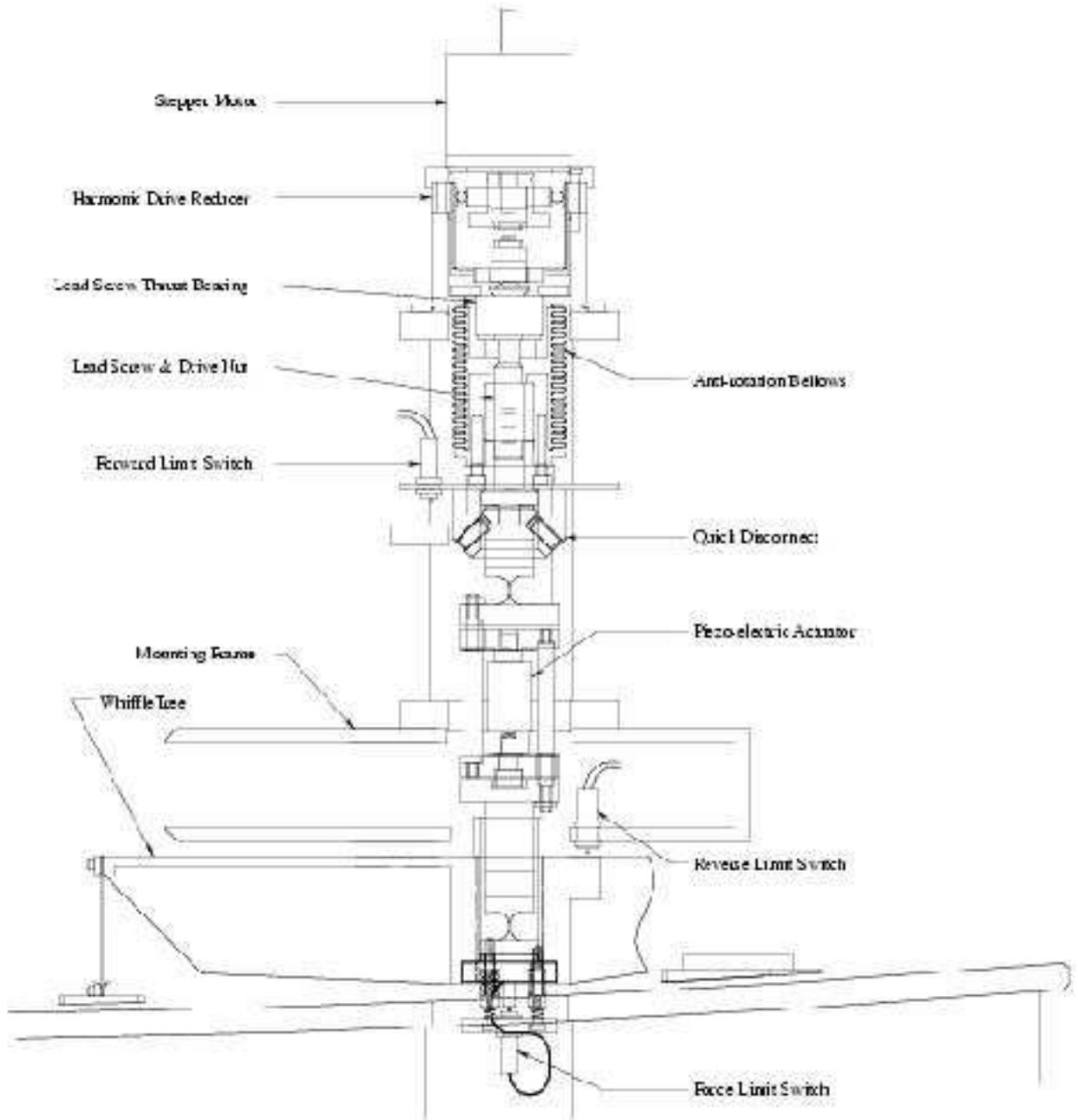}
\figcaption{Secondary mirror axial actuator assembly
\label{fig15}
}
\end{figure}
\clearpage
\begin{figure}[tb]
\figurenum{16}
\epsscale{0.9}
\dfplot{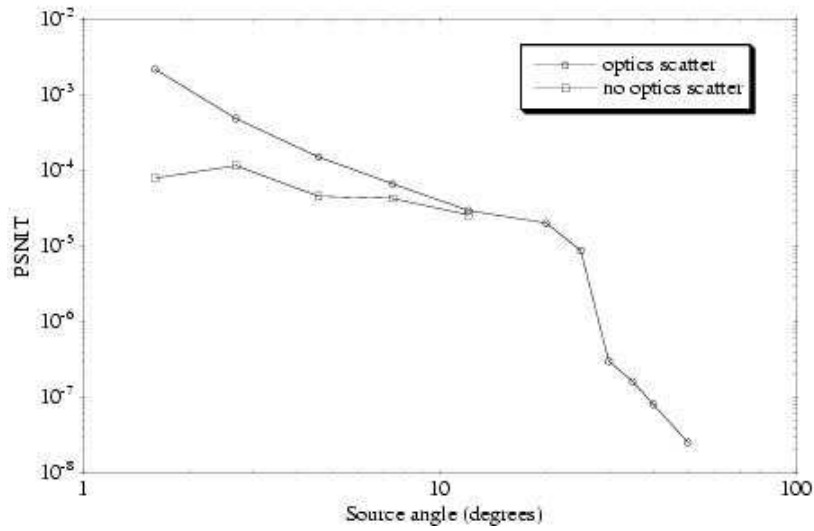}
\figcaption{Calculated point source normalized irradiance transmission (PSNIT)
versus source off-axis angle for the fully baffled SDSS 2.5 m telescope.
The two curves for angles less than $\rm 20^{\circ}$ show the illumination
with and without scattering from clean optics. The steep drop at $\rm
25^{\circ}$ is due to the source moving so far off axis that the 
aperture stop at the primary mirror is no longer directly illuminated.
\label{fig16}
}
\end{figure}
\clearpage
\begin{figure}[tb]
\figurenum{17}
\epsscale{0.9}
\dfplot{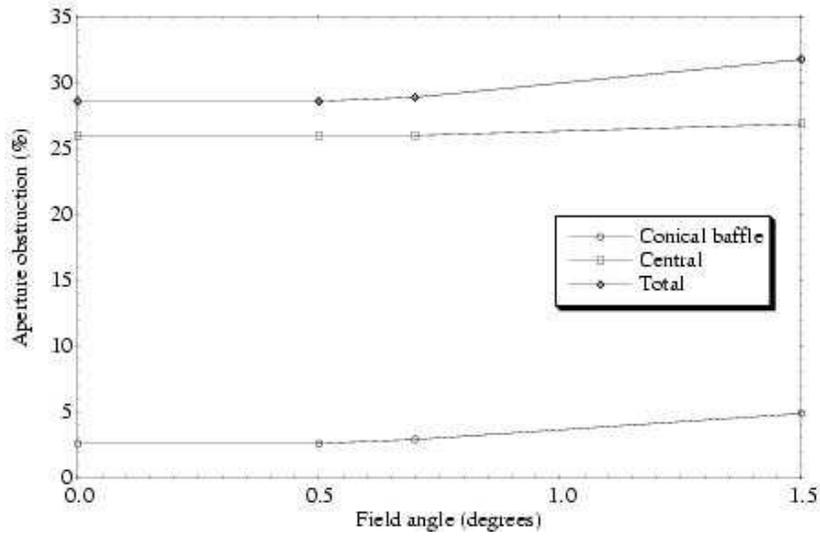}
\figcaption{Obstruction of the fully-baffled SDSS 2.5 m aperture. The conical 
baffle contributes most of the differential vignetting.
\label{fig17}
}
\end{figure}
\clearpage
\begin{figure}[tb]
\figurenum{18}
\epsscale{0.3}
\dfplot{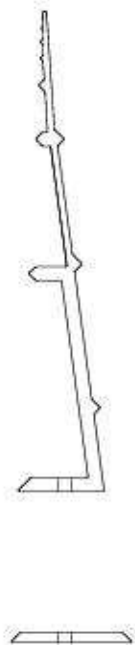}
\figcaption{Section through the right side of the tip of the primary
baffle. The tip is a machined aluminum alloy truncated cone with vanes.
It is 858 mm in diameter at its upper edge and 222 mm high. Below the
tip are a series of progressively larger aluminum annuli each 4.8 mm
thick.
\label{fig18}
}
\end{figure}
\clearpage
\begin{figure}[tb]
\figurenum{19}
\epsscale{0.9}
\dfplot{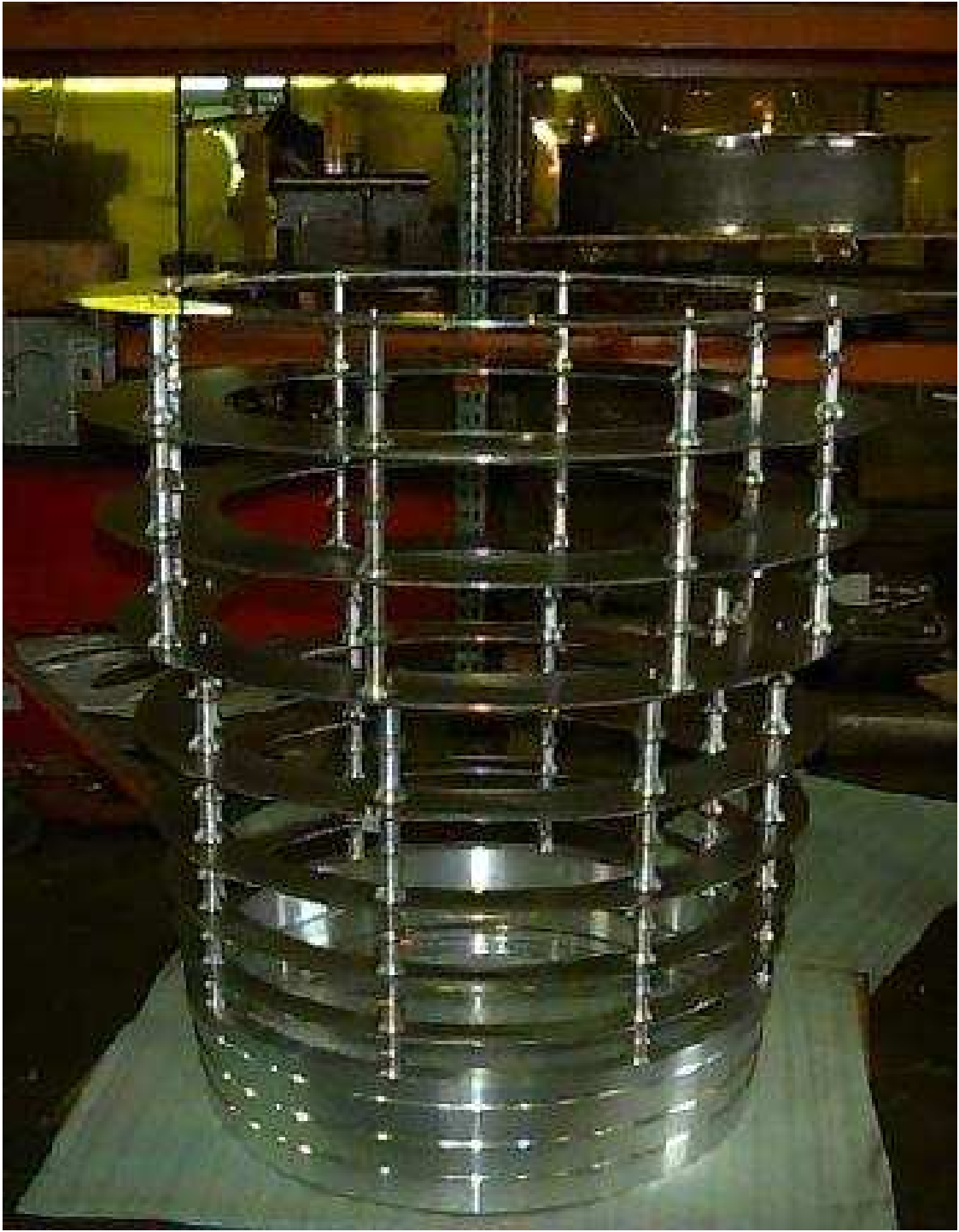}
\figcaption{The assembled primary baffle (here upside-down in the shop) 
for the SDSS 2.5 m telescope
before painting.
\label{fig19}
}
\end{figure}
\clearpage
\begin{figure}[tb]
\figurenum{20}
\epsscale{0.2}
\dfplot{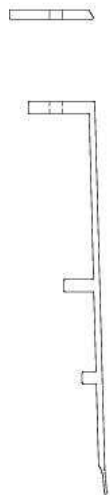}
\figcaption{Section through the right side of the secondary baffle showing
the baffle tip with the vanes that are necessary to interrupt grazing 
scattered light paths.  The baffle tip, machined from aluminum alloy, is 
1285 mm in outside diameter and 206 mm high.  Above the tip are a series of 
smaller diameter aluminum annuli each 4.8 mm thick.
\label{fig20}
}
\end{figure}
\clearpage
\begin{figure}[tb]
\figurenum{21}
\epsscale{0.6}
\dfplot{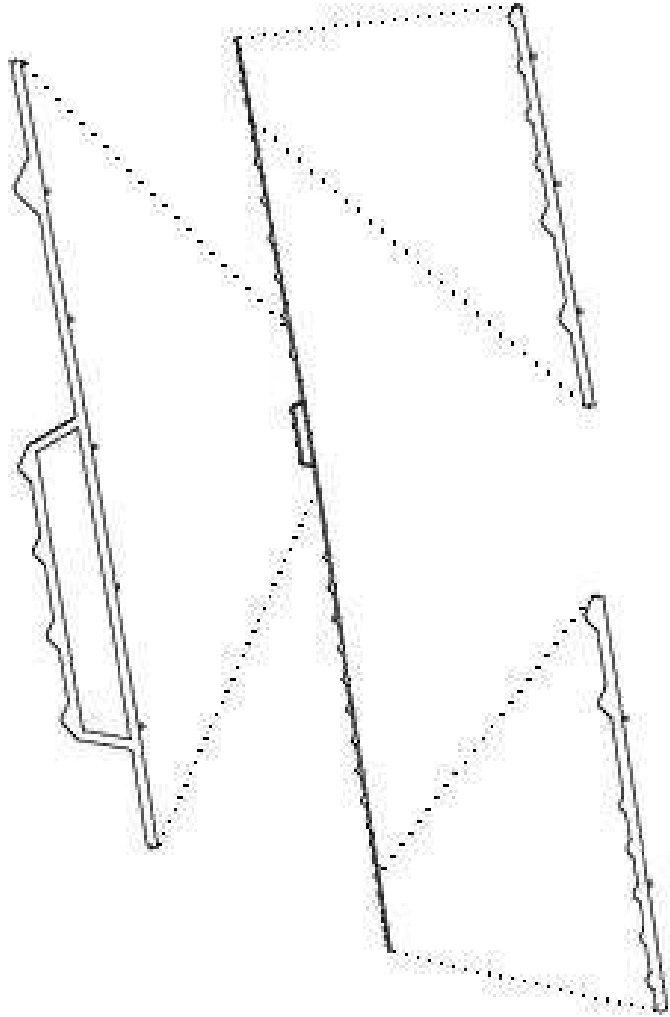}
\figcaption{Section through the right side of the conical baffle. The vanes
are necessary to interrupt grazing scattered light. The details are magnified
by a factor of 5.  The deep central vane provides attachment points and
bending stiffness. The vanes are small near the tips to minimize field
edge vignetting. The baffle is made of graphite fiber epoxy, has an
outside diameter of 1239 mm, is 725 mm high, and weighs 10 kg. 
\label{fig21}
}
\end{figure}
\clearpage
\begin{figure}[tb]
\figurenum{22}
\epsscale{0.9}
\dfplot{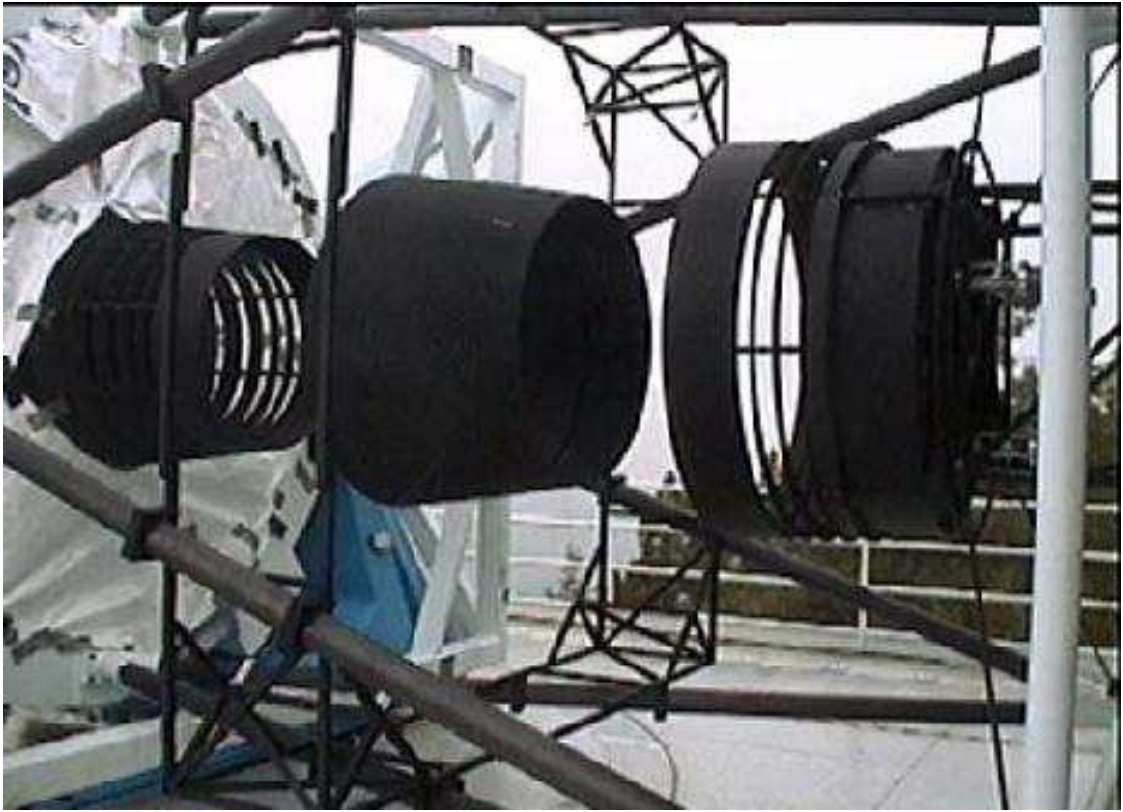}
\figcaption{The completed internal baffling system installed in the
SDSS telescope.  From left to right: primary baffle, conical baffle and
secondary baffle.
\label{fig22}
}
\end{figure}
\clearpage
\begin{figure}[tb]
\figurenum{23}
\epsscale{1.1}
\dfplot{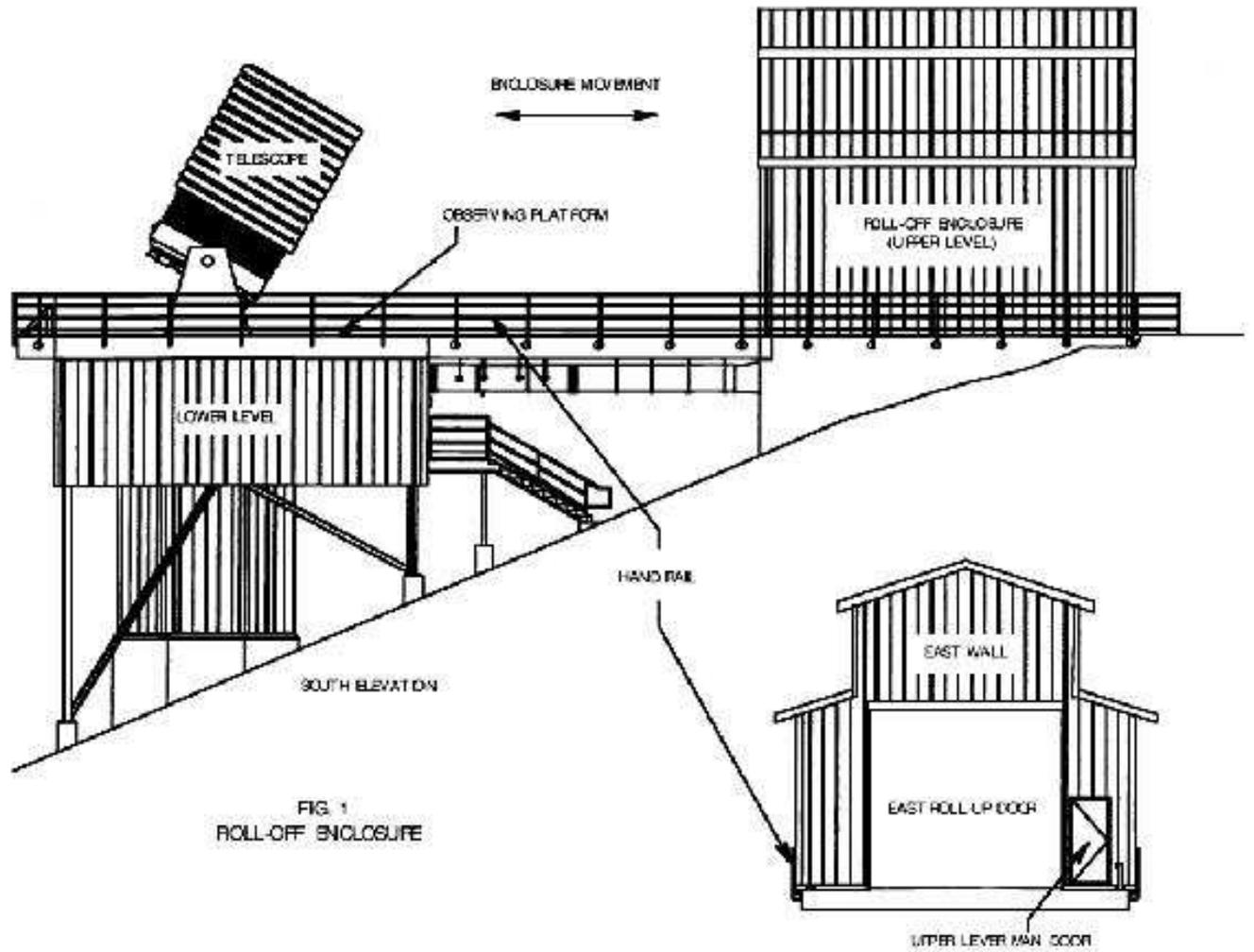}
\figcaption{The SDSS telescope and the new enclosure.
\label{fit23}
}
\end{figure}
\clearpage
\begin{figure}[tb]
\figurenum{24}
\epsscale{0.9}
\dfplot{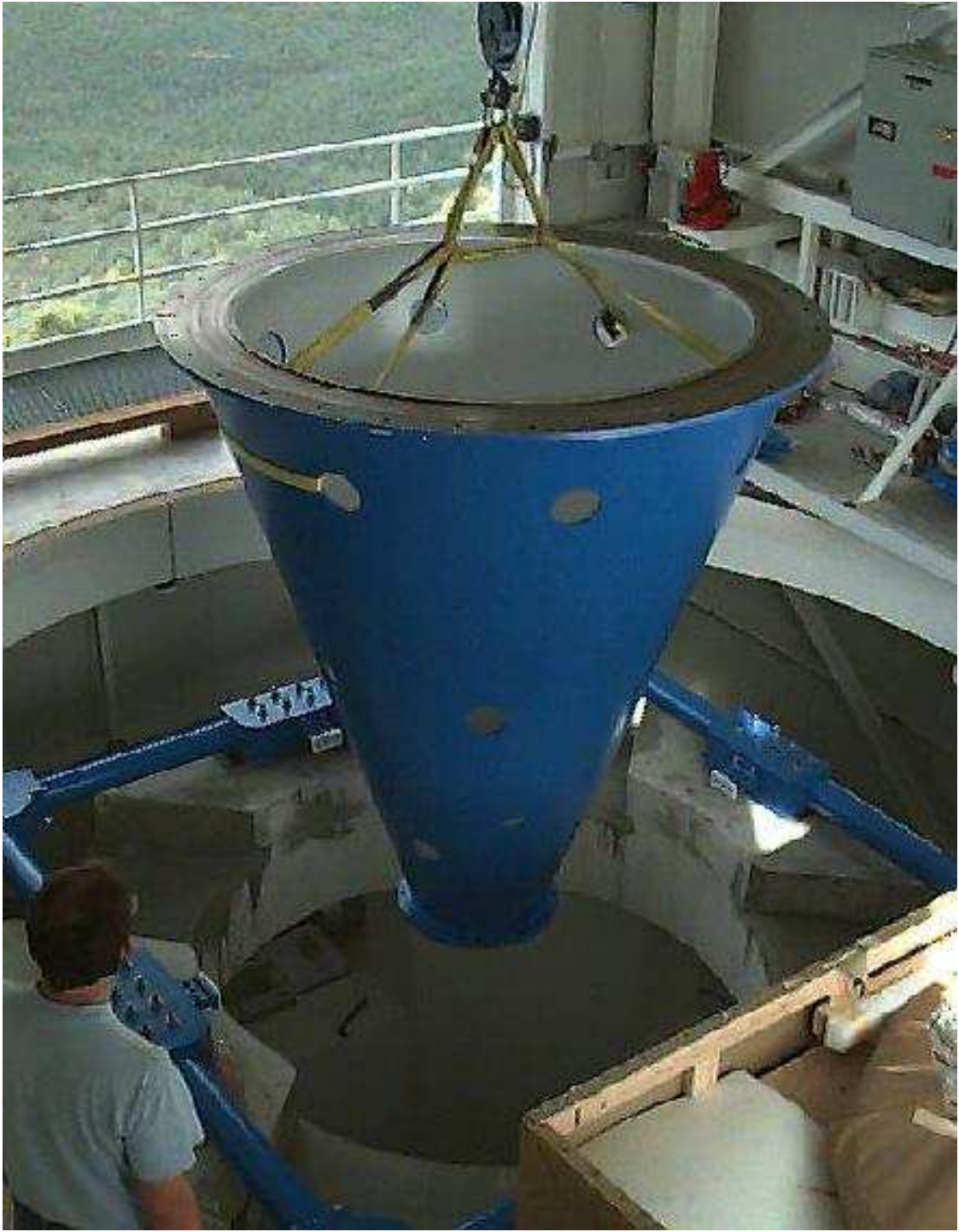}
\figcaption{Installation of the SDSS 2.5 m telescope begins in earnest.
On August 15 1995 the azimuth bearing and support is lowered into
place through the floor of the support platform (see Figure~12).
\label{fig24}
}
\end{figure}
\clearpage
\begin{figure}[tb]
\figurenum{25}
\epsscale{0.9}
\dfplot{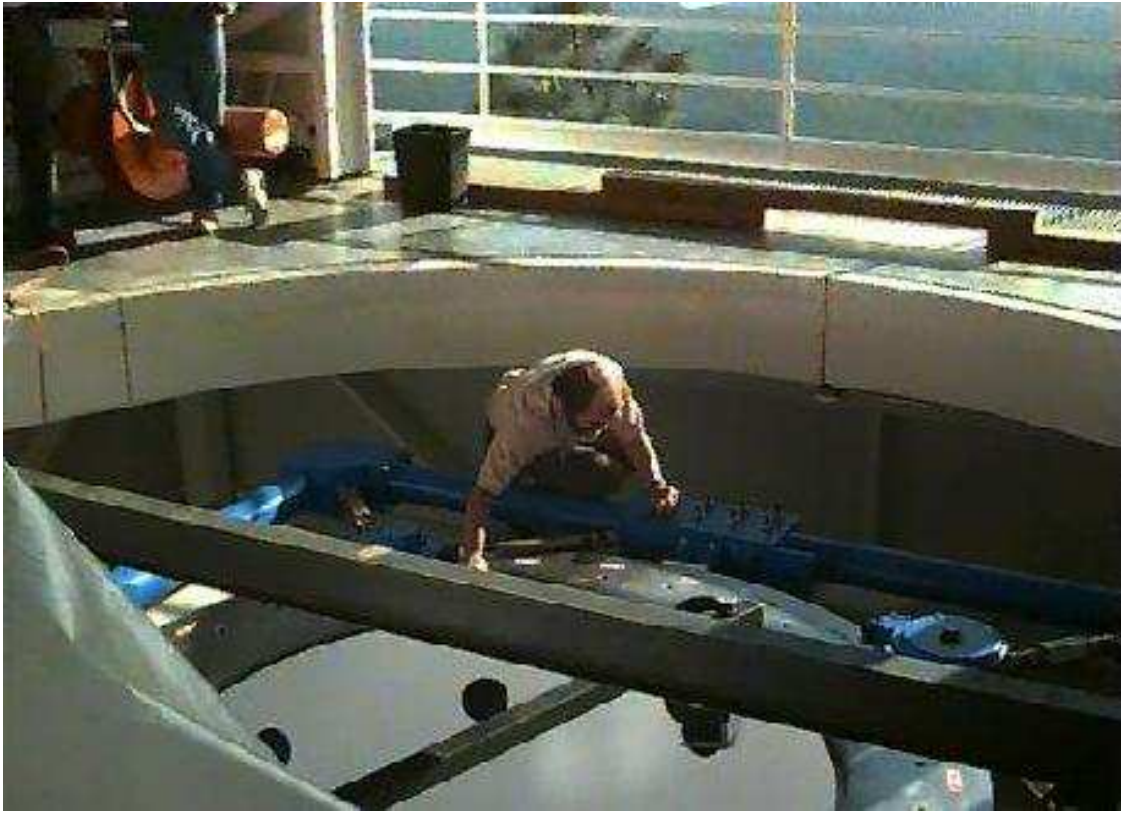}
\figcaption{The azimuth support is installed and tested. This image shows
the inspection of the azimuth drive disk.
\label{fig25}
}
\end{figure}
\clearpage
\begin{figure}[tb]
\figurenum{26}
\epsscale{0.9}
\dfplot{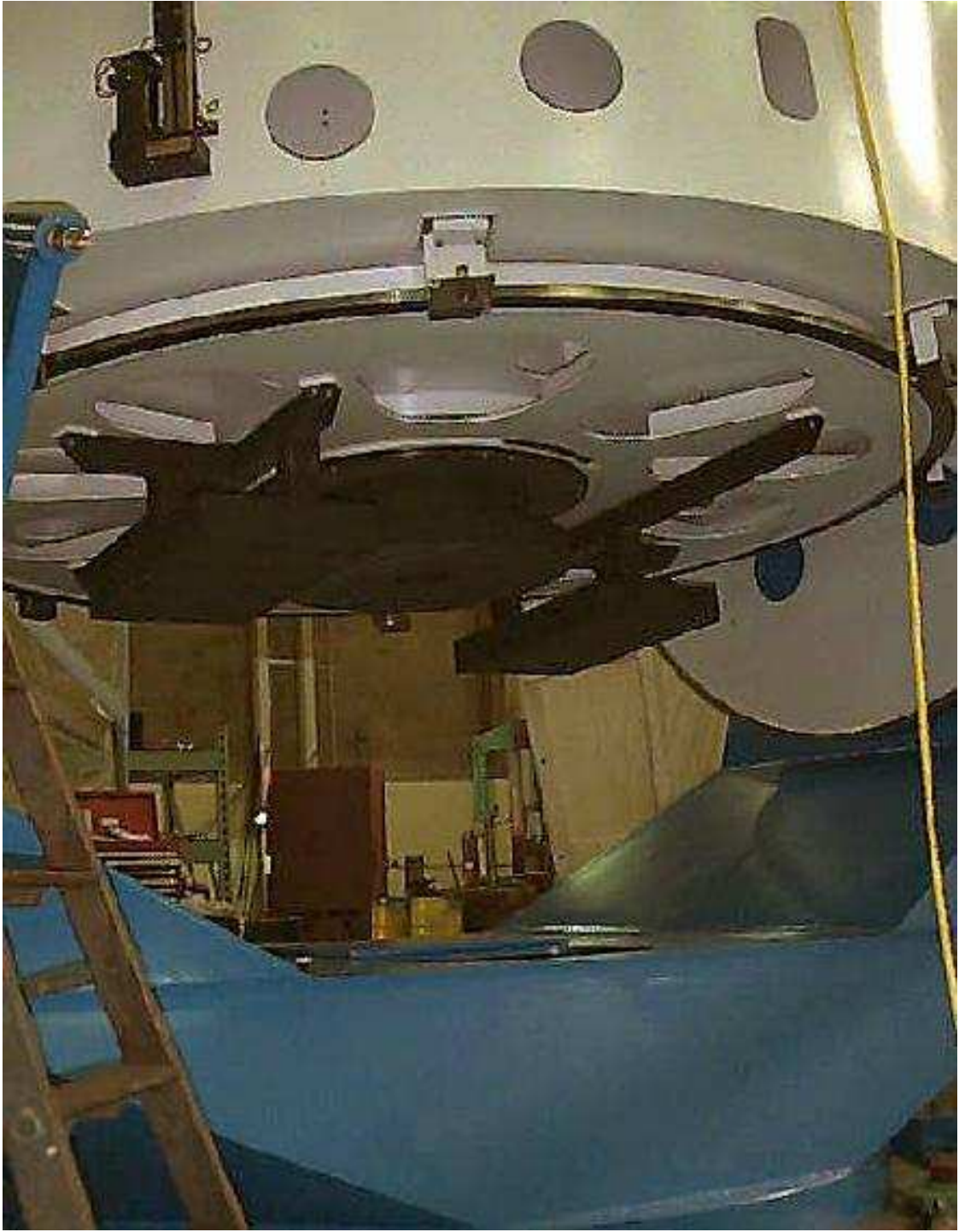}
\figcaption{Installation of the image rotator on the back of the SDSS
telescope, under the primary mirror and covering the rear of the mirror
cell. The camera mounts to the central
hole during imaging. The spectrographs are permanently mounted on
the back of the image rotator (the black mounting fixtures are seen here)
and a spectroscopic plug plate replaces the camera in spectroscopic 
observing mode.
\label{fig26}
}
\end{figure}
\clearpage
\begin{figure}[tb]
\figurenum{27}
\epsscale{0.9}
\dfplot{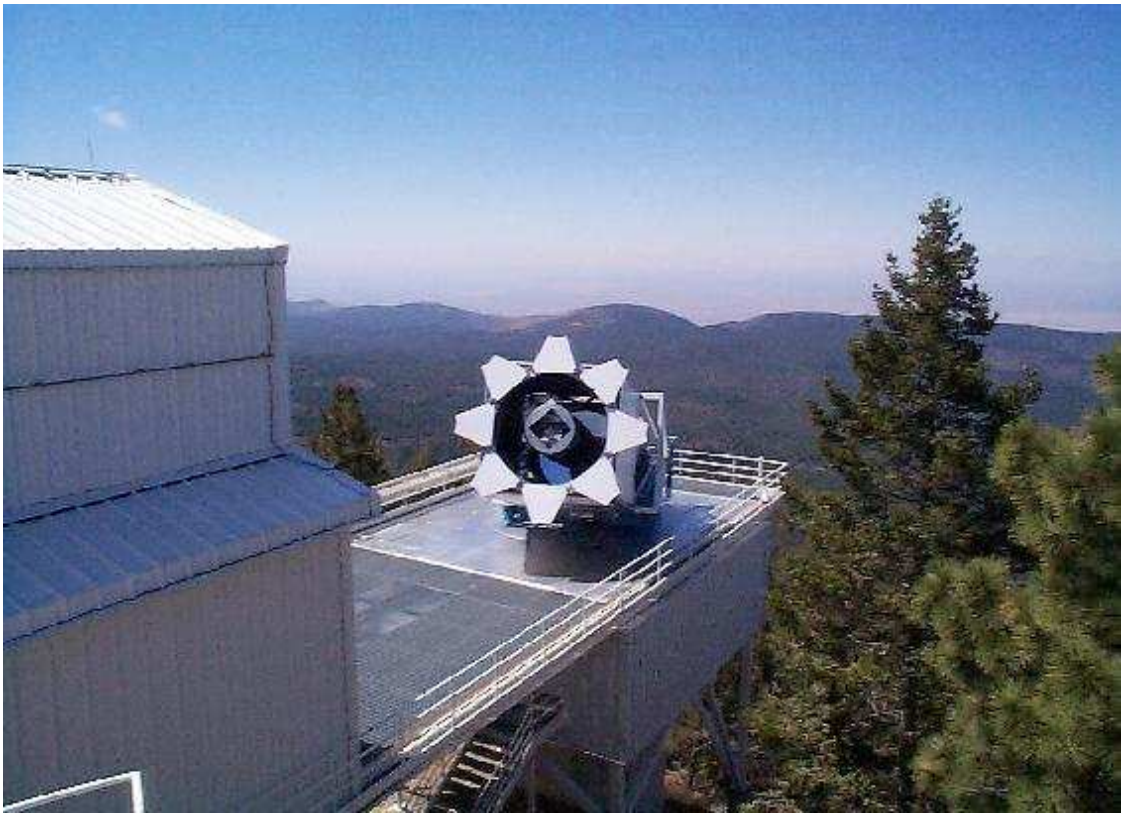}
\figcaption{The SDSS telescope in the open.
The rebuilt enclosure is seen on the left. 
\label{fig27}
}
\end{figure}
\clearpage
\end{document}